\documentclass[aps,prd,superscriptaddress,nofootinbib,longbibliography]{revtex4-2}
\usepackage{amsmath,amssymb,graphicx,bm}
\usepackage{tikz}
\usepackage{pgfplots}
\usetikzlibrary{patterns,positioning,decorations.markings}
\pgfplotsset{compat=1.17}
\usepackage[colorlinks=true,linkcolor=blue,citecolor=blue,urlcolor=blue,breaklinks=true]{hyperref}

\newcommand{\e}{\varepsilon}
\newcommand{\MS}{\overline{\rm MS}}
\newcommand{\MZ}{M_Z}
\newcommand{\Imt}{{\rm Im}\,\tau}
\newcommand{\Ret}{{\rm Re}\,\tau}

\begin{document}

\title{Level-Four Modular Symmetry Selected by the Harmonic Pattern
  of Quark and Lepton Masses}

\author{Vernon Barger}
\affiliation{Department of Physics, University of Wisconsin--Madison,
Madison, WI 53706, USA}

\date{September 8, 2026}

\begin{abstract}
The mass ratios of the quarks and charged leptons follow a simple
pattern. The logarithmic generation steps come in a two to one ratio,
a note to its octave, and the ordering is mirrored between the up-down
and down-lepton comparisons. The pattern, read as charge counting on a
$Z_N$ clock, is realized at level four and at no level that is not a
multiple of four, which motivates modular level four and the finite
group $S_4$, with the generations in the triplet at $T$ charges
$(3,1,0)$ and the mirror in the sign singlet. The charged-lepton mass
ratios lie on a theta-dressed lattice at $0.02\%$, which sets the base
$\e=0.18664$, taken as $14/75$, and the harmonic modulus
$\tau=1.0685i$. Five postulates in the weight one-half theta
constants, identified on the same data, reproduce the six quark masses
from three lepton masses and the electroweak scale. Two more and one
empirical relation for $|V_{td}|$, itself equivalent to a half-unit
insertion, close the CKM matrix, giving
$|V_{us}|=\sqrt{m_d/m_s}=0.2255$, $|V_{cb}|=0.04197$,
$|V_{ub}|=0.003721$, and $\delta=1.138$ against the measured
$1.139\pm0.023$, with no continuous parameter. The unitarity triangle
follows, with apex $(0.161,0.348)$ against the measured
$(0.161\pm0.010,0.347\pm0.010)$, and $J=3.11\times10^{-5}$ against
$(3.09\pm0.07)\times10^{-5}$. The same structure carries the
neutrinos. A single theta insertion gives
$\sin\theta_{13}=\theta_2(\tau)m_2/m_3$, which holds at $0.7\sigma$ on
the 207-day JUNO data, and the harmonic condition completes a
normal-ordered spectrum with $m_1=0.25$~meV, $\Sigma m_\nu=0.0589$~eV,
a CP-conserving phase $\delta_{CP}=\pi$, and effective Majorana masses
quantized at $1.3$ to $3.9$~meV. The ultraviolet completion belongs to
the magnetized torus class, with Pati--Salam the natural gauge
embedding.
\end{abstract}

\maketitle

\section{Introduction}
\label{sec:intro}

Take the nine charged-fermion masses at the $Z$ scale and form
their double ratios.  These combinations survive every rescaling of
a sector or a generation.  What remains is the part of the mass
matrix that no choice of units can touch.  Each double ratio is a
plaquette on the sector-generation grid, the four masses at the
corners of a rectangle with one diagonal pair over the other, and
the plaquette is the rescaling-invariant object because every row
and column factor cancels around it.

These plaquettes are the two-over-two ratios of
Ref.~\cite{Barger:2026tot}, where they organize the quark and
charged-lepton masses at $\MZ$ on a lattice of ninths in powers of
$B=75/14$, the inverse of the base $\e$ used here, within a
single-flavon Froggatt--Nielsen framework~\cite{Froggatt:1978nt}
in which each Yukawa entry carries an $O(1)$ coefficient left free.  That work supplies
the phenomenological ground of the present one, the same lattice,
the same base, and the same running masses.  It leaves open the
origin of the lattice, the values of the coefficients, which there
are fitted, and the two to one pattern the lattice contains but
does not explain.  The present paper closes those gaps by replacing
the flavon with a modulus.  The coefficients become values of
modular forms fixed by the modulus, the lattice becomes a set of
predictions, and the pattern selects the symmetry.  A
subconstituent reading of the same lattice, in which each unit of
charge counts one scalar exchanged along a messenger chain, is
developed in Ref.~\cite{Barger:2026sub}.

The result is a pattern.  The generation steps come in a ratio of
two to one on both of the independent logarithmic axes, and the ordering flips
between them.  A two to one ratio of intervals is the ratio of a
note to its first harmonic, an octave, so we call the pattern
harmonic.

A pattern this clean invites an arithmetic reading.  Read as charge
counting on a $Z_N$ clock, with the two axes related by a uniform
shift, the pattern is realized at level four and at no level that is
not a multiple of four.  That reading fixes the symmetry
group, the representation of the generations, and the identity of
the mirror, and Sec.~\ref{sec:pattern} states both the reading and
its hypotheses.  The charged-lepton masses then fix the two
remaining constants of the framework, a rational base and a purely
imaginary modulus, at the $0.02\%$ precision of the lepton data.

The rest is built on eight postulates, seven of them theta-constant
insertions identified on the data of Table~\ref{tab:data} and one an
adopted empirical relation.  The six quark masses are reproduced
from the three lepton masses and the electroweak scale.  The full CKM
matrix follows from the quark masses, including the CP phase and
the unitarity triangle.  The reactor angle follows from the
neutrino mass ratio, the neutrino spectrum closes with the same two
to one step that names the model, and the lepton CP phase is fixed
at $\pi$.  Each claim is a number against a measurement, and the
comparisons are collected in the tables and figures throughout.

The paper is organized to be read straight through, in four
parts.  Sections~\ref{sec:theory} to~\ref{sec:base} set up the
theory and its inputs: level-four modular symmetry from the modulus
to the metaplectic cover on which the theta constants are defined, so that
what follows is the implementation of a definite theory rather than
a fit; the pattern and its reading; the modulus with its Jacobi
insertions; and the base.  Sections~\ref{sec:masses}
and~\ref{sec:ckm} deliver the quark sector, masses and then the CKM
matrix.  Section~\ref{sec:leptons} delivers the neutrinos.
Sections~\ref{sec:evidence} to~\ref{sec:uv} weigh the evidence,
list the tests, and address the ultraviolet completion.
Throughout, masses are $\MS$ running values at $\MZ$ from
Ref.~\cite{Antusch:2025rqp}, collected in Table~\ref{tab:data},
particle properties and world averages
follow the 2026 Review of Particle Physics~\cite{PDG:2026}, the
unitarity-triangle comparisons use the UTfit global
analysis~\cite{UTfit:2022hsi}, with the four CKM parameters and
the Jarlskog invariant taken in its 2023 update as compiled in
Ref.~\cite{Antusch:2025rqp}, and the oscillation
inputs are NuFIT~6.1~\cite{Esteban:2024eli} with the 207-day JUNO
determinations~\cite{JUNO:2025,JUNO:DPF2026}.  Deviations quoted
throughout are measured minus predicted in units of the
uncertainty, computed on the corresponding ratios with correlated
uncertainties.

\begin{table}[t]
\caption{Charged-fermion Yukawa couplings and running masses at
$\MZ$ from Ref.~\cite{Antusch:2025rqp}.  The single-mass
uncertainties shown are not independent, and the analysis
propagates the correlated inputs, so uncertainties on mass ratios
differ from naive combinations of these entries.  The Yukawa
uncertainties of the reference are rounded to two figures; the
propagated uncertainty on $m_s$ used in the deviations quoted
throughout is $0.58$~MeV, below the rounded $0.70$~MeV shown
here.}
\label{tab:data}
\begin{ruledtabular}
\begin{tabular}{lll}
Fermion & $y_f(\MZ)$ & $m_f(\MZ)$\\
\hline
$u$    & $(7.04\pm0.15)\times10^{-6}$      & $1.237\pm0.026$~MeV\\
$c$    & $(3.56\pm0.06)\times10^{-3}$      & $625.3\pm10.5$~MeV\\
$t$    & $0.967\pm0.004$                   & $169.85\pm0.70$~GeV\\
$d$    & $(1.54\pm0.02)\times10^{-5}$      & $2.705\pm0.035$~MeV\\
$s$    & $(3.06\pm0.04)\times10^{-4}$      & $53.75\pm0.70$~MeV\\
$b$    & $(1.630\pm0.009)\times10^{-2}$    & $2.863\pm0.016$~GeV\\
$e$    & $(2.77713\pm0.00036)\times10^{-6}$ & $0.487792\pm0.000089$~MeV\\
$\mu$  & $(5.85042\pm0.00075)\times10^{-4}$ & $102.760\pm0.019$~MeV\\
$\tau$ & $(0.99378\pm0.00014)\times10^{-2}$ & $1.74554\pm0.00033$~GeV\\
\end{tabular}
\end{ruledtabular}
\end{table}

\section{Level-four modular symmetry}
\label{sec:theory}

The construction rests on level-four modular symmetry.  This
section sets out the framework before the data enter.  It is
self-contained for a reader who knows quantum field theory but not
the modular approach, and it follows the order in which the ideas
are used later, from the modulus to the finite group $S_4$, to
modular forms as couplings, to the half-integral weights of the
metaplectic cover and the theta constants that supply the
phenomenology.  The modular approach to
flavor~\cite{Feruglio:2017spp}, reviewed in
Refs.~\cite{KobayashiTanimoto,DingKing}, attributes the structure
of the fermion masses and mixings to a single complex field $\tau$,
the modulus, together with a discrete symmetry that acts on $\tau$
and on the generations at the same time; once the value of $\tau$
is known, every symmetry-controlled coupling is a computable number
times an $O(1)$ coefficient, and the construction of this paper sets
that coefficient to one.  A reader fluent in modular flavor symmetry can pass
to Sec.~\ref{sec:pattern}, returning to Table~\ref{tab:charges} for
the $T$-charge content of the $S_4$ representations and to
Sec.~\ref{sec:metaplectic} for the metaplectic conventions.

\subsection{The modular group and the modulus}
\label{sec:modular}

The modulus $\tau$ takes values in the upper half of the complex
plane, $\Imt>0$.  The modular group $SL(2,\mathbb{Z})$, the group of
$2\times2$ integer matrices of unit determinant, acts on it by
fractional linear transformations,
\begin{equation}
  \tau \;\to\; \gamma\tau = \frac{a\tau+b}{c\tau+d},
  \qquad
  \gamma=\begin{pmatrix} a & b\\ c & d\end{pmatrix},
  \quad ad-bc=1 .
\label{eq:mobius}
\end{equation}
Two elements generate the whole group,
\begin{equation}
  S:\ \tau\to-\frac{1}{\tau},
  \qquad
  T:\ \tau\to\tau+1 ,
\label{eq:ST}
\end{equation}
an inversion and a unit translation.  Since $\gamma$ and $-\gamma$
act identically on $\tau$, the group acting faithfully on the
modulus is the projective quotient $\bar\Gamma=PSL(2,\mathbb{Z})$.
In string compactifications $\tau$ is the complex-structure
parameter of a torus, the shape of the compact space, and
Eq.~(\ref{eq:mobius}) relates tori of identical geometry.  For
phenomenology one may equally treat $\tau$ as a spurion whose vacuum
value breaks the symmetry.  Every point of the upper half plane is
equivalent under Eq.~(\ref{eq:mobius}) to one point of a
fundamental domain, the familiar keyhole region $|\tau|\ge1$,
$|\Ret|\le\tfrac12$.

\subsection{From the infinite group to a finite one}
\label{sec:finite}

The framework takes the matter fields to transform under a finite
quotient of the modular group rather than under $SL(2,\mathbb{Z})$
itself, so that the generations fill a finite-dimensional
representation of a finite group.  For each integer $N\ge2$, the
principal congruence subgroup of level $N$ collects the matrices
congruent to the identity modulo $N$,
\begin{equation}
  \Gamma(N)=\left\{\gamma\in SL(2,\mathbb{Z})\ :\
  \gamma\equiv\begin{pmatrix}1&0\\0&1\end{pmatrix} \bmod N\right\} .
\label{eq:GammaN}
\end{equation}
Matter fields and modular forms of level $N$ are taken to transform
trivially under $\Gamma(N)$, up to the automorphy factor of
Sec.~\ref{sec:forms}.  What acts nontrivially on them is the finite
quotient
\begin{equation}
  \Gamma_N=\bar\Gamma/\bar\Gamma(N) ,
\label{eq:quotient}
\end{equation}
the finite modular group of level $N$.  The low levels reproduce
familiar permutation groups, $\Gamma_2\simeq S_3$ of order six,
$\Gamma_3\simeq A_4$ of order twelve, $\Gamma_4\simeq S_4$ of order
twenty-four, and $\Gamma_5\simeq A_5$ of order sixty.  In terms of
the generators the level-four group is presented as
\begin{equation}
  S^2=(ST)^3=T^4=1 ,
\label{eq:presentation}
\end{equation}
and the relation $T^4=1$ is the fingerprint of the level.  The
translation $T$ becomes an element of order four, generating a $Z_4$
subgroup written $Z_4^T$ below (at level three the analogous
relation is $T^3=1$).  The level fixes the order of $T$, and much of
the arithmetic of a level-four model traces back to counting modulo
four.

Modular $S_4$ models were first developed for the lepton
sector~\cite{PenedoPetcov,NovichkovS4}.  Quark sectors were
addressed at level three~\cite{OkadaTanimotoQ}, joint descriptions
of quarks and leptons followed at level
three~\cite{OkadaTanimotoU,LuLiuDing} and, on the double cover
$S_4'$, at level four~\cite{NPPdouble,LiuYaoDingS4p}, with modular
weights serving as generators of hierarchy~\cite{KingKing} and
grand-unified embeddings in place~\cite{DingKingYao}.  Those
constructions fit $O(1)$ coefficients to the data.

The
construction of this paper differs in three declared respects.  The
data select the level
(Sec.~\ref{sec:pattern}), the forms carry half-integral weight so
that the theta constants themselves are the couplings
(Sec.~\ref{sec:metaplectic}), and every insertion enters with unit
coefficient, so the outputs are numbers, not fits.  The
lineage is continuous, since the weight-one forms of the double
cover are already built from two Jacobi theta
constants~\cite{NPPdouble}, and the metaplectic step of
Sec.~\ref{sec:metaplectic} completes it.

\subsection{\texorpdfstring{$S_4$ and the $Z_4$ of the $T$ generator}{S4 and the Z4 of the T generator}}
\label{sec:S4}

$S_4$ has five irreducible representations, two singlets
$\mathbf{1}$ and $\mathbf{1}'$, one doublet $\mathbf{2}$, and two
triplets $\mathbf{3}$ and $\mathbf{3}'$, with
$1^2+1^2+2^2+3^2+3^2=24$.  Three generations can sit in either
triplet or in $\mathbf{2}\oplus\mathbf{1}$, and the primed
representations track a $Z_2$ parity, the sign of the permutation.
The sign singlet $\mathbf{1}'$ takes the value $-1$ on odd
permutations, and $\mathbf{3}'=\mathbf{3}\otimes\mathbf{1}'$.  This
parity governs which couplings exist.  For example
$\mathbf{3}\otimes\mathbf{3}'$ contains $\mathbf{1}'$ but not
$\mathbf{1}$, so an invariant requires one further unit of the sign.

For level-four model building the most useful bookkeeping is the set
of $T$ eigenvalues within each representation.  Since $T^4=1$, every
eigenvalue is a fourth root of unity $e^{2\pi i c/4}$ with a charge
$c\in\{0,1,2,3\}$, and each representation has a definite
multiset of charges, listed in Table~\ref{tab:charges}.  Two
readings of the table matter here.  First, the four-object
permutation representation decomposes as $\mathbf{1}\oplus\mathbf{3}'$
with charges $\{0\}\cup\{1,2,3\}$, so the four charges of the $Z_4$
clock are distributed once each.  Second, the sign singlet
$\mathbf{1}'$ carries charge $2$, because a four-cycle is an odd
permutation, so multiplying any representation by $\mathbf{1}'$ shifts
every $T$-charge by $2$ modulo $4$, a half turn of the clock.
Triplet naming is convention dependent, both assignments of the
labels $\mathbf{3}$ and $\mathbf{3}'$ appear in the literature, and
the charge multisets themselves do not change.  This paper attaches
the label $\mathbf{3}$ to the charge set $\{0,1,3\}$.

\begin{table}[t]
\caption{$Z_4^T$ charges of the $S_4$ representations, i.e., the
eigenvalues $e^{2\pi ic/4}$ of the order-four generator $T$, in the
triplet-naming convention of this paper.}
\label{tab:charges}
\begin{ruledtabular}
\begin{tabular}{ll}
Representation & $T$-charges\\
\hline
$\mathbf{1}$  & $\{0\}$\\
$\mathbf{1}'$ & $\{2\}$\\
$\mathbf{2}$  & $\{0,2\}$\\
$\mathbf{3}$  & $\{0,1,3\}$\\
$\mathbf{3}'$ & $\{1,2,3\}$\\
\end{tabular}
\end{ruledtabular}
\end{table}

The triplet charge sets have a consequence that matters for
hierarchies.  The charges $\{0,1,3\}$ of $\mathbf{3}$ have
consecutive gaps of $1$ and $2$, so once the three generations are
assigned to a triplet, one adjacent generation step is twice the
other, a fixed consequence of the representation content rather
than a choice.  This is the two to one step that names the model,
and Sec.~\ref{sec:pattern} shows that the data select it, together
with its orientation and its mirror.

\subsection{Modular forms as Yukawa couplings}
\label{sec:forms}

The replacement of flavons by functions is the central move of the
framework.  In a conventional discrete-symmetry model the couplings
are constants and the symmetry is broken by scalar flavon fields
whose vacuum alignment must be arranged by a dedicated potential.
In a modular-invariant model the couplings are modular forms of
level $N$, holomorphic functions $Y(\tau)$ transforming with a
definite weight $k$ and a definite representation $\mathbf{r}$ of
$\Gamma_N$,
\begin{equation}
  Y(\gamma\tau)=(c\tau+d)^{k}\,\rho_{\mathbf{r}}(\gamma)\,Y(\tau) ,
\label{eq:formtrafo}
\end{equation}
where $(c\tau+d)^k$ is the automorphy factor and $\rho_{\mathbf{r}}$
is a representation matrix.  In a supersymmetric realization the
matter superfields have a weight, conventionally $-k_i$, and a
representation as well, and a superpotential term is invariant when
the weights sum to zero and the product of representations contains
a singlet.  The symmetry thereby dictates both which operators exist
and what functions of $\tau$ multiply them.

Two structural facts follow.  First, at fixed level and weight the
forms span a finite-dimensional space.  At level four the space of
weight-$k$ forms has dimension $2k+1$, so the five forms of weight
two arrange into a doublet and a triplet of $S_4$, written
$\mathbf{2}\oplus\mathbf{3}$ in the naming convention adopted here,
with leading $q$-powers spanning all four $Z_4^T$ charges.  Higher
weights are products of lower ones.  Second, each form component has
a $q$-expansion in the nome
\begin{equation}
  q=e^{2\pi i\tau},\qquad |q|=e^{-2\pi\Imt} ,
\label{eq:q}
\end{equation}
and a level-$N$ component of $T$-charge $c$ transforms under $T$ by
the phase $e^{2\pi ic/N}$, so its expansion opens at order
$q^{c/N}$.  A charge assignment is therefore a hierarchy
prediction.  A coupling that compensates a matter charge $c$ is
suppressed by $|q|^{c/N}$, powers of $|q|^{1/4}$ at level four, and
since $|q|$ is exponentially small already at moderate $\Imt$,
hierarchies among matrix elements arise from charge counting rather
than from tuned couplings, a modular analogue of the
Froggatt--Nielsen picture~\cite{Froggatt:1978nt} with the
$T$-charge in the role of the horizontal charge.

One further ingredient connects holomorphic forms to physical
couplings.  The K\"ahler potential contributes wavefunction
normalizations, and after canonical normalization a physical
coupling of total modular weight $w$ carries the real factor
\begin{equation}
  \bigl(2\,\Imt\bigr)^{w/2} ,
\label{eq:automorphy}
\end{equation}
one factor of $(2\,\Imt)^{1/2}$ for each unit of weight and
$(2\,\Imt)^{1/4}$ for each half unit, which compensates the
automorphy factor of Eq.~(\ref{eq:formtrafo}) and renders the
physical coupling invariant.  This is the coupling
dictionary of magnetized toroidal compactifications, where the
Yukawa couplings are theta constants dressed by exactly such
factors~\cite{Cremades:2004wa}.  In ratios of couplings the absolute
normalization cancels and weight differences survive as powers of
$(2\,\Imt)^{1/2}$, and at $\tau=i$ these are powers of $\sqrt2$, one
route by which the coefficients $2$ and $\sqrt2$ enter otherwise
parameter-free relations.  Equation~(\ref{eq:automorphy}) is a
statement at the level of such ratios.  In a supergravity
realization the K\"ahler potential of the modulus itself
contributes a further power of $(2\,\Imt)$ through $e^{K/2}$,
common to every coupling, which cancels from all ratios and so
from every mass ratio and mixing in this paper, and enters only an
absolute normalization such as the top anchor P4.  This is the
counting used in the fourth rule of Sec.~\ref{sec:rules}.

\subsection{Half-integral weight, the metaplectic cover and the
theta constants}
\label{sec:metaplectic}

The weight $k$ in Eq.~(\ref{eq:formtrafo}) need not be an integer,
and the construction of this paper uses the half-integral
extension.  The word metaplectic, which recurs throughout, names
that extension.  For half-integral $k$ the automorphy factor
requires a square root $(c\tau+d)^{1/2}$ whose sign is not fixed by
the matrix $\gamma$ alone.  The metaplectic group is the double
cover of $SL(2,\mathbb{Z})$ whose elements are the pairs
$\bigl(\gamma,\pm(c\tau+d)^{1/2}\bigr)$, so that the branch of the
root is carried as part of the group element and the composition of
half-integral automorphy factors is single valued.  The group and
its name are due to Weil, who introduced it for the unitary
representation of the symplectic group now called the Weil
representation~\cite{Weil:1964}.  The theory of modular forms of
half-integral weight on it is due to Shimura~\cite{Shimura:1973}.
Three consequences are used throughout.

First, modular forms of half-integral weight exist, and the
weight-one-half forms that concern us are the three Jacobi theta
constants~\cite{WhittakerWatson,Mumford:1983},
\begin{equation}
  \theta_2(\tau)=\!\sum_{n\in\mathbb{Z}}q^{(n+\frac12)^2/2},\quad
  \theta_3(\tau)=\!\sum_{n\in\mathbb{Z}}q^{n^2/2},\quad
  \theta_4(\tau)=\!\sum_{n\in\mathbb{Z}}(-1)^n q^{n^2/2},
\label{eq:thetadef}
\end{equation}
with leading expansions
\begin{equation}
\begin{aligned}
  \theta_2&=2\,q^{1/8}\bigl(1+q+q^3+\cdots\bigr) ,\\
  \theta_3&=1+2q^{1/2}+2q^2+\cdots ,\\
  \theta_4&=1-2q^{1/2}+2q^2-\cdots .
\end{aligned}
\label{eq:thetaexp}
\end{equation}

Second, the finite group acting on the generations is enlarged.
At level four the weight-one-half forms span a two-dimensional
space, the doublet built from $\theta_3(2\tau)$ and
$\theta_2(2\tau)$, transforming irreducibly under the finite
metaplectic group $\widetilde\Gamma_4\cong\widetilde S_4$ of order
$96$, in which the level-four framework with half-integral weights
is developed~\cite{Liu:2020msy}.  Metaplectic flavor groups arise
from magnetized torus compactifications, where the theta constants
are the Yukawa couplings
themselves~\cite{Almumin:2021fbk,Cremades:2004wa}.

Third, the
$T$-phase can carry order eight.  The constants at argument $\tau$
are the same doublet evaluated at the halved modulus, so their
expansions open at $q^{1/8}$, half the level-four unit $q^{1/4}$,
and an operator carrying one such insertion transforms under the
metaplectic cover at level eight, of order
$768$~\cite{Liu:2020msy}.  The $q^{1/8}$ is the leading power of
$\theta_2$ in Eq.~(\ref{eq:thetaexp}), and it arrives with the
exact leading coefficient $2$, supplied by the form rather than by
hand.  A construction may thus carry two interleaved lattices,
one in $|q|^{1/4}$ and one in $|q|^{1/8}$, and in this paper it
does.  The mass ratios sit on the coarser lattice and the mixing
magnitudes on the finer one (Sec.~\ref{sec:ckm}), the coefficient
$2$ appears wherever a $\theta_2$ insertion acts, and the odd steps
of the finer lattice are where CP appears
(Secs.~\ref{sec:ckm} and~\ref{sec:uv}).  In the language of
Sec.~\ref{sec:forms}, one unit of $T$ charge is one power
$q^{1/4}$ and one metaplectic half unit is one power $q^{1/8}$, the
counting used in every insertion below.

\subsection{Symmetric points, hierarchies, and CP}
\label{sec:points}

Symmetry breaking is controlled entirely by where the vacuum value
of $\tau$ sits.  A generic $\tau$ breaks $\Gamma_N$ completely,
while three inequivalent points leave residual subgroups unbroken.  The
cusp $\tau=i\infty$ preserves the $Z_N$ generated by $T$, a genuine
$Z_4$ at level four.  The self-dual point $\tau=i$, the fixed point
of the inversion, preserves the subgroup generated by $S$, a $Z_2$
of the finite group and a $Z_4$ in $SL(2,\mathbb{Z})$, where
$S^2=-1$.  The point $\tau=\omega=e^{2\pi i/3}$ preserves the $Z_3$
generated by $ST$.  A modulus near, but not at, a symmetric point
yields hierarchies governed by the small
departure~\cite{FeruglioTauI,NovichkovHier}.  Near the cusp the
small parameter is $|q|^{1/N}$, and near the self-dual point it is
the displacement
\begin{equation}
  u=\frac{\tau-i}{\tau+i} ,
\label{eq:u}
\end{equation}
which vanishes at $\tau=i$.  The two descriptions are compatible
when $\Imt$ is of order one.  A modulus slightly above $\tau=i$ has
both a small $u$ and a usefully small $|q|^{1/4}$, and a model may
use the fixed point to set normalizations while charge counting in
$q$ sets the hierarchy, which is the situation of
Sec.~\ref{sec:base}.

CP has a clean geometric realization.  In modular-invariant theories
with generalized CP the coupling coefficients can be taken real, and
CP is conserved whenever $\tau$ lies on the imaginary axis or
elsewhere on the boundary of the fundamental domain, where all theta
values are real.  A nonzero $\Ret$ breaks CP
spontaneously~\cite{NovichkovCP}.  The CP phase then sits in the
vacuum value of the modulus rather than in an independent coupling.
Under $\tau\to\rho+i\,\Imt$ at fixed $|q|$, each coupling acquires
the phase $2\pi\rho$ times its $q$ exponent.  The construction of
this paper stays on the imaginary axis, and the origin of the quark
phase it requires is taken up in Sec.~\ref{sec:ckm}.

\subsection{What level four offers the flavor sector}
\label{sec:why4}

Level four has a doublet, absent at level three, two inequivalent
triplets, a residual $Z_4$ at the cusp with hierarchies in $q^{1/4}$
and the metaplectic refinement to $q^{1/8}$, and a form content that
grows as $2k+1$ with the weight against $k+1$ at level three.  None
of this makes level four inevitable.  Whether the data motivate it is
the question the next section answers.  Table~\ref{tab:glossary}
collects the vocabulary that recurs from here on.

\begin{table}[t]
\caption{Glossary of recurring terms.}
\label{tab:glossary}
\begin{ruledtabular}
\begin{tabular}{p{0.22\textwidth}p{0.70\textwidth}}
Term & Meaning\\
\hline
Modulus $\tau$ & Complex field in the upper half plane whose vacuum
value breaks the modular symmetry, in string settings the shape of
a compactification torus.\\
Level $N$ & Integer selecting the congruence subgroup $\Gamma(N)$
and hence the finite group $\Gamma_N$.  Level four gives
$\Gamma_4\simeq S_4$ and $T^4=1$.\\
Weight $k$ & Power of the automorphy factor $(c\tau+d)^k$ in
Eq.~(\ref{eq:formtrafo}), half-integral on the metaplectic cover.\\
$T$-charge & Eigenvalue label $c\in\{0,\dots,N-1\}$ under
$T{:}\ \tau\to\tau+1$.  A component of charge $c$ opens at
$q^{c/N}$.\\
Cusp & The point $\tau=i\infty$, where $q\to0$, with residual symmetry
$Z_N^T$.\\
Self-dual point & The point $\tau=i$, fixed under
$S{:}\ \tau\to-1/\tau$.\\
Automorphy dressing & The K\"ahler factor $(2\,\Imt)^{w/2}$ carried
by a physical coupling of total weight $w$, one factor of
$(2\,\Imt)^{1/2}$ per unit of weight.\\
Metaplectic cover & Double cover of $SL(2,\mathbb{Z})$ on which
the square root of the automorphy factor, and hence half-integral
weight, is defined.  At level four its weight-one-half forms are the
doublet $\theta_3(2\tau)$, $\theta_2(2\tau)$, the constants of
Eq.~(\ref{eq:thetadef}) are that doublet at the halved modulus,
and the $q^{1/8}$ unit belongs to the level-eight cover.\\
Plaquette & A double ratio $m_{ik}m_{jl}/(m_{il}m_{jk})$ of four
masses at the corners of a rectangle on the sector-generation grid,
the two-over-two ratios of Ref.~\cite{Barger:2026tot}.\\
\end{tabular}
\end{ruledtabular}
\end{table}

\section{The harmonic pattern selects level four symmetry}
\label{sec:pattern}

\subsection{The measured pattern}

For sectors $i,j$ among up, down, and lepton, and generations $k$
and $l$, the double ratios
\begin{equation}
  P^{ij}_{kl}=\frac{m_{ik}\,m_{jl}}{m_{il}\,m_{jk}}
\label{eq:pdef}
\end{equation}
isolate the non-factorizable content of the masses.  Each is a
plaquette, the product around the rectangle with corners $(i,k)$,
$(i,l)$, $(j,k)$, and $(j,l)$ on the grid of sectors against
generations, and the plaquettes are the two-over-two ratios of
Ref.~\cite{Barger:2026tot}.  Four unit entries in rectangle position
are the smallest integer pattern whose powers sum to zero along every
row and every column; those cancellations remove the sector
normalizations and the common generation factors.  Every invariant
of the grid, including any in which a mass enters squared, is a
product of plaquettes.  Equation~(\ref{eq:pdef}) is therefore the
elementary object and not a selection.  The measured
exponent ratios on the two adjacent sector axes are
\begin{equation}
  \frac{\ln P^{UD}_{12}}{\ln P^{UD}_{23}}=+1.987\pm0.040 ,\qquad
  \frac{\ln P^{DL}_{12}}{\ln P^{DL}_{23}}=-2.066\pm0.024 .
\label{eq:harmonic}
\end{equation}
Here $\ln$ is the natural logarithm, and the ratios are unchanged
in any other base, since the conversion factor cancels in the
quotient.

The logarithm is not a convenience but the inverse of the
mechanism.  If hierarchy is generated by exponentiating charges, so
that each mass is an order-one coefficient times a power of a small
parameter, then the charge lives additively while the mass lives
multiplicatively, and $\ln P$ reads off the lattice content of the
plaquette, an integer times the log of the base plus the log of an
order-one dressing.  The ratio of two such logarithms is then the
unique base-free pure number, since a single $\ln P$ still contains
the unknown base, while the quotient is a ratio of integers up to the
dressings, independent of the base's value, of the choice of unit,
and of the logarithm's own base, so Eq.~(\ref{eq:harmonic}) is a
statement about the charge pattern that requires no knowledge of the
small parameter at all.

Written with each mass as its particle symbol, $u=m_u$ and so on,
the full collection reads
\begin{equation}
\begin{aligned}
P^{UD}_{12}&=\frac{us}{cd} , &
P^{UD}_{23}&=\frac{cb}{ts} , &
P^{UD}_{13}&=\frac{ub}{td} ,\\[2pt]
P^{DL}_{12}&=\frac{d\mu}{se} , &
P^{DL}_{23}&=\frac{s\tau}{b\mu} , &
P^{DL}_{13}&=\frac{d\tau}{be} ,\\[2pt]
P^{UL}_{12}&=\frac{u\mu}{ce} , &
P^{UL}_{23}&=\frac{c\tau}{t\mu} , &
P^{UL}_{13}&=\frac{u\tau}{te} .
\end{aligned}
\label{eq:plist}
\end{equation}
The collection holds nine products, three on each of the three
sector axes $UD$, $DL$, and $UL$, and equally three of each
generation type $12$, $23$, and $13$.  Same-sector products equal
one identically.  Each $13$ entry is the product of its $12$ and
$23$ partners, and each $UL$ entry is the product of its $UD$ and
$DL$ partners, so four of the nine are independent, and
Eq.~(\ref{eq:harmonic}) is built from exactly those
four.\footnote{In this shorthand the up-down products are labeled by
the index pairs $us$, $cb$, and $ub$ over $td$, the same labels as the
off-diagonal CKM magnitudes.  The coincidence is one of naming,
since each position pairs one symbol from each axis, though the
same four elements return in Eq.~(\ref{eq:tower}) on the lattice
the mass ratios define.}

The first-to-second generation step is twice the second-to-third
step on both logarithmic axes.  The sign flip says the ordering is
mirrored.  The two axes are made explicit by the per-generation
sector ratios whose logarithmic differences the products in
Eq.~(\ref{eq:plist}) record.  On the up-down axis the generations
march in order,
\begin{equation}
  \frac{m_u}{m_d}=0.457\ <\ \frac{m_c}{m_s}=11.6\ <\
  \frac{m_t}{m_b}=59.3 ,
\label{eq:udorder}
\end{equation}
each up quark heavier relative to its down partner than the
generation before, with the logarithmic gap of $3.24$ between the
first two ratios twice the gap of $1.63$ between the last two.  On
the down-lepton axis the second generation sits at an end,
\begin{equation}
  \frac{m_s}{m_\mu}=0.523\ <\ \frac{m_b}{m_\tau}=1.64\ <\
  \frac{m_d}{m_e}=5.55 ,
\label{eq:dlorder}
\end{equation}
so the muon outweighs the strange quark while the electron and the
tau fall below the down and bottom quarks, and the third generation
sits between the other two.  The gap of $2.36$ from the first ratio
to the second is twice the gap of $1.14$ from the second to the
third, with the opposite sense.  These are the placements
$(0,2,3)$ and $(2,0,1)$ that the selection theorem below reads as
charges on the $Z_4$ clock.

The pattern passes a null test.  On hierarchy-matched
unstructured spectra, each of the nine measured masses multiplied
by an independent factor drawn log-uniformly from
$[e^{-0.35},e^{0.35}]$, both ratios land as close to $\pm2$ as
observed in $0.14\%$ of draws
(Appendix~\ref{app:stats}).

\subsection{The selection theorem}

Suppose one charge vector $a=(a_1,a_2,a_3)$ governs the first axis
and the second axis carries the shifted image $b_g=(a_g+c)\bmod N$.
The measured steps require $(2,1)$ before reduction and $(-2,+1)$
after.  A one-unit step keeps its sense under any shift.  A
two-unit step reverses only through a wrap-around, which turns
$+2$ into $2-N$, so reversal requires
\begin{equation}
  2-N=-2 ,\qquad\text{that is,}\qquad 2\equiv-2 \pmod N ,
\label{eq:select}
\end{equation}
and the only level above two that satisfies it is $N=4$.  Level
three, the $A_4$ of most modular flavor models, and level five,
the $A_5$, fail the mirror at the first step, since $-2$ reduces to
$1$ modulo three and to $3$ modulo five, neither equal to $2$.
Working through the remaining wrap-around conditions fixes the
charges, $a=(0,2,3)$ and $b=a+2\bmod4=(2,0,1)$, and an exhaustive
scan over $3\le N\le12$ confirms that no other level admits any
solution.

Two hypotheses enter.  The first is that the
second axis is a uniform shift of the first, $b_g=a_g+c$, the
statement that one group element relates the two comparisons.  The
second is that the smaller measured step is one unit of charge.  With
the steps taken as $(2m,m)$ the same wrap-around gives $N=4m$, so the
selection is of level four up to a rescaling of the unit, and the case
$m=2$, charges in eighths with $T^8=1$, is the metaplectic lattice of
Sec.~\ref{sec:metaplectic} on which the construction below is built.
What is excluded is a level that is not a multiple of four.

The finite modular group is therefore the
$\Gamma_4\simeq S_4$ of
Sec.~\ref{sec:finite}~\cite{Feruglio:2017spp,PenedoPetcov,NovichkovS4},
selected by the data, not assumed.  Geometrically $S_4$ is
the rotation group of the cube and the octahedron.  Its triplets
hold three generations, its sign singlet $\mathbf{1}'$
flips under odd permutations, which is the mirror the pattern
requires, and its natural mixing textures are the trimaximal
patterns that return in the lepton sector of
Sec.~\ref{sec:leptons}.

\subsection{The mirror is the sign singlet}

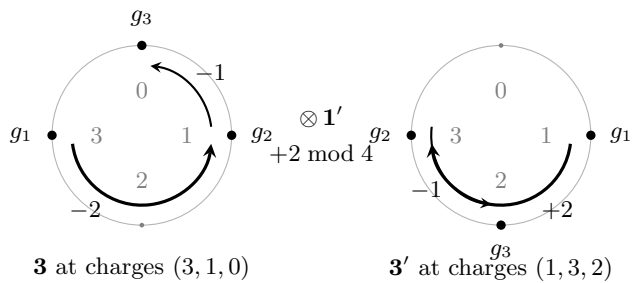
\begin{figure}[t]
\begin{tikzpicture}[>=stealth,scale=0.88]
\begin{scope}[shift={(0,0)}]
  \draw[gray!60] (0,0) circle (1.35);
  \foreach \c/\ang in {0/90,1/0,2/-90,3/180}{
    \fill[gray] (\ang:1.35) circle (0.035);
    \node[gray] at (\ang:0.68) {\small $\c$};}
  \fill (180:1.35) circle (0.07) node[left=4pt] {$g_1$};
  \fill (0:1.35)  circle (0.07) node[right=4pt] {$g_2$};
  \fill (90:1.35)  circle (0.07) node[above=4pt] {$g_3$};
  \draw[->,very thick] (187:1.05) arc (187:353:1.05)
        node[pos=0.32,below left=-2pt] {\small $-2$};
  \draw[->,thick] (7:1.05) arc (7:83:1.05)
        node[pos=0.5,above right=-3pt] {\small $-1$};
  \node at (0,-2.0) {$\mathbf{3}$ at charges $(3,1,0)$};
\end{scope}
\node at (2.7,0.28) {$\otimes\,\mathbf{1}'$};
\node at (2.7,-0.28) {\small $+2\bmod 4$};
\begin{scope}[shift={(5.4,0)}]
  \draw[gray!60] (0,0) circle (1.35);
  \foreach \c/\ang in {0/90,1/0,2/-90,3/180}{
    \fill[gray] (\ang:1.35) circle (0.035);
    \node[gray] at (\ang:0.68) {\small $\c$};}
  \fill (0:1.35)  circle (0.07) node[right=4pt] {$g_1$};
  \fill (180:1.35) circle (0.07) node[left=4pt] {$g_2$};
  \fill (-90:1.35) circle (0.07) node[below=4pt] {$g_3$};
  \draw[->,very thick] (-7:1.05) arc (-7:-173:1.05)
        node[pos=0.32,below right=-2pt] {\small $+2$};
  \draw[->,thick] (173:1.05) arc (173:262:1.05)
        node[pos=0.5,below left=-3pt] {\small $-1$};
  \node at (0,-2.0) {$\mathbf{3}'$ at charges $(1,3,2)$};
\end{scope}
\end{tikzpicture}
\caption{The mirror as the sign singlet on the $Z_4$ clock.
Multiplying the triplet by $\mathbf{1}'$ rotates the clock by half
a turn.  The doubled step reverses and the single step does not,
reproducing the measured pattern.  Half the group equals two units
only at level four.}
\label{fig:clock}
\end{figure}

The charge set $(3,1,0)$ is the $T$-charge content of the
$S_4$ triplet $\mathbf{3}$ in Table~\ref{tab:charges}, with heavier
generations at lower charge.  The shift by two units is multiplication by the sign
singlet $\mathbf{1}'$, whose $T$ charge is two.
Figure~\ref{fig:clock} shows the action.  A two-unit step is its
own reverse modulo four and flips under the half turn, while a
one-unit step keeps its sense.  The doubled step mirrors and the
single step does not, which is the measured pattern.  In the
forced assignment of Appendix~\ref{app:assignment} the twist is
carried by a reducible $e^c$ rather than by a uniform half turn of a
triplet, so the identification of the mirror with $\mathbf{1}'$ is a
statement about the clock reading rather than about the realized
right-handed charges.

The mirror therefore sits where Georgi--Jarlskog Clebsch factors
act in unified fits~\cite{Georgi:1979df}, as a single unit of
$\mathbf{1}'$ separating the lepton line from the down line.  In
the Georgi--Jarlskog texture the charged-lepton and down-quark mass
matrices coincide except for a Clebsch factor of $-3$ on the
second-generation entry, supplied by a $\mathbf{45}$-dimensional
Higgs representation of $SU(5)$.  The sign drops out of the mass
eigenvalues, so that at the unification scale
$m_b=m_\tau$, $m_\mu=3m_s$, and the plaquette
$m_bm_\mu/(m_sm_\tau)$ equals $3$.  That plaquette is the inverse
of the down-lepton product $P^{DL}_{23}$ in Eq.~(\ref{eq:plist}),
and the common QCD rescaling of the two down-quark masses cancels
within it, so it can be read at $\MZ$, where the measured value is
$3.136\pm0.027$.

The half turn of the clock does the same job by a
different mechanism.  It moves the second generation to an end of
the ordering on the down-lepton axis, singling that generation out
as the Clebsch does, and its numerical effect is a ratio of modular
form values rather than a group-theoretic rational number.  Once
the insertions P1 and P5 of Sec.~\ref{sec:masses}
[Eqs.~(\ref{eq:t1}) and~(\ref{eq:p5})] are in place, the
construction returns the plaquette as
$\bigl(2/(\e\,\Imt)\bigr)^{1/2}=3.167$ [the inverse of
$P^{DL}_{23}$ in Eq.~(\ref{eq:closedP})], with the $2$ the leading
coefficient of $\theta_2$, the $\e$ its leading power, and $\Imt$
the automorphy dressing of the strange quark, at $1.1\sigma$ from
the measured value.

The measured Georgi--Jarlskog parameter thus
becomes a quantity the model computes at $\MZ$.  Its comparison
with the rational Clebsch at a unification scale is a question of
renormalization-group running, taken up in Sec.~\ref{sec:uv}.

\subsection{The inversion and its control}

The inversion is the older fact.  That the muon outweighs the
strange quark has been known since the two masses could first be
compared, and it is the anomaly the Clebsch was built to
accommodate.  Here it is the fingerprint that selects the level,
and the construction fixes its size as well as its sense.  The
inversion factor is the octave dressed by the automorphy factor,
$m_\mu/m_s=2\,(\Imt)^{-1/2}=1.935$ against the measured
$1.912\pm0.021$, at $-1.1\sigma$, so the factor that reverses the
ordering is the same two to one that names the model.

The up-down axis, by contrast, carries no inversion, and the two
sequences sit side by side.  The down-lepton ratios order as
$m_s/m_\mu<m_b/m_\tau<m_d/m_e$, the second generation at an end,
while the up-down ratios march in generation order,
$m_u/m_d<m_c/m_s<m_t/m_b$ [Eqs.~(\ref{eq:udorder})
and~(\ref{eq:dlorder})].  On the clock the reason is where the
twist sits.  The single unit of $\mathbf{1}'$ enters only between
the down and lepton lines, so the down-lepton comparison is the
axis turned by half a revolution and the up-down comparison is not.
A doubled step reverses under the half turn and a single step does
not, so the reversal that puts the second generation at an end
appears on the down-lepton axis alone.

The construction below realizes the two axes differently from
this clock reading.  In
the operator gradings of Appendix~\ref{app:assignment} the up and
down columns are identical, so the up-down double ratios carry no
power of $q$ at all, and their two to one pattern is borne by the
coefficients of Eq.~(\ref{eq:closedP}), $16\sqrt2\,\theta_3^{2}=25.9$
and $4\theta_3^{3}(\Imt)^{1/2}=5.06$, which sit near $\e^{-2}=28.7$
and $\e^{-1}=5.36$ without being powers of the base.  The lepton
column differs from the down column by $(2,0,1)$ in eighths, so the
down-lepton steps $-2$ and $+1$ are charge steps, but the sequence
$2\to0\to1$ involves no wrap-around and would be admissible at any
level.  The clock reading of the selection theorem is therefore the
motivation for level four; its hypotheses, the uniform shift and the
unit step, are not those of the realized assignment, whose exponents
the finite group fixes only modulo eight
(Appendix~\ref{app:assignment}), and the case for the construction
rests on what the assignment then delivers.
The one inversion inside the quark grid, the up quark lighter than
the down, follows from normalizations rather than from the
clock, and the postulates deliver it quantitatively, since the
postulates give $m_u/m_d=0.454$ against the
measured $0.457\pm0.009$, at $+0.4\sigma$, the ordering that makes
the neutron heavier than the proton.

\section{The harmonic modulus and the Jacobi insertions}
\label{sec:jacobi}

\subsection{One field behind the couplings}

The framework has a single flavor field, the modulus $\tau$ of
Sec.~\ref{sec:modular}.  Matter fields are inert under the
level-four congruence subgroup $\Gamma(4)$, so the finite group
acting on the generations is the $\Gamma_4\simeq S_4$ of
Sec.~\ref{sec:finite}, selected by the data in
Sec.~\ref{sec:pattern}, and the Yukawa couplings are modular forms
expanded in the nome $q=e^{2\pi i\tau}$ as in Sec.~\ref{sec:forms}.
The working value of the modulus, fixed by the charged leptons in
Sec.~\ref{sec:base}, is the harmonic modulus of the title.

\subsection{The Jacobi theta constants}

The couplings of the construction are built from the three Jacobi
theta constants of Eq.~(\ref{eq:thetadef}), weight one-half forms
on the metaplectic cover of Sec.~\ref{sec:metaplectic}, at the
arguments $\tau$ and $2\tau$.  Their expansions in
Eq.~(\ref{eq:thetaexp}) supply the two facts the model uses
everywhere.  The leading coefficient of $\theta_2$ is exactly $2$,
and its leading power is one metaplectic unit $q^{1/8}$.  Every
recurring coefficient in this paper is one of these expansion
coefficients, a form value, or the automorphy factor
$(2\Imt)^{1/2}$ of Eq.~(\ref{eq:automorphy}) that dresses a
physical coupling for each unit of weight.  At any common
modulus the three constants satisfy the Jacobi identity
\begin{equation}
  \theta_2^{4}(\tau)+\theta_4^{4}(\tau)=\theta_3^{4}(\tau) ,
\label{eq:jacobi}
\end{equation}
one exact relation that later eliminates the modulus from the lepton
mixings.  Theta Yukawa
couplings with half-integral weights are the coupling dictionary
of magnetized toroidal
compactifications~\cite{Cremades:2004wa,Antoniadis:2009bg}, the
zero modes and Yukawa couplings on a magnetized torus transform as
weight one-half forms of the double cover~\cite{Kobayashi:2018rad,
Kikuchi:2020frc,Ohki:2020frc}, and the level-four framework with
half-integral weights is developed in Ref.~\cite{Liu:2020msy}.

\subsection{Insertions as Froggatt--Nielsen spurions}

The construction writes no superpotential.  Its statements are
about the physical Yukawa couplings, and an operator in this paper
means the effective coupling that generates one mass or one
transition between generations, one for each charged fermion and
one for each off-diagonal entry, with an insertion being a form
that multiplies that coupling.

The construction uses the forms
the way a Froggatt--Nielsen model uses a
flavon~\cite{Froggatt:1978nt}.  There, an operator
with charge $c$ picks up $c$ powers of a small vacuum ratio.  Here
the $T$ charge plays the Froggatt--Nielsen charge and the theta
constant plays the spurion, one factor per unit of charge, so a
charge assignment is a hierarchy prediction, with one unit of
charge one leading power $q^{1/4}$ and one metaplectic half unit
one power $q^{1/8}$.  The difference is decisive.  A
Froggatt--Nielsen spurion is an independent vacuum value, so its
coefficient and its size are free in every operator.  A theta
insertion is a fixed function of the one modulus, so its
coefficient is the expansion coefficient $2$, its size is set by
$\tau$, and relations between sectors become predictions rather
than accidents.

\subsection{The insertion rules}
\label{sec:rules}

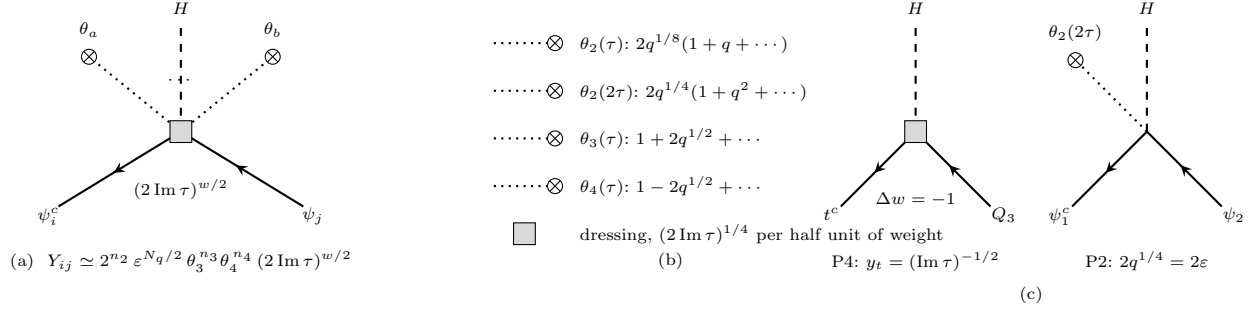
\begin{figure*}[t]
\centering
\begin{tikzpicture}[>=stealth,scale=0.9,every node/.style={transform shape},
  ferm/.style={thick,postaction={decorate},decoration={markings,mark=at position 0.55 with {\arrow{>}}}},
  higgs/.style={thick,dashed},
  spur/.style={thick,dotted},
  dress/.style={draw,fill=black!15,minimum size=3.2mm,inner sep=0pt},
  lab/.style={font=\scriptsize},
  ptxt/.style={font=\scriptsize,align=center}]
\tikzset{spurend/.pic={\draw[solid,thin,fill=white] (0,0) circle (0.11); \draw[solid,thin] (-0.078,-0.078)--(0.078,0.078); \draw[solid,thin] (-0.078,0.078)--(0.078,-0.078);}}
\begin{scope}[shift={(0,0)}]
\coordinate (V) at (0,0);
\draw[ferm] (V) -- (-1.8,-1.1) node[lab,pos=0.95,below left=-2pt] {$\psi^c_i$};
\draw[ferm] (1.8,-1.1) -- (V) node[lab,pos=0.05,below right=-2pt] {$\psi_j$};
\draw[higgs] (V) -- (0,1.6) node[lab,above] {$H$};
\draw[spur] (V) -- (-1.35,1.1) pic{spurend} ;
\draw[spur] (V) -- (1.35,1.1) pic{spurend};
\node[lab] at (-1.35,1.5) {$\theta_{a}$};
\node[lab] at (1.35,1.5) {$\theta_{b}$};
\node[lab] at (0,0.75) {$\cdots$};
\node[dress] at (V) {};
\node[lab] at (0,-0.85) {$(2\,\Imt)^{w/2}$};
\node[ptxt] at (0,-1.9) {(a)\ \ $Y_{ij}\simeq2^{n_2}\,\e^{N_q/2}\,\theta_3^{\,n_3}\theta_4^{\,n_4}\,(2\,\Imt)^{w/2}$};
\end{scope}
\begin{scope}[shift={(4.6,1.3)}]
\foreach \y/\nm/\fac in {0/{\theta_2(\tau)}/{2q^{1/8}(1+q+\cdots)},
  -0.7/{\theta_2(2\tau)}/{2q^{1/4}(1+q^2+\cdots)},
  -1.4/{\theta_3(\tau)}/{1+2q^{1/2}+\cdots},
  -2.1/{\theta_4(\tau)}/{1-2q^{1/2}+\cdots}}{
  \draw[spur] (0,\y) -- (0.9,\y) pic{spurend};
  \node[lab,anchor=west] at (1.15,\y) {$\nm$:\ $\fac$};}
\node[dress] at (0.45,-2.8) {};
\node[lab,anchor=west] at (1.15,-2.8) {dressing, $(2\,\Imt)^{1/4}$ per half unit of weight};
\node[ptxt] at (2.6,-3.2) {(b)};
\end{scope}
\begin{scope}[shift={(12.4,0)}]
\begin{scope}[shift={(-1.6,0)}]
\coordinate (V) at (0,0);
\draw[ferm] (V) -- (-1.1,-1.1) node[lab,pos=0.95,below left=-2pt] {$t^c$};
\draw[ferm] (1.1,-1.1) -- (V) node[lab,pos=0.05,below right=-2pt] {$Q_3$};
\draw[higgs] (V) -- (0,1.6) node[lab,above] {$H$};
\node[dress] at (V) {};
\node[lab] at (0,-1.0) {$\Delta w=-1$};
\node[ptxt] at (0,-1.9) {P4:\ $y_t=(\Imt)^{-1/2}$};
\end{scope}
\begin{scope}[shift={(1.8,0)}]
\coordinate (V) at (0,0);
\draw[ferm] (V) -- (-1.1,-1.1) node[lab,pos=0.95,below left=-2pt] {$\psi^c_1$};
\draw[ferm] (1.1,-1.1) -- (V) node[lab,pos=0.05,below right=-2pt] {$\psi_2$};
\draw[higgs] (V) -- (0,1.6) node[lab,above] {$H$};
\draw[spur] (V) -- (-1.05,1.05) pic{spurend};
\node[lab] at (-1.05,1.45) {$\theta_2(2\tau)$};
\node[ptxt] at (0,-1.9) {P2:\ $2q^{1/4}=2\e$};
\end{scope}
\node[ptxt] at (0.1,-2.4) {(c)};
\end{scope}
\end{tikzpicture}
\caption{The vertex rules of Sec.~\ref{sec:rules}.  (a)~A generic
effective coupling.  Solid lines are the two fermion fields, with
arrows marking the fermion-number flow through the vertex, in on
$\psi_j$ and out on $\psi^c_i$ as in four-component notation, the
dashed line is the Higgs, each dotted leg ending in a cross is one
theta insertion, and the shaded square at the vertex is the
dressing of Eq.~(\ref{eq:automorphy}) for the total weight $w$;
the coupling is read off as in Eq.~(\ref{eq:rulemag}).  (b)~The
four insertion legs with their factors from Table~\ref{tab:rules},
and the dressing.  (c)~Two examples.  The top anchor P4 has no
insertion and one inverse unit of weight measured from the fixed
point, and the down-lepton $1$--$2$ transition P2 has one leg
of $\theta_2(2\tau)$.}
\label{fig:rules}
\end{figure*}

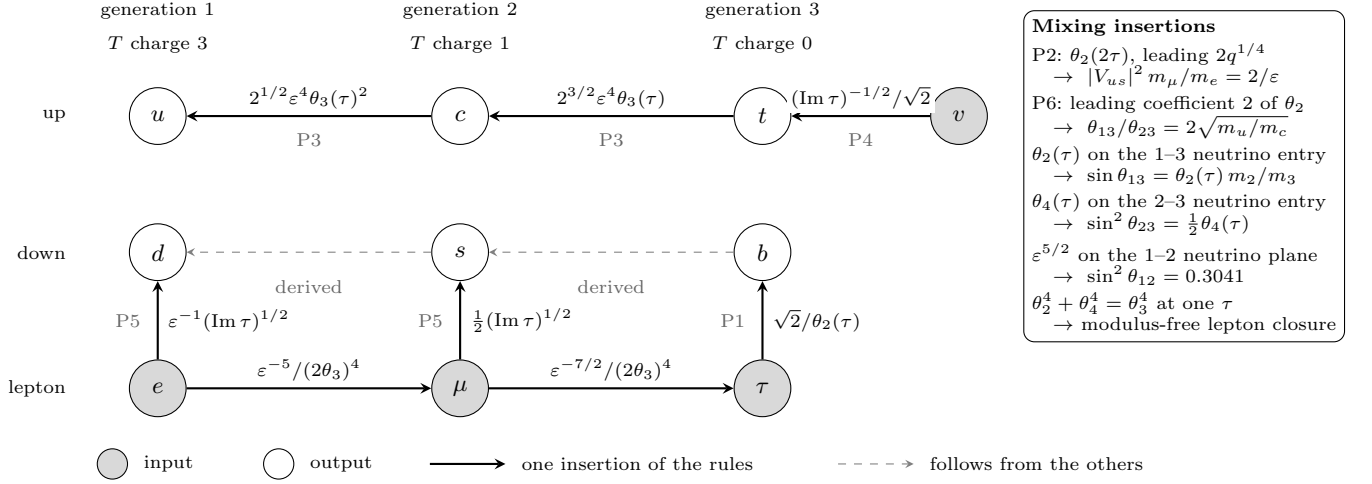
\begin{figure*}[t]
\centering
\begin{tikzpicture}[>=stealth,
  inp/.style={circle,draw,fill=black!15,minimum size=7.5mm,inner sep=0pt,font=\small},
  pred/.style={circle,draw,minimum size=7.5mm,inner sep=0pt,font=\small},
  ins/.style={->,thick},
  der/.style={->,dashed,gray},
  lab/.style={font=\scriptsize,fill=white,inner sep=1.5pt},
  plab/.style={font=\scriptsize,gray!80!black},
  mix/.style={draw,rounded corners,font=\scriptsize,inner sep=3pt,align=left}]
\node[font=\scriptsize] at (0,5.0) {generation 1};
\node[font=\scriptsize] at (4,5.0) {generation 2};
\node[font=\scriptsize] at (8,5.0) {generation 3};
\node[font=\scriptsize] at (0,4.55) {$T$ charge $3$};
\node[font=\scriptsize] at (4,4.55) {$T$ charge $1$};
\node[font=\scriptsize] at (8,4.55) {$T$ charge $0$};
\node[font=\scriptsize,anchor=east] at (-1.1,3.6) {up};
\node[font=\scriptsize,anchor=east] at (-1.1,1.8) {down};
\node[font=\scriptsize,anchor=east] at (-1.1,0.0) {lepton};
\node[pred] (u) at (0,3.6) {$u$};
\node[pred] (c) at (4,3.6) {$c$};
\node[pred] (t) at (8,3.6) {$t$};
\node[pred] (d) at (0,1.8) {$d$};
\node[pred] (s) at (4,1.8) {$s$};
\node[pred] (b) at (8,1.8) {$b$};
\node[inp] (e) at (0,0.0) {$e$};
\node[inp] (mu) at (4,0.0) {$\mu$};
\node[inp] (tau) at (8,0.0) {$\tau$};
\node[inp] (v) at (10.6,3.6) {$v$};
\draw[ins] (e) -- (mu) node[lab,midway,above=1pt] {$\e^{-5}/(2\theta_3)^4$};
\draw[ins] (mu) -- (tau) node[lab,midway,above=1pt] {$\e^{-7/2}/(2\theta_3)^4$};
\draw[ins] (tau) -- (b) node[lab,midway,right=2pt] {$\sqrt2/\theta_2(\tau)$}
  node[plab,midway,left=3pt] {P1};
\draw[ins] (mu) -- (s) node[lab,midway,right=2pt] {$\tfrac12(\Imt)^{1/2}$}
  node[plab,midway,left=3pt] {P5};
\draw[ins] (e) -- (d) node[lab,midway,right=2pt] {$\e^{-1}(\Imt)^{1/2}$}
  node[plab,midway,left=3pt] {P5};
\draw[ins] (v) -- (t) node[lab,midway,above=1pt] {$(\Imt)^{-1/2}/\sqrt2$}
  node[plab,midway,below=3pt] {P4};
\draw[ins] (t) -- (c) node[lab,midway,above=1pt] {$2^{3/2}\e^{4}\theta_3(\tau)$}
  node[plab,midway,below=3pt] {P3};
\draw[ins] (c) -- (u) node[lab,midway,above=1pt] {$2^{1/2}\e^{4}\theta_3(\tau)^2$}
  node[plab,midway,below=3pt] {P3};
\draw[der] (b) -- (s);
\draw[der] (s) -- (d);
\node[plab,anchor=north] at (6,1.55) {derived};
\node[plab,anchor=north] at (2,1.55) {derived};
\node[mix,anchor=north west] at (11.45,5.0) {%
  \textbf{Mixing insertions}\\[2pt]
  P2:\ $\theta_2(2\tau)$, leading $2q^{1/4}$\\
  \quad$\to\ |V_{us}|^2\,m_\mu/m_e=2/\e$\\[2pt]
  P6:\ leading coefficient $2$ of $\theta_2$\\
  \quad$\to\ \theta_{13}/\theta_{23}=2\sqrt{m_u/m_c}$\\[2pt]
  $\theta_2(\tau)$ on the $1$--$3$ neutrino entry\\
  \quad$\to\ \sin\theta_{13}=\theta_2(\tau)\,m_2/m_3$\\[2pt]
  $\theta_4(\tau)$ on the $2$--$3$ neutrino entry\\
  \quad$\to\ \sin^2\theta_{23}=\tfrac12\theta_4(\tau)$\\[2pt]
  $\e^{5/2}$ on the $1$--$2$ neutrino plane\\
  \quad$\to\ \sin^2\theta_{12}=0.3041$\\[2pt]
  $\theta_2^4+\theta_4^4=\theta_3^4$ at one $\tau$\\
  \quad$\to$ modulus-free lepton closure};
\node[inp,minimum size=4mm] at (-0.6,-1.0) {};
\node[font=\scriptsize,anchor=west] at (-0.3,-1.0) {input};
\node[pred,minimum size=4mm] at (1.6,-1.0) {};
\node[font=\scriptsize,anchor=west] at (1.9,-1.0) {output};
\draw[ins] (3.6,-1.0) -- (4.6,-1.0);
\node[font=\scriptsize,anchor=west] at (4.7,-1.0) {one insertion of the rules};
\draw[der] (9.0,-1.0) -- (10.0,-1.0);
\node[font=\scriptsize,anchor=west] at (10.1,-1.0) {follows from the others};
\end{tikzpicture}
\caption{The theta insertions on the mass grid and in the mixings.
Left, the nine charged-fermion masses by sector (rows) and
generation (columns), the generations at $T$ charges $3$, $1$, and
$0$.  Filled nodes are the inputs, the three charged-lepton masses
and the electroweak scale $v$, and open nodes are outputs.  Each
solid arrow carries one insertion, labeled with the ratio of the
mass at its head to the mass at its tail and with the postulate
that supplies it.  The down-quark steps (dashed) follow from the
arrows that reach them.  Right, the insertions that fix the
mixings, with the relations they produce.}
\label{fig:insertions}
\end{figure*}

The placements are fixed by six rules, so nothing about an
insertion is chosen operator by operator.  They can be read as
vertex factors in the manner of a spurion analysis, with one limit
stated once.  In a Froggatt--Nielsen model the powers of the small
parameter arise from heavy messengers integrated out, so the
diagrammatic reading is literal, whereas here each theta constant is
an exact function of the one modulus and the analogy supplies
vertices without propagators.  Table~\ref{tab:rules} lists
the factors, Table~\ref{tab:jacobi} their insertion sites,
Fig.~\ref{fig:rules} draws them as spurion legs, and
Fig.~\ref{fig:insertions} places the insertions on the mass grid,
one arrow per insertion, together with the five that fix the
mixings.

First, charge counting.  An operator with $n$ units of $T$ charge
carries the leading power $q^{n/4}$, one metaplectic half unit
carries $q^{1/8}$, and charges add across the fields in an
operator, so every entry exponent is a sum of one left charge and
one right charge.  Weight balance fixes the number of insertions,
$2k$ of weight one half for fields of total weight $-k$, and the
$T$ transformation fixes the choice among the constants, since
$\theta_2(\tau+1)=e^{i\pi/4}\theta_2(\tau)$ while $T$ exchanges
$\theta_3$ and $\theta_4$; the $T$ charge therefore fixes the number
of $\theta_2$ factors modulo eight, and the $\theta_3$ and
$\theta_4$ content must combine to the required eigenvalue.  This
is the one point at which the scheme has more freedom than its
Froggatt--Nielsen counterpart, where an integer charge fixes the
power uniquely, and it is where the insertion sites of
Table~\ref{tab:postulates} enter as input.

Second, the half-unit carrier.  A half-unit step is carried by
$\theta_2$, the one form in the triple whose expansion begins at a
power of $q$, so every half-unit hierarchy arrives with the
coefficient $2$.  The same form supplies the $2$ together with
$\sqrt\e$ at argument $\tau$ and together with $\e$ at argument
$2\tau$, where its expansion begins at $2q^{1/4}$, and the two
arguments are the two quark targets P1 and P2.  The unit spurion is
not an independent object, since the duplication formulas
\begin{equation}
\begin{aligned}
  \theta_2(2\tau)^2&=\tfrac12\bigl[\theta_3(\tau)^2-\theta_4(\tau)^2\bigr],\\
  \theta_3(2\tau)^2&=\tfrac12\bigl[\theta_3(\tau)^2+\theta_4(\tau)^2\bigr],\\
  \theta_4(2\tau)^2&=\theta_3(\tau)\,\theta_4(\tau) ,
\end{aligned}
\label{eq:duplication}
\end{equation}
express the doublet at argument $2\tau$ in the constants at
argument $\tau$; at the harmonic values
$\tfrac12(\theta_3^2-\theta_4^2)=0.1394$ gives
$\theta_2(2\tau)=0.373$, against the leading value $2\e=0.373$.

Third, the neutral dressings.  The charge-zero forms $\theta_3$
and $\theta_4$ begin at $1$ and dress an operator without moving
it on the lattice.  One factor of $\theta_3$ dresses each unit of
up-sector charge in P3, and one $(2\theta_3)^4$ dresses each
charged-lepton step in the anchors of Sec.~\ref{sec:base}.  The
form $\theta_4$, the image of $\theta_3$ under $T$, carries the
sign-flipped dressing and enters where the observable is a small
deficit below its symmetric value, the atmospheric angle of
Eq.~(\ref{eq:atm}).  The Jacobi identity of Eq.~(\ref{eq:jacobi})
removes the modulus from ratios of four-insertion products, which
is the modulus-free lepton closure.

Fourth, the weight dressing.  After the product is formed, the
physical coupling carries the real factor $(2\Imt)^{w/2}$ of
Eq.~(\ref{eq:automorphy}) for its total weight $w$, one factor
$(2\Imt)^{1/2}$ per unit of weight and $(2\Imt)^{1/4}$ per half
unit, so suppression can sit in weight rather than in charge.
Three points make the rule unambiguous.  The count is by the net
weight of the completed coupling, taken after the identities above
have been applied, not by the number of insertions written.  The
dressing is applied once to the completed coupling, so in a ratio
of two couplings only the weight difference $\Delta w$ survives as
$(2\Imt)^{\Delta w/2}$, which measured from the fixed point
$\tau=i$, where $2\Imt=2$, is $(\Imt)^{\Delta w/2}$; this is the
form the top anchor P4 and the light down line P5 take, with
$\Delta w=-1$ for the anchor, whose reference is the electroweak
vacuum expectation value through $y_t=\sqrt2\,m_t/v$, and
$\Delta w=+1$ for $d$ and $s$, whose references are $e$ and $\mu$.
And the factor is real and carries no phase.

Fifth, unit coefficients.  Every insertion enters with coefficient
one, a postulate of the construction and its point of departure
from the Froggatt--Nielsen case, where each vertex carries an
independent coefficient of order one.  Each number in the
construction is then an expansion coefficient, a form value, or an
automorphy factor, and no operator carries a free continuous
constant; what remains is a discrete selection among such factors,
stated below.  Collected, a coupling with $n_2$ insertions of
$\theta_2$ type at either argument, $n_3$ of $\theta_3$, $n_4$ of
$\theta_4$, total $q$-charge $N_q$ in eighths, and total weight $w$
has leading magnitude
\begin{equation}
  |Y|\simeq 2^{n_2}\,\e^{N_q/2}\,\theta_3^{\,n_3}\,\theta_4^{\,n_4}\,
  (2\,\Imt)^{w/2} .
\label{eq:rulemag}
\end{equation}
On the imaginary axis every factor is real.  For a displaced
modulus $\tau=\rho+i\,\Imt$ each holomorphic factor $q^{n/8}$
carries the phase $2\pi\rho\,n/8$, so the leading phase of the
coupling is $2\pi\rho N_q/8$, with subleading phases
$a|q|\sin2\pi\rho$ from a $\theta_2$ bracket $1+aq+\cdots$ and
$2|q|^{1/2}\sin\pi\rho$ from each $\theta_3$ or $\theta_4$ factor;
summed over the diagonal these give the $\bar\theta$ accounting of
Sec.~\ref{sec:uv}.

Sixth, the half lattice.  Half-integral weights generate the
mixing lattice at half the quantum of the mass lattice, and the
odd half steps are where CP appears, the content of
Secs.~\ref{sec:ckm} and~\ref{sec:uv}.  The rules produce the
entries of a Yukawa matrix; masses are its singular values and
mixings come from its diagonalization, and that second stage is
where square roots of entry ratios appear, a
Gatto--Sartori--Tonin block giving $|V_{us}|=\sqrt{m_d/m_s}$ and
the texture route P6 giving $\sqrt{m_u/m_c}$.  The mass postulates
P1, P3, P4, and P5 constrain single entries and their
ratios, while P2, P6, P7, and P8 and the Cabibbo relation act
at the level of the diagonalization.
Appendix~\ref{app:matrices} exhibits one pair of matrices whose
entries obey the rules and whose singular values and mixings
reproduce Tables~\ref{tab:masses} and~\ref{tab:ckm}.

\begin{table}[t]
\caption{Vertex factors for the theta insertions.  The $q$-charge
is in metaplectic eighths, and the magnitude is the leading term
at the harmonic modulus, where $|q|=\e^4$.}
\label{tab:rules}
\begin{ruledtabular}
\begin{tabular}{lcccl}
Insertion & Weight & $q$-charge & Factor & Magnitude\\
\hline
$\theta_2(\tau)$  & $\tfrac12$ & $1$ & $2q^{1/8}(1+q+q^3+\cdots)$ & $2\sqrt\e$\\
$\theta_2(2\tau)$ & $\tfrac12$ & $2$ & $2q^{1/4}(1+q^2+\cdots)$   & $2\e$\\
$\theta_3(\tau)$  & $\tfrac12$ & $0$ & $1+2q^{1/2}+2q^2+\cdots$   & $1.06969$\\
$\theta_4(\tau)$  & $\tfrac12$ & $0$ & $1-2q^{1/2}+2q^2-\cdots$   & $0.93031$\\
\end{tabular}
\end{ruledtabular}
\end{table}

What the rules derive and what they take as input separate
cleanly.  Derived from the pattern of Sec.~\ref{sec:pattern}
are the level, the triplet with its charges $(3,1,0)$, and the
identity of the mirror as the sign singlet.  Derived from the
charged-lepton masses are the base and the modulus.  Taken as input
are the insertion sites, the eight postulates of
Table~\ref{tab:postulates} and the four mixing insertions of
Fig.~\ref{fig:insertions}, which the six rules constrain but do
not select.

Taken as input as well are the coefficients that the
rules do not produce: the $\sqrt2$ of P1, which is the fixed-point
value of one unit of weight without its $\Imt$ dependence, the
$2^{3/2}$ and $2^{1/2}$ of P3, which are not integer powers of the
leading coefficient of $\theta_2$, and the $\tfrac12\theta_3^{2}$
of P8, two neutral dressings on a three-unit step.
Appendix~\ref{app:assignment} identifies these as selections within
the weight-$k$ spaces and Appendix~\ref{app:stats} assigns a chance
probability to each landing.  The rules assign no $S_4$ representation or modular
weight field by field, and they write no modular-invariant Yukawa
matrices; the insertions are statements about effective couplings.
Appendix~\ref{app:assignment} derives the operator charges the
postulates imply, exhibits the field assignment they force at
leading order, and lists what remains open, and the matrices that
realize the insertions are the first requirement on the completion
in Sec.~\ref{sec:uv}.  What the paper claims is that the
insertions, once stated, are parameter-free and overdetermined, and
that their outputs are the numbers of Secs.~\ref{sec:masses}
to~\ref{sec:leptons}.

\begin{table}[t]
\caption{The Jacobi factors and where they enter the construction.
Postulate labels P1 to P8 are defined in
Table~\ref{tab:postulates}.}
\label{tab:jacobi}
\begin{ruledtabular}
\begin{tabular}{lll}
Factor & Leading series & Enters at\\
\hline
$\theta_2(\tau)$ & $2q^{1/8}(1+q+\cdots)$ &
P1 and the reactor identity Eq.~(\ref{eq:identity})\\
$\theta_2(2\tau)$ & $2q^{1/4}(1+q^2+\cdots)$ &
the Cabibbo target P2\\
$\theta_3(\tau)$ & $1+2q^{1/2}+\cdots$ &
the lepton anchors Eq.~(\ref{eq:anchor}), the up charges P3, and
the $1$--$3$ step P8\\
$\theta_4(\tau)$ & $1-2q^{1/2}+\cdots$ &
the atmospheric angle Eq.~(\ref{eq:atm})\\
$(2\Imt)^{1/2}$ & automorphy dressing, one unit of weight &
the top anchor P4 and the light down line P5\\
$\theta_2^4+\theta_4^4=\theta_3^4$ & exact at any $\tau$ &
the modulus-free lepton closure\\
\end{tabular}
\end{ruledtabular}
\end{table}

\section{The base and the modulus from the charged leptons}
\label{sec:base}

\subsection{The anchor identities}

\begin{figure}[t]
\centering
\begin{tikzpicture}[>=stealth,scale=0.9]
  \draw[gray!60] (0,0) circle (1.35);
  \foreach \c/\ang in {0/90,1/0,2/-90,3/180}{
    \fill[gray] (\ang:1.35) circle (0.035);
    \node[gray] at (\ang:0.68) {\small $\c$};}
  \fill (180:1.35) circle (0.07) node[left=4pt] {$e$};
  \fill (0:1.35)  circle (0.07) node[right=4pt] {$\mu$};
  \fill (90:1.35)  circle (0.07) node[above=4pt] {$\tau$};
  \draw[->,very thick] (187:1.05) arc (187:353:1.05)
        node[pos=0.35,below=2pt,font=\small] {$\e^{-5}$};
  \draw[->,thick] (7:1.05) arc (7:83:1.05)
        node[pos=0.5,above right=-3pt,font=\small] {$\e^{-7/2}$};
  \node at (0,-2.05) {$e,\mu,\tau$ at $T$ charges $3,1,0$};
\end{tikzpicture}
\caption{The three charged leptons on the $Z_4$ clock, at $T$
charges $3$, $1$, and $0$, with the anchor ratios of
Eq.~(\ref{eq:anchor}) attached to the steps.  The arcs measure
the clock steps in quarter turns, two for $e\to\mu$ and one for
$\mu\to\tau$, and not the exponents, which the labels record.
The two-unit step carries $\e^{-5}$ and the one-unit step carries
$\e^{-7/2}$, the ordered placement behind Eqs.~(\ref{eq:anchor})
and~(\ref{eq:lepbase}).}
\label{fig:leptonclock}
\end{figure}
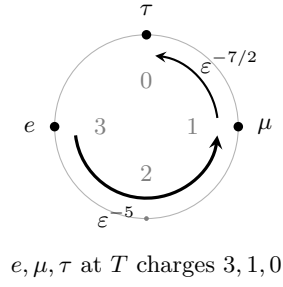

The charged-lepton mass ratios sit on a dressed theta lattice,
\begin{equation}
  \frac{m_\tau}{m_\mu}=\frac{\e^{-7/2}}{(2\theta_3)^4}=16.9868 ,
  \qquad
  \frac{m_\mu}{m_e}=\frac{\e^{-5}}{(2\theta_3)^4}=210.626 ,
\label{eq:anchor}
\end{equation}
against the measured $16.9865$ and $210.664$, each known to
$0.02\%$.  The deviations are $-0.1\sigma$ and
$+1.0\sigma$.\footnote{The Koide relation
$(m_e+m_\mu+m_\tau)/(\sqrt{m_e}+\sqrt{m_\mu}+\sqrt{m_\tau})^2=2/3$~\cite{Koide:1982wm}
holds with pole masses at the $10^{-5}$ level and evaluates to
$0.6678$ on the running masses of Table~\ref{tab:data}, so it is
a pole-mass relation~\cite{Xing:2006vk} and the identities here,
which hold at $\MZ$, are not a rewriting of it.  The structures
also differ.  Koide's relation is a permutation-symmetric
quadratic in $\sqrt{m}$, naturally parametrized on a $Z_3$
circle, while Eqs.~(\ref{eq:anchor}) and~(\ref{eq:lepbase}) are
ordered power laws on the $Z_4$ clock, with the second generation
at the definite displacement $m_\mu/\sqrt{m_em_\tau}=\e^{-3/4}$ of
Eq.~(\ref{eq:base}).}
Figure~\ref{fig:leptonclock} places the three charged leptons on
the clock with the anchor ratios attached to the steps.

\subsection{One pair of identities, three constants}

The identities in Eq.~(\ref{eq:anchor}) determine the framework's
constants three times over.  Dividing one by the other cancels the
common dressing $(2\theta_3)^4$ exactly and leaves a relation among
the lepton masses and the base alone,
\begin{equation}
  \frac{m_em_\tau}{m_\mu^2}=\e^{3/2}=0.08065 ,
\label{eq:lepbase}
\end{equation}
against the measured $0.08063\pm0.00004$.  Inverted, this
$\theta_3$-free relation fixes the base,
\begin{equation}
  \e=\left(\frac{m_em_\tau}{m_\mu^2}\right)^{2/3}
   =0.186642\pm0.000040 ,
\label{eq:base}
\end{equation}
and the rational value $\e=14/75=0.186667$ sits at $-0.6\sigma$.
The fraction is a continued-fraction convergent of the measured
number, with the next convergent requiring denominator $509$.  A
convergent this close at this denominator is not itself evidence,
since any real number has one with probability of order tens of
percent; the rational value is adopted as a convenient reference, and
every result of the paper is unchanged at the $10^{-4}$ level if the
measured value of Eq.~(\ref{eq:base}) is used instead
(Appendix~\ref{app:fits}).

Their $\e$-free closure fixes the modulus through
$|q^{1/4}|=\e$.  And the common dressing $(2\theta_3)^4$ pins it onto
the imaginary axis, since a real part shifts $|\theta_3|$ at second
order in $\Ret$, by $2.4$ standard deviations at $\Ret=0.02$ and by
fifteen at $0.05$, so the anchors hold $|\Ret|$ below about $0.02$.  The modulus is
therefore
\begin{equation}
  \tau=\frac{2i}{\pi}\ln\frac{1}{\e}=1.0685\,i ,\qquad q=\e^{4} ,
\label{eq:tau}
\end{equation}
and every form value is real.  It sits seven
percent above the self-dual point of Sec.~\ref{sec:points}, with
the displacement of Eq.~(\ref{eq:u}) equal to $u=0.0331$, within
five percent of $\e^2$, so the fixed point sets the normalizations
while charge counting in $q$ sets the hierarchy.

At this modulus the theta triple takes the values
$\theta_2=0.86515$, $\theta_3=1.06969$, and $\theta_4=0.93031$,
and one metaplectic unit has the magnitude
$|q^{1/8}|=\sqrt\e$.

\section{The quark masses}
\label{sec:masses}

\subsection{Two exact targets}

Two identities free of the light-quark uncertainties hold at half a
standard deviation,
\begin{align}
  \frac{m_b}{m_\tau} &= \frac{1}{\sqrt{2\e}} = 1.6366 ,
   & &\text{measured } 1.6402\pm0.0091 ,
\label{eq:t1}\\
  |V_{us}|^2\,\frac{m_\mu}{m_e} &= \frac{2}{\e}=10.714 ,
   & &\text{measured } 10.674\pm0.076 .
\label{eq:t2}
\end{align}
Both follow exactly at leading order from a single insertion of the
form $\theta_2$ with unit coefficient.  Its leading coefficient
supplies the $2$, and its leading powers at arguments $\tau$ and
$2\tau$ supply the $\sqrt\e$ and the $\e$.  The exact form of the
first is $m_b/m_\tau=\sqrt2/\theta_2(\tau)=1.6346$, at
$+0.6\sigma$, not yet distinguishable from the leading order.  In a
Froggatt--Nielsen setting~\cite{Froggatt:1978nt} these relations
would be accidents,
because the expansion parameter is an independent vacuum value.  In
a modular setting there is no independent parameter, and relations
of this type are the expected output.

\subsection{Five postulates, six masses}

Five postulates, each one insertion permitted by the level-four rules
with its $\e$ power fixed by charge counting and its coefficient drawn
from the declared dictionary of Appendix~\ref{app:stats}, extend the
two targets into a complete quark spectrum.  Table~\ref{tab:postulates} lists them together with the
three that close the CKM matrix in Sec.~\ref{sec:ckm}, one of them
empirical, and
Fig.~\ref{fig:insertions} places each insertion on the mass grid.

The third
postulate dresses each unit of up-sector generation charge with one
factor of $\theta_3$, which derives the axis unit and the up-quark
spectrum from the top anchor.  The fourth reads the top Yukawa
coupling as the automorphy displacement of the modulus,
$y_t=(\Imt)^{-1/2}=0.9674$ against the measured $0.967\pm0.004$.
Throughout, the electroweak scale is the $\MS$ vacuum expectation
value at $\MZ$ of Ref.~\cite{Antusch:2025rqp}, $v=248.40$~GeV, with
$m_t=y_tv/\sqrt2$.
The top mass thereby determines the modulus a second time, in
agreement with the lattice matching at $0.1\sigma$.  The fifth
completes the down line,
\begin{equation}
  m_d=\frac{m_e}{\e}\,(\Imt)^{1/2},
  \qquad
  m_s=\tfrac12\,m_\mu\,(\Imt)^{1/2} ,
\label{eq:p5}
\end{equation}
one automorphy factor on each light down generation, measured from
its fixed-point value $\sqrt2$ as in the fourth rule of
Sec.~\ref{sec:rules}.  The coefficient $\tfrac12$ is the inverse
of the leading coefficient $2$ of $\theta_2$, and the power
$\e^{-1}$ on $m_d$ is one unit of $T$ charge, $q^{-1/4}$ at the
harmonic modulus, and both relations carry no phase.  The strange quark, at $+1.1\sigma$, is
the one mass in the sector that departs from its relation by more
than half a standard deviation, and Eq.~(\ref{eq:p5}) is the
relation the Georgi--Jarlskog plaquette of Secs.~\ref{sec:pattern}
and~\ref{sec:uv} tests.

Table~\ref{tab:masses} gives the result, with a joint $\chi^2$
near $3$ and no fitted parameter, and the statement it supports is
the central one of the quark sector.
\begin{center}
\emph{All six quark masses are reproduced from three lepton masses
and the electroweak scale.}
\end{center}

\begin{table*}[t]
\caption{The eight postulates and their outputs.  Each mass
postulate is one insertion, as Sec.~\ref{sec:rules} records.  In P2
the coefficient $2$ and the power $q^{1/4}$
are the leading coefficient and power of $\theta_2(2\tau)$, and in
P3 the power $q^{-1}$ equals $\e^{-4}$ at the harmonic modulus.
The deviation quoted for P1 is that of the leading order
$(2\e)^{-1/2}$; the exact form $\sqrt2/\theta_2(\tau)$ sits at
$+0.6\sigma$.  Deviations follow the convention of
Sec.~\ref{sec:intro}.}
\label{tab:postulates}
\begin{ruledtabular}
\begin{tabular}{l p{0.34\textwidth} p{0.30\textwidth} l}
Postulate & Insertion & Relation & Dev.\\
\hline
P1 & one $\theta_2(\tau)$ between the $b$ and $\tau$ Yukawa operators &
$m_b/m_\tau=\sqrt2/\theta_2(\tau)=(2\e)^{-1/2}[1-q+\cdots]$ & $+0.4\sigma$\\
P2 & one $\theta_2$-type entry on the down-lepton $1$--$2$ transition &
$|V_{us}|^2\,m_\mu/m_e=2\,q^{-1/4}=2/\e$ & $-0.5\sigma$\\
P3 & one $\theta_3(\tau)$ per unit of up-sector charge &
$m_t/m_c=q^{-1}/(2^{3/2}\theta_3)$, $m_c/m_u=q^{-1}/(2^{1/2}\theta_3^{2})$ &
$-0.1\sigma$, $-0.3\sigma$\\
P4 & one inverse automorphy factor on the charge-zero anchor &
$y_t=(\Imt)^{-1/2}$ & $-0.1\sigma$\\
P5 & one automorphy factor on the light down generations &
$m_d=(m_e/\e)(\Imt)^{1/2}$, $m_s=\tfrac12 m_\mu(\Imt)^{1/2}$ &
$+0.1\sigma$, $+1.1\sigma$\\
P6 & the factor-two ratio on the $1$--$3$ texture route &
$\theta_{13}/\theta_{23}=2\sqrt{m_u/m_c}$ & $-0.2\sigma$\\
P7 & the Grossman--Ruderman relation $|V_{td}|^2=|V_{cb}|^3$, adopted; equivalent at $0.04\%$ to one half-unit insertion, $|V_{td}|=2|V_{ub}|/\theta_2(\tau)$ &
$|V_{td}|=|V_{cb}|^{3/2}=0.008599$; closure gives $\delta=1.138$ & $0.0\sigma$\\
P8 & two $\theta_3$ dressings on the three-unit $1$--$3$ step, with the inverse leading coefficient of $\theta_2$ &
$|V_{ub}|=\tfrac12\theta_3^2\,\e^3$; with P6, $|V_{cb}|=|V_{ub}|/(2\sqrt{m_u/m_c})$ &
$-0.3\sigma$, $-0.1\sigma$\\
\end{tabular}
\end{ruledtabular}
\end{table*}

\begin{table}[t]
\caption{Six quark masses reproduced from three lepton masses and
the electroweak scale, $v=248.40$~GeV at $\MZ$.  Data from
Ref.~\cite{Antusch:2025rqp}.}
\label{tab:masses}
\begin{ruledtabular}
\begin{tabular}{lllll}
Quantity & Relation & Predicted & Measured & Dev.\\
\hline
$m_t$ & $(\Imt)^{-1/2}v/\sqrt2$ & $169.92$~GeV & $169.85\pm0.70$ & $-0.1\sigma$\\
$m_b$ & $m_\tau/\sqrt{2\e}$ & $2.857$~GeV & $2.863\pm0.016$ & $+0.4\sigma$\\
$m_c$ & $2^{3/2}\e^{4}\theta_3\,m_t$ & $0.6239$~GeV & $0.6253\pm0.0105$ & $+0.1\sigma$\\
$m_s$ & $\tfrac12 m_\mu(\Imt)^{1/2}$ & $53.11$~MeV & $53.75\pm0.58$ & $+1.1\sigma$\\
$m_d$ & $(m_e/\e)(\Imt)^{1/2}$ & $2.701$~MeV & $2.705\pm0.034$ & $+0.1\sigma$\\
$m_u$ & $4\e^{8}\theta_3^{3}\,m_t$ & $1.226$~MeV & $1.237\pm0.026$ & $+0.4\sigma$\\
\end{tabular}
\end{ruledtabular}
\end{table}

\section{The CKM matrix and the unitarity triangle}
\label{sec:ckm}

\subsection{Two elements derived directly}

The Cabibbo element follows from the spectrum through the
Gatto--Sartori--Tonin texture~\cite{Gatto:1968ss}, obtained again
from discrete flavor symmetry and from a texture zero in
Refs.~\cite{Wilczek:1977uh,Fritzsch:1977vd},
\begin{equation}
  |V_{us}|=\sqrt{\frac{m_d}{m_s}}=0.2255 ,
\label{eq:vus}
\end{equation}
against the measured $0.2251\pm0.0008$.  The renormalization-stable
ratio follows from the up sector,
\begin{equation}
  \frac{\theta_{13}}{\theta_{23}}=2\sqrt{\frac{m_u}{m_c}}=0.08865 ,
\label{eq:ratio}
\end{equation}
against the measured $0.08824\pm0.00209$.  The textbook relation
lacks the factor of two.  The rules supply it as the leading
coefficient of $\theta_2$.

\subsection{Closing the matrix}

Two elements remain, $|V_{ub}|$ and $|V_{td}|$, and with them the
phase.  The first is one more insertion of the rules.  The
$1$--$3$ transition sits at three units of charge, the charge the
triplet assignment fixes for it and the order at which a single
multiplet of that charge opens (Appendix~\ref{app:matrices}), and
dressed with two factors of the charge-zero form and the inverse of
the leading coefficient of $\theta_2$, a selection within the
dictionary, not a count, since one factor per unit would give
$0.00398$ at $3.5\sigma$, it reads
\begin{equation}
  |V_{ub}|=\tfrac12\,\theta_3^{2}(\tau)\,\e^{3}=0.003721 ,
\label{eq:P8main}
\end{equation}
against the measured $0.00370\pm0.00008$, at
$-0.3\sigma$.\footnote{The dressing is permitted, not
determined.  Six forms of the dictionary
$2^{k/2}\theta_2^{a}\theta_3^{b}\theta_4^{c}$ lie within $0.3\%$ of
the coefficient an exact fit would take, and by the accident budget
of Appendix~\ref{app:stats} a single percent-level landing is worth
little on its own.  What supports Eq.~(\ref{eq:P8main}) is the
closure it produces below, in which the second relation of
Refs.~\cite{Grossman:2020qrp,Grossman:2022ehc} is returned rather
than adopted.}  With the ratio of P6 it fixes the $2$--$3$ element,
\begin{equation}
  |V_{cb}|=\frac{|V_{ub}|}{2\sqrt{m_u/m_c}}=0.04197 ,
\label{eq:vcb}
\end{equation}
against the measured $0.04193\pm0.00041$, at $-0.1\sigma$.

The second is empirical.  Grossman and Ruderman identified on 2020
data~\cite{Grossman:2020qrp,Grossman:2022ehc} the relation
\begin{equation}
  |V_{td}|^{2}=|V_{cb}|^{3} ,
\label{eq:GR}
\end{equation}
which holds at the $0.2\%$ level on the 2024 values at $\MZ$.  Its
provenance matters.  It was proposed before the data set used
here, and the data have since moved onto it.  Adopted, it gives
$|V_{td}|=0.008599$.  Unitarity then closes the matrix with no
further input.  In the standard parametrization
$|V_{td}|^{2}=(s_{12}s_{23})^{2}+(c_{12}c_{23}s_{13})^{2}
-2s_{12}s_{23}c_{12}c_{23}s_{13}\cos\delta$, and with the three
magnitudes fixed this gives
\begin{equation}
  \delta=1.138 ,
\label{eq:delta}
\end{equation}
against the measured $1.139\pm0.023$.  Every CKM parameter follows
from masses and form values, with one empirical
relation supplying the fourth magnitude.  The companion relation
of Refs.~\cite{Grossman:2020qrp,Grossman:2022ehc},
$|V_{ub}|^{2}|V_{us}|=|V_{cb}|^{4}$, is returned at $0.6\%$ and is
no longer an input, and the closed form the two relations jointly
imply, $\cos\delta=(x^{2}+y^{2}-y\sqrt{x})/2xy$ with
$x=\sqrt{m_d/m_s}$ and $y=2\sqrt{m_u/m_c}$, is returned with it,
giving $\delta=1.142$ against the $1.138$ of
Eq.~(\ref{eq:delta}); both are consequences of the closure rather
than its statement.  Table~\ref{tab:ckm} collects the matrix.

The adopted relation has an insertion reading.  On the tower of
Eq.~(\ref{eq:tower}) below, $|V_{td}|$ sits nine eighteenths above
$|V_{ub}|$, one half unit, so the closure can be written as
\begin{equation}
  \frac{|V_{ub}|}{|V_{td}|}=\frac{\theta_2(\tau)}{2}
  =\sqrt\e\,(1+q+\cdots)=0.4326 ,
\label{eq:halfunit}
\end{equation}
one insertion of the half-unit carrier $\theta_2(\tau)$ with the
inverse of its leading coefficient, the combination P5 and P8 also
use.  Adopted
in place of Eq.~(\ref{eq:GR}) it gives $|V_{td}|=0.008602$ against
$0.008599$, a split of $0.04\%$, and moves $\delta$ from $1.138$ to
$1.139$; the two closures are one closure at this precision, so the
matrix closes on insertions alone with the Grossman--Ruderman
relation returned at $0.04\%$, and the independent provenance of
that relation then stands as a check on the insertion rather than
as an input.  The ratio is independent of $\lambda$ and of $A$, so
it tests the third column with no reference to the Cabibbo sector,
and it is measured at $0.431\pm0.016$ on the global-fit values, on
$\sqrt\e$ within three tenths of a percent, and at $0.444$ with the
direct semileptonic $|V_{ub}|$, the spread being that of the
$|V_{ub}|$ determinations.

\begin{table}[t]
\caption{The CKM matrix, all four parameters derived from the
quark spectrum and the form values at the harmonic modulus, with
one empirical relation supplying $|V_{td}|$, and consequences from
unitarity.}
\label{tab:ckm}
\begin{ruledtabular}
\begin{tabular}{lll}
Quantity & Value & Measured\\
\hline
$\sin\theta_{12}$ & $0.2255$ & $0.2251\pm0.0008$\\
$\sin\theta_{23}$ & $0.04197$ & $0.04193\pm0.00041$\\
$\sin\theta_{13}$ & $0.003721$ & $0.00370\pm0.00008$\\
$\delta$ & $1.138$ & $1.139\pm0.023$\\
\hline
$|V_{td}|$ & $0.008599$ & \\
$J$ & $3.11\times10^{-5}$ & $(3.09\pm0.07)\times10^{-5}$\\
\end{tabular}
\end{ruledtabular}
\end{table}

\subsection{The eighteenths tower}

\begin{figure}[t]
\begin{tikzpicture}
\begin{axis}[width=8.4cm,height=4.6cm,
  xmin=0.5, xmax=4.5, ymin=-0.62, ymax=0.62,
  xtick={1,2,3,4},
  xticklabels={$|V_{us}|$,$|V_{cb}|$,$|V_{td}|$,$|V_{ub}|$},
  ylabel={$18\ln|V|/\ln\e\; -\; n$},
  ytick={-0.5,0,0.5},
  tick label style={font=\small}, label style={font=\small}]
\addplot[dashed,domain=0.5:4.5] {0.5};
\addplot[dashed,domain=0.5:4.5] {-0.5};
\addplot[thick,domain=0.5:4.5] {0};
\addplot+[only marks,black,mark=*,mark size=1.6pt,
  error bars/.cd,y dir=both,y explicit,error bar style={black}]
  coordinates {
  (1,-0.008) +- (0,0.038)
  (2, 0.015) +- (0,0.105)
  (3, 0.005) +- (0,0.185)
  (4, 0.050) +- (0,0.232)};
\node[font=\scriptsize,anchor=west] at (axis cs:0.55,0.44) {neighboring rung};
\node[font=\scriptsize,anchor=north] at (axis cs:1,-0.10) {$n=16$};
\node[font=\scriptsize,anchor=north] at (axis cs:2,-0.13) {$n=34$};
\node[font=\scriptsize,anchor=north] at (axis cs:3,-0.24) {$n=51$};
\node[font=\scriptsize,anchor=north] at (axis cs:4,-0.24) {$n=60$};
\end{axis}
\end{tikzpicture}
\caption{The four CKM magnitudes on the eighteenths lattice.  Each
sits on its rung within $0.2$ standard deviations while the
neighboring rung lies half a unit away.}
\label{fig:ladder}
\end{figure}
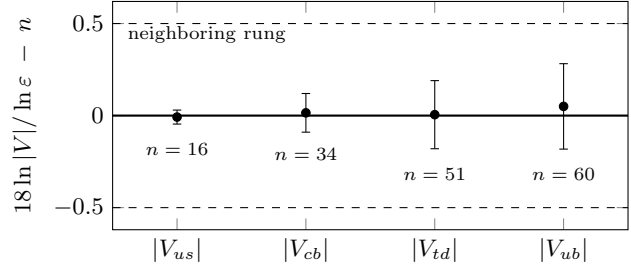

The four off-diagonal magnitudes sit individually on one lattice,
\begin{equation}
  |V_{us}|=\e^{16/18},\ \ |V_{cb}|=\e^{34/18},\ \
  |V_{ub}|=\e^{60/18},\ \ |V_{td}|=\e^{51/18},
\label{eq:tower}
\end{equation}
each within $0.2\sigma$ of its rung, with the neighboring rung half
a unit away (Fig.~\ref{fig:ladder}).  The tower embeds the relation of Eq.~(\ref{eq:GR}) exactly and
returns its companion at $0.6\%$.  The exponent quantum is an
eighteenth, half the ninth that quantizes the mass ratios.  Half a
step is exactly what half-integral weights on the metaplectic
cover supply, so the doubling between the mass and mixing lattices
is a structural feature of the rules.  The one odd exponent, $51$
in $|V_{td}|$, sits in the CP sector, though its oddness follows
from P7 as $51=\tfrac32\times34$ and the realization of
Appendix~\ref{app:matrices} places the phase on the $(1,3)$ entry of
exponent $60$, so the association of CP with the odd step is
suggestive, not structural.

Two features of the exponents are worth stating.  The exponents of
$|V_{us}|$ and $|V_{cb}|$, $16$ and $34$, are $18-2$ and $36-2$: each
sits one ninth below an integer power, so the pair reads
$|V_{us}|=r\,\e$ and $|V_{cb}|=r\,\e^{2}$ with the one coefficient
$r=\e^{-1/9}=1.205$, which is $1/A$ and the reason $A\lambda=\e$
holds exactly on the tower in Eq.~(\ref{eq:wolfenstein}) below.  And
the differences $34-16=18$ and $60-51=9$ are one unit and one half
unit, so the tower embeds, besides Eq.~(\ref{eq:GR}), the two
relations
\begin{equation}
  \frac{|V_{cb}|}{|V_{us}|}=\e ,\qquad
  \frac{|V_{ub}|}{|V_{td}|}=\sqrt\e ,
\label{eq:towerpair}
\end{equation}
exactly, the second being Eq.~(\ref{eq:halfunit}) at leading order,
together with their product,
$|V_{cb}|/|V_{us}|=(|V_{ub}|/|V_{td}|)^{2}$, which in Wolfenstein
language and to leading order in $\lambda$ reads
$A\lambda=R_b^{2}/R_t^{2}=\e$: the squared ratio of the two sides of
the unitarity triangle is the base.  The closed matrix gives
$0.1861$, $0.4327$, and $0.1873$ for the three, and the global-fit
values give $0.186$, $0.431$, and $0.185$ to $0.19$, against
$0.1867$, $0.4320$, and $0.1867$.  Any two of the three imply the
third; the first fixes the coefficient the Cabibbo sector carries,
the second is free of it.

\subsection{CP violation geometrized}

With the matrix closed, the sides of the unitarity triangle are
outputs, $R_b=|V_{ud}V_{ub}^{*}|/|V_{cd}V_{cb}^{*}|=0.3832$ and
$R_t=|V_{td}V_{tb}^{*}|/|V_{cd}V_{cb}^{*}|=0.9082$, against the
UTfit values $0.383\pm0.010$ and
$0.908\pm0.010$~\cite{UTfit:2022hsi}.  The lattice readings
$R_b=\e^{10/18}=0.3936$ and $R_t=\e^{1/18}=0.9110$ of the
eighteenths tower remain the coarse statement, within $1.1\sigma$
and $0.3\sigma$, and the exact sides refine them.  The triangle is
rigid.  Its sides are $1$, $R_b$, and $R_t$, so the angles follow
from the law of cosines,
\begin{equation}
\begin{aligned}
  \cos\gamma&=\frac{1+R_b^2-R_t^2}{2R_b} ,\qquad
  \cos\beta=\frac{1+R_t^2-R_b^2}{2R_t} ,\\
  \alpha&=\pi-\beta-\gamma ,
\end{aligned}
\label{eq:angles}
\end{equation}
and each is a function of masses and form values alone.

The angles come out
\begin{equation}
  \beta=22.5^\circ ,\qquad \gamma=65.2^\circ ,\qquad \alpha=92.3^\circ ,
\label{eq:anglevalues}
\end{equation}
against the measured $22.4\pm0.7$, $65.1\pm1.3$, and $92.4\pm1.4$
from the UTfit global analysis~\cite{UTfit:2022hsi}, at
$-0.2\sigma$, $0.0\sigma$, and $+0.1\sigma$.  In the standard parametrization $\gamma$ coincides
with the phase $\delta$ of Eq.~(\ref{eq:delta}) to $0.03^\circ$.
The apex lands at $(\bar\rho,\bar\eta)=(0.161,\,0.348)$ against
the measured $(0.161\pm0.010,\,0.347\pm0.010)$, and the Jarlskog
invariant is $J=3.11\times10^{-5}$ against
$(3.09\pm0.07)\times10^{-5}$.  In the Wolfenstein
parametrization~\cite{Wolfenstein:1983yz,Buras:1994ec} the four
parameters are $\lambda=|V_{us}|$, $A=|V_{cb}|/\lambda^2$, and
the apex.  The construction gives $\lambda=\sqrt{m_d/m_s}=0.2255$
and $A=0.825$ against the measured $0.2251\pm0.0008$ and
$0.828\pm0.010$.  On the lattice the whole set is a set of powers
of the base,
\begin{equation}
  \lambda=\e^{8/9} ,\qquad
  A=\e^{1/9} ,\qquad
  |\bar\rho+i\bar\eta|=R_b=\e^{5/9} ,
\label{eq:wolfenstein}
\end{equation}
with the two exponents of $\lambda$ and $A$ summing to one, so
that $A\lambda=\e$ is the stable ratio $|V_{cb}|/|V_{us}|$ in
Wolfenstein language, one of the routes to the base collected
below.
Table~\ref{tab:triangle} gives the triangle and the Wolfenstein
parameters.

\begin{table}[t]
\caption{The unitarity triangle and the Wolfenstein parameters
from the closed matrix.  Measured sides are derived from the
UTfit apex~\cite{UTfit:2022hsi}.}
\label{tab:triangle}
\begin{ruledtabular}
\begin{tabular}{llll}
Quantity & Form & Predicted & Measured\\
\hline
$\lambda$ & $\sqrt{m_d/m_s}$, lattice $\e^{8/9}$ & $0.2255$ & $0.2251\pm0.0008$\\
$A$ & $|V_{cb}|/\lambda^2$, lattice $\e^{1/9}$ & $0.825$ & $0.828\pm0.010$\\
$R_b$ & closed matrix, lattice $\e^{10/18}$ & $0.3832$ & $0.383\pm0.010$\\
$R_t$ & closed matrix, lattice $\e^{1/18}$ & $0.9082$ & $0.908\pm0.010$\\
$\beta$ & Eq.~(\ref{eq:angles}) & $22.5^\circ$ & $22.4\pm0.7$\\
$\gamma$ & Eq.~(\ref{eq:angles}) & $65.2^\circ$ & $65.1\pm1.3$\\
$\alpha$ & $\pi-\beta-\gamma$ & $92.3^\circ$ & $92.4\pm1.4$\\
$\bar\rho$ & $R_b\cos\gamma$ & $0.161$ & $0.161\pm0.010$\\
$\bar\eta$ & $R_b\sin\gamma$ & $0.348$ & $0.347\pm0.010$\\
$J$ & from the closed matrix & $3.11\times10^{-5}$ & $(3.09\pm0.07)\times10^{-5}$\\
\end{tabular}
\end{ruledtabular}
\end{table}

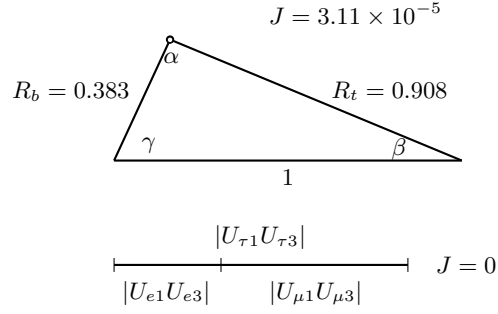
\begin{figure}[t]
\begin{tikzpicture}[scale=4.6]
\draw[thick] (0,0) -- (1,0);
\draw[thick] (0,0) -- (0.1610,0.3478);
\draw[thick] (1,0) -- (0.1610,0.3478);
\fill (0.1610,0.3478) circle (0.012);
\draw (0.161,0.337) -- (0.161,0.357);
\draw (0.151,0.347) -- (0.171,0.347);
\fill[white] (0.161,0.347) circle (0.008);
\draw (0.161,0.347) circle (0.008);
\node[below] at (0.5,0) {\small $1$};
\node[left] at (0.07,0.20) {\small $R_b=0.383$};
\node[right] at (0.60,0.20) {\small $R_t=0.908$};
\node at (0.10,0.045) {\small $\gamma$};
\node at (0.82,0.033) {\small $\beta$};
\node at (0.165,0.295) {\small $\alpha$};
\node[anchor=west] at (0.42,0.42)
  {\small $J=3.11\times10^{-5}$};
\begin{scope}[shift={(0,-0.30)}]
\draw[thick] (0,0) -- (0.845,0);
\draw (0.3075,-0.02) -- (0.3075,0.02);
\draw (0,-0.02) -- (0,0.02);
\draw (0.845,-0.02) -- (0.845,0.02);
\node[below] at (0.15,-0.02) {\small $|U_{e1}U_{e3}|$};
\node[below] at (0.58,-0.02) {\small $|U_{\mu1}U_{\mu3}|$};
\node[above] at (0.42,0.02) {\small $|U_{\tau1}U_{\tau3}|$};
\node[anchor=west] at (0.90,0.0) {\small $J=0$};
\end{scope}
\end{tikzpicture}
\caption{The two unitarity triangles.  Top, the quark triangle built from the sides $R_b$ and $R_t$ of
the closed matrix, with the predicted apex (dot) inside the measured
apex (open circle with error bars).
Bottom, the leptonic $(1,3)$-column triangle at
$\delta_{CP}=\pi$, where the inequality saturates,
$0.338=0.123+0.215$, the area vanishes, and $J=0$.}
\label{fig:triangles}
\end{figure}

Figure~\ref{fig:triangles} draws the triangle.  The reading is
geometric.  A triangle with sides $1$, $R_b$, and $R_t$ lies flat
on the real axis, with vanishing area and a phase of $0$ or $\pi$,
only if the sides satisfy $R_b+R_t=1$ or $|1-R_b|=R_t$.  The
closed values $R_b=0.383$ and $R_t=0.908$ satisfy neither, so the
apex is forced off the axis, and Eq.~(\ref{eq:angles}) returns
$\cos\gamma=0.42$, not $\pm1$.  The phase measures how far
the two sides are from a degenerate triangle, and the Jarlskog
invariant, twice the area of the triangle in the unrescaled matrix,
measures the same failure as an area.

This is the sense in which quark CP violation is not an input
phase but the closure defect of the magnitudes, and the side that
drives the area is $R_t$, one lattice step below unity and the
odd-parity member of the tower.

The angles are the sharpest near-term test of that reading, since
$\beta$ from $B\to J/\psi K_S$ and $\gamma$ from $B\to DK$ are
measured directly and both will reach sub-degree precision at LHCb
Upgrade~II and Belle~II, against predictions with no
uncertainty beyond the base, the form values, and the one adopted
relation.  The direct determination of $\gamma$ is independent of
the loop-level inputs that enter the global fit.  The current
LHCb combination of tree-level $B\to DK$ and related decays,
$\gamma=(62.8\pm2.6)^\circ$~\cite{LHCb:2021dcr,LHCb:gamma2025},
sits at $-0.9\sigma$ from the prediction, and the combined Belle
and Belle~II determination, $\gamma\equiv\phi_3=(75.2\pm7.6)^\circ$
in that collaboration's notation~\cite{Belle:2024gamma}, at
$+1.3\sigma$.
The quantized second torus of Sec.~\ref{sec:cporigin} would put
$\gamma$ at $67.5^\circ$ instead, and the same measurement
separates that reading from the closure by $2.3^\circ$.

The magnitudes fix $\cos\delta$ and leave open only the sign of
$\delta$, the orientation of the triangle, which is taken from the
data.  Every
form value in this paper is real, the modulus lying on the
imaginary axis, so the phase the closure requires must be supplied by
the completion rather than by a displacement of $\tau$, which
the charged-lepton anchors of Sec.~\ref{sec:base} exclude above
$|\Ret|\simeq0.02$.  Section~\ref{sec:cporigin} locates it in a
second torus of the completion, and Sec.~\ref{sec:torusUT}
identifies the phase that torus supplies as the angle $\gamma$.

\subsection{The base returned by the matrix}

The rephasing-invariant ratio of the closed matrix ties the
construction back to the leptons,
\begin{equation}
  \frac{|V_{cb}||V_{ub}|}{|V_{us}||V_{td}|}=0.08054
  \simeq\e^{3/2}=\frac{m_em_\tau}{m_\mu^{2}}=0.08065 ,
\label{eq:beta}
\end{equation}
a split of $0.13\%$ between the closed matrix and the base the leptons
fix at $0.02\%$.  The exponent $3/2$ is an identity of the tower
assignment, $34+60-16-51=27$ eighteenths, so the ratio is a
consistency check between the closed matrix and the tower rather than
an independent recovery of the base; in the split the $0.3\%$ by which $A\lambda$ falls below
$\e$ and the $0.6\%$ at which the second Grossman--Ruderman relation is
returned partly cancel; both are scheduled tests of Sec.~\ref{sec:tests}.
Table~\ref{tab:base} collects the lattice-matching definition and
three determinations.  One rational number
threads the quark masses, the top Yukawa coupling, the CKM matrix,
and the charged leptons.

\begin{table}[t]
\caption{The base and the modulus, one definition and three determinations.}
\label{tab:base}
\begin{ruledtabular}
\begin{tabular}{lll}
Route & Result & Dev.\\
\hline
lattice matching $|q^{1/4}|=\e$ & defines $\Imt=1.0685$ & \\
top anchor $y_t^2\,\Imt=1$ & $\Imt=1.0694\pm0.0089$ & $0.1\sigma$\\
CKM ratio $|V_{cb}|/|V_{us}|$ & $0.18627\pm1.0\%$ & $-0.2\sigma$\\
leptons $(m_em_\tau/m_\mu^2)^{2/3}$ & $0.186642\pm0.000040$ & $-0.6\sigma$\\
\end{tabular}
\end{ruledtabular}
\end{table}

\section{Neutrino masses and lepton mixing}
\label{sec:leptons}

\subsection{The reactor identity}

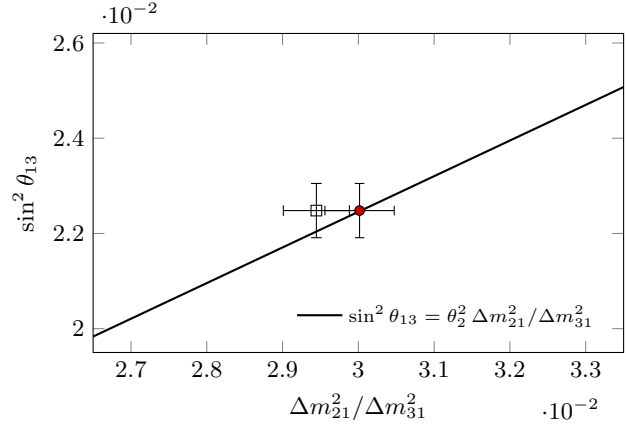
\begin{figure}[t]
\begin{tikzpicture}
\begin{axis}[width=8.6cm,height=5.8cm,
  xmin=0.0265, xmax=0.0335, ymin=0.0195, ymax=0.0262,
  xlabel={$\Delta m^2_{21}/\Delta m^2_{31}$},
  ylabel={$\sin^2\theta_{13}$},
  legend style={font=\scriptsize,at={(0.97,0.05)},anchor=south east,
    draw=none,fill=none},
  tick label style={font=\small}, label style={font=\small},
  xticklabel style={/pgf/number format/fixed,
    /pgf/number format/precision=3},
  yticklabel style={/pgf/number format/fixed,
    /pgf/number format/precision=3}]
\addplot[thick,domain=0.026:0.034] {0.748481*x};
\addlegendentry{$\sin^2\theta_{13}=\theta_2^2\,\Delta m^2_{21}/\Delta m^2_{31}$}
\addplot+[only marks,black,mark=*,mark size=1.8pt,
  error bars/.cd, x dir=both, x explicit, y dir=both, y explicit,
  error bar style={black}]
  coordinates {(0.030016,0.02248) +- (0.000457,0.00057)};
\addplot+[only marks,black,mark=square,mark size=2.0pt,
  error bars/.cd, x dir=both, x explicit, y dir=both, y explicit,
  error bar style={black}]
  coordinates {(0.029446,0.02248) +- (0.000436,0.00057)};
\end{axis}
\end{tikzpicture}
\caption{The reactor identity as a parameter-free slope through
the origin.  The filled point is the global fit and the open
square uses the 207-day JUNO splittings.  Improved precision on
either axis tests the slope directly.}
\label{fig:slope}
\end{figure}

The neutrino sector is described with no adjustable parameter.
The central result relates the one small PMNS angle to the one
small neutrino mass ratio through a single $\theta_2$ insertion,
\begin{equation}
  \sin\theta_{13}=\theta_2(\tau)\,\frac{m_2}{m_3}
  =2\sqrt\e\,\frac{m_2}{m_3}\,\bigl[1+q\bigr].
\label{eq:identity}
\end{equation}
The measured combination on the 207-day JUNO splittings is
$\sin\theta_{13}/(m_2/m_3)=0.8738\pm0.0128$ against
$\theta_2(\tau)=0.86515$, an agreement at $0.7\sigma$.  On the
NuFIT global splittings the agreement is exact to the quoted
digits.  The identity is inert under running, it requires the normal
ordering (Sec.~\ref{sec:octant}), and it bounds the lightest mass
below $2$~meV.
Figure~\ref{fig:slope} shows the identity as a slope through the
origin, which advancing JUNO precision tests directly and which
separates it from every fixed-number alternative.

\subsection{The harmonic spectrum}

The same two to one step that names the model completes the
spectrum.  With the first-to-second logarithmic step twice the
second-to-third,
\begin{equation}
  m_1=m_2\left(\frac{m_2}{m_3}\right)^2=0.25~{\rm meV},
\label{eq:m1}
\end{equation}
inside the identity's bound.  The measured splitting ratio sits on
a charged-lepton fixed number,
\begin{equation}
  \frac{\Delta m^2_{21}}{\Delta m^2_{31}}=\frac{m_\mu}{2m_\tau}=0.02944 ,
\label{eq:splitratio}
\end{equation}
against the measured $0.02945\pm0.00044$.  This is a single landing
at $1.5\%$ precision with no insertion rule behind it, and
Appendix~\ref{app:stats} assesses it accordingly; it is retained
because it closes the spectrum and because JUNO will test it at the
per-mille level.  Combined with the reactor
identity of Eq.~(\ref{eq:identity}), which gives
$m_2/m_3=\sin\theta_{13}/\theta_2$, it predicts the reactor angle
outright from the charged leptons,
\begin{equation}
  \sin^2\theta_{13}=\theta_2^{2}(\tau)\,\frac{m_\mu}{2m_\tau}=0.02203 ,
\label{eq:th13pred}
\end{equation}
that is $\theta_{13}=8.54^\circ$, with no neutrino input and no free
parameter, up to the $m_1^2$ correction below the per-mille level.

The shape of the spectrum is therefore fixed by the charged leptons
alone, $m_2/m_3=(m_\mu/2m_\tau)^{1/2}$ from
Eq.~(\ref{eq:splitratio}) and $m_1/m_3=(m_\mu/2m_\tau)^{3/2}$ from
Eq.~(\ref{eq:m1}), and one measured splitting sets the scale.  With
$m_3$ from $\Delta m^2_{31}$, the lightest mass is the $0.25$~meV
above and the mass sum follows,
$\Sigma m_\nu=m_3\bigl[1+(m_\mu/2m_\tau)^{1/2}+(m_\mu/2m_\tau)^{3/2}\bigr]
=0.0589\pm0.0003$~eV, the minimal normal-ordering value.  The cosmological bounds it is compared with all assume
$\Lambda$CDM with three degenerate neutrino species and a prior that
extends to $\Sigma m_\nu=0$.  On that assumption it sits at the DESI
DR2 bound of $0.064$~eV~\cite{DESI:2025} and above the $0.055$~eV
that the combination of Planck, ACT, and SPT-3G with DESI gives,
where the minimal normal-ordering value itself is disfavored at
$96.6\%$ confidence, a $\Delta\chi^2$ of $5.1$ or about
$2.3\sigma$~\cite{SPT:2026}.  The model shares that tension with
every normal-ordered spectrum and inherits whatever resolves it,
whether a systematic or a departure from $\Lambda$CDM, since the
same data with a dynamical dark-energy equation of state relax the
bound above $0.1$~eV.  The prediction is two-sided.  A
cosmological determination above or below $0.059$~eV at the
few-per-cent level excludes it, and the preference of the
current data for a small or vanishing sum makes this the nearest
cosmological test the model faces.  In dark-matter terms the cosmic
neutrino background is fixed at
$\Omega_\nu h^2=\Sigma m_\nu/(93.14~{\rm eV})=6.3\times10^{-4}$,
half a per cent of the cold component, with no freedom.

\subsection{The large angles}

The large angles are structural, not hierarchical.  The
natural level-four structures are the trimaximal patterns of
$S_4$~\cite{HPS,KingZhou,Krishnan2022}, and the data select the
TM$_1$ column broken by one small real rotation.  The rules allow
a rotation of the size the data require,
$\theta_{12}^\nu=\e^{5/2}=0.0150$, five metaplectic units on the
fine lattice.  The identification is weak on its own, since the
measured solar angle fixes the rotation only to $0.015\pm0.007$ and
Appendix~\ref{app:stats} records it as such; what it supplies is a
definite value for JUNO to test.  It acts in the neutrino sector, in the
$1$--$2$ plane of the neutrino mass basis, multiplying the
trimaximal pattern from the right with the sign that subtracts from
the solar angle, so it breaks the fixed TM$_1$ column at order
$\theta_{12}^\nu$ and leaves the reactor and atmospheric angles
untouched,
\begin{equation}
  \sin^2\theta_{12}
  =\frac{1-3\sin^2\theta_{13}}{3\cos^2\theta_{13}}
  -2\sin\theta_{12}\cos\theta_{12}\,\theta_{12}^\nu
  +O\bigl((\theta_{12}^\nu)^{2}\bigr)
  =0.3041 ,
\label{eq:solar}
\end{equation}
against the 207-day $0.3036\pm0.0064$, at $0.1\sigma$.  The first
term is the exact TM$_1$ relation, $0.3180$ at the predicted
reactor angle, the shift evaluates to $-0.0140$, and the full
rotation reproduces the same $0.3041$.

The neutrino placement is
forced, not chosen.  In $U=U_e^\dagger U_\nu$ a charged-lepton
rotation acts on the rows, and a row rotation of any size shifts
the solar angle by $2\theta\,U_{12}U_{22}/\cos^2\theta_{13}$ and
cannot reproduce the column breaking, and the bound this places on the
charged-lepton sector is made quantitative in Sec.~\ref{sec:uv}.
The three PMNS insertions thus all act in the
neutrino sector, $\theta_2$ on the $1$--$3$ entry, $\theta_4$ on
the $2$--$3$ entry, and $\e^{5/2}$ on the $1$--$2$ plane.

The atmospheric angle takes the third theta
constant,
\begin{equation}
  \sin^2\theta_{23}=\tfrac12\,\theta_4(\tau)=\tfrac12-\e^2=0.4652 ,
\label{eq:atm}
\end{equation}
against the NuFIT~6.1 normal-ordering value
$0.470^{+0.017}_{-0.014}$~\cite{Esteban:2024eli}, at $+0.3\sigma$.
The deviation from maximality is one half power of the nome, and
its sign, the first octant, is a prediction of the assignment
(Sec.~\ref{sec:octant}).

Each theta constant then
carries one leptonic assignment, and at a common modulus
the three satisfy the Jacobi identity Eq.~(\ref{eq:jacobi}), which
eliminates the modulus
and leaves one parameter-free condition on the mixings.  The data
satisfy it within half a standard deviation.  Inverted, with
$\theta_3$ fixed by the lepton anchors of Eq.~(\ref{eq:anchor}) at
$0.02\%$ and $\theta_2$ read from the measured reactor combination of
Eq.~(\ref{eq:identity}), the identity predicts the atmospheric
angle without the modulus,
\begin{equation}
  \sin^2\theta_{23}=\tfrac12\bigl(\theta_3^{4}-\theta_2^{4}\bigr)^{1/4}
  =0.4616\pm0.0054 ,
\label{eq:atmpred}
\end{equation}
on the 207-day JUNO combination, and $0.4651\pm0.0052$ on the
global splittings, three times sharper than the direct
determination and in the first octant in either case.

\subsection{Normal ordering with the first octant predicted}
\label{sec:octant}

Two discrete features of the lepton sector are outputs, not
inputs.  The ordering is normal.  The reactor identity of
Eq.~(\ref{eq:identity}) admits no inverted spectrum, since there
$m_3$ is the lightest state, $m_2/m_3$ exceeds unity, and
$\theta_2(\tau)=0.865$ cannot carry that ratio to a small reactor
angle.  The harmonic spectrum of Eq.~(\ref{eq:m1}) then places
$m_1$ at $0.25$~meV.  The octant is the first.  On the imaginary
axis the nome is real and positive, so
$\theta_4=1-2q^{1/2}+2q^{2}-\cdots$ is an alternating series below
one for every modulus of the construction, and the assignment of
$\theta_4$ to the atmospheric angle in Eq.~(\ref{eq:atm}) places
$\sin^2\theta_{23}$ below one half by exactly $q^{1/2}=\e^2$.  The
second octant would require $\theta_3$ in that slot, but
$\theta_3$ is committed to the charged-lepton anchors and the
up-sector charges.  Neither feature is adjustable.

\begin{figure}[t]
\begin{tikzpicture}
\begin{axis}[width=8.6cm,height=5.6cm,
  xmin=0.42, xmax=0.60, ymin=0, ymax=10,
  xlabel={$\sin^2\theta_{23}$}, ylabel={$\Delta\chi^2$},
  legend style={font=\scriptsize,at={(0.5,0.97)},anchor=north,
    draw=none,fill=none},
  tick label style={font=\small}, label style={font=\small},
  xticklabel style={/pgf/number format/fixed,
    /pgf/number format/precision=2}]
\fill[gray!30] (axis cs:0.4562,0) rectangle (axis cs:0.4670,10);
\addplot[thick] coordinates {(0.380,91.810) (0.385,79.359) (0.390,68.076) (0.395,57.864) (0.400,48.678) (0.405,40.492) (0.410,33.253) (0.415,26.912) (0.420,21.379) (0.425,16.532) (0.430,12.417) (0.435,9.029) (0.440,6.274) (0.445,4.111) (0.450,2.445) (0.455,1.247) (0.460,0.498) (0.465,0.116) (0.470,0.000) (0.475,0.100) (0.480,0.394) (0.485,0.804) (0.490,1.279) (0.495,1.739) (0.500,2.119) (0.505,2.411) (0.510,2.585) (0.515,2.600) (0.520,2.494) (0.525,2.282) (0.530,1.946) (0.535,1.551) (0.540,1.186) (0.545,0.885) (0.550,0.756) (0.555,0.824) (0.560,1.130) (0.565,1.780) (0.570,2.875) (0.575,4.482) (0.580,6.633) (0.585,9.354) (0.590,12.730) (0.595,16.789) (0.600,21.578) (0.605,27.187) (0.610,33.600) (0.615,40.904) (0.620,49.164) (0.625,58.451) (0.630,68.817) (0.635,80.242) (0.640,92.797) (0.645,106.567) (0.650,121.605)};
\addlegendentry{with atmospheric tables}
\addplot[thick,dashed] coordinates {(0.380,71.680) (0.385,62.363) (0.390,53.814) (0.395,46.006) (0.400,38.948) (0.405,32.575) (0.410,26.898) (0.415,21.890) (0.420,17.457) (0.425,13.627) (0.430,10.348) (0.435,7.581) (0.440,5.300) (0.445,3.470) (0.450,2.094) (0.455,1.097) (0.460,0.452) (0.465,0.099) (0.470,0.000) (0.475,0.103) (0.480,0.399) (0.485,0.782) (0.490,1.238) (0.495,1.708) (0.500,2.116) (0.505,2.428) (0.510,2.634) (0.515,2.700) (0.520,2.614) (0.525,2.423) (0.530,2.127) (0.535,1.781) (0.540,1.450) (0.545,1.192) (0.550,1.034) (0.555,1.080) (0.560,1.392) (0.565,1.928) (0.570,2.808) (0.575,4.083) (0.580,5.753) (0.585,7.915) (0.590,10.555) (0.595,13.726) (0.600,17.465) (0.605,21.781) (0.610,26.737) (0.615,32.319) (0.620,38.594) (0.625,45.610) (0.630,53.348) (0.635,61.873) (0.640,71.165) (0.645,81.269) (0.650,92.184)};
\addlegendentry{without}
\addplot[thick,gray] coordinates {(0.4652,0) (0.4652,10)};
\addlegendentry{$\tfrac12\theta_4(\tau)$}
\addplot[dotted] coordinates {(0.42,1) (0.60,1)};
\addplot[dotted] coordinates {(0.42,4) (0.60,4)};
\node[font=\scriptsize,anchor=east] at (axis cs:0.598,1.4) {$1\sigma$};
\node[font=\scriptsize,anchor=east] at (axis cs:0.598,4.4) {$2\sigma$};
\node[font=\scriptsize,anchor=south] at (axis cs:0.470,3.0) {first octant};
\node[font=\scriptsize,anchor=south] at (axis cs:0.545,3.0) {second octant};
\end{axis}
\end{tikzpicture}
\caption{The two normal-ordering octant solutions and the model.
The curves are the one-dimensional $\Delta\chi^2$ projections of
NuFIT~6.1~\cite{Esteban:2024eli} for normal ordering, with (solid)
and without (dashed) the tabulated Super-Kamiokande and IceCube
atmospheric samples.  Both variants place the global minimum in the
first octant at $0.470$ and a second-octant local minimum at
$0.55$, higher by $\Delta\chi^2=0.76$ and $1.03$ respectively,
with a barrier near $2.5$ at maximal mixing.  The vertical line is
the prediction $\sin^2\theta_{23}=\tfrac12\theta_4(\tau)$, which
lies in the first octant for every modulus on the imaginary axis,
and the band is the modulus-free inversion of the Jacobi identity,
Eq.~(\ref{eq:atmpred}), $0.4616\pm0.0054$, with $\theta_2$ from the
207-day JUNO reactor
combination and $\theta_3$ from the lepton anchors.}
\label{fig:octant}
\end{figure}
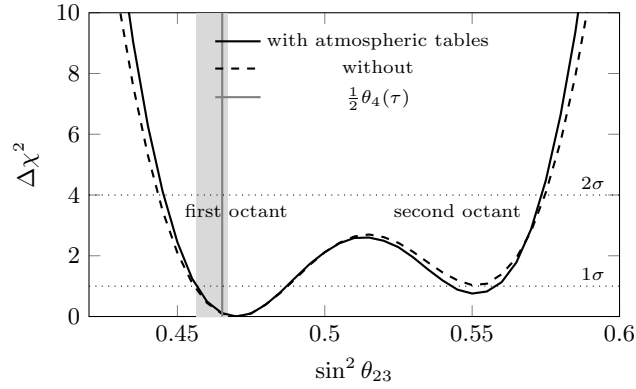

The global fits leave both open, and they leave them correlated.
NuFIT~6.1 places the normal-ordering best fit in the first octant
at $0.470^{+0.017}_{-0.014}$ whether or not the tabulated
Super-Kamiokande and IceCube atmospheric samples are included, with
a second-octant local minimum at $0.55$ higher by only
$\Delta\chi^2=0.76$ and $1.03$ in the two variants
(Fig.~\ref{fig:octant}).  The first-octant preference is recent,
the previous release having placed the variant without those
samples in the second octant, and the $3\sigma$ range still reaches
$0.587$.  For the inverted ordering both variants prefer the second
octant, $0.550$ to $0.555$.  The model therefore predicts the
pairing the fits do not yet resolve, normal ordering with the first
octant, against the pairing that an inverted ordering would bring.
JUNO decides the ordering on its own, and DUNE and Hyper-Kamiokande
decide the octant, so one prediction is tested from two sides, and
either an inverted ordering or a second-octant angle would falsify
the assignment.

\subsection{CP in the lepton sector}

With the modulus on the imaginary axis every form value is real, so
the Dirac phase is CP conserving and the construction sits on
$\delta_{CP}=\pi$, consistent with the measured
$(212^{+26}_{-36})^\circ$ at $0.9\sigma$ in the NuFIT~6.1 variant
with the tabulated atmospheric samples.  The variant without them
sits near $177^\circ$, essentially on the CP-conserving point, so
the comparison quoted is the conservative one.  The Jarlskog
invariant vanishes.  In modulus language the $(1,3)$-column triangle
inequality saturates, $0.338=0.123+0.215$, the flattened triangle
of Fig.~\ref{fig:triangles}.  Real matrices quantize the Majorana
phases, so the effective mass of neutrinoless double beta decay
takes discrete values,
\begin{equation}
  m_{\beta\beta}\in\{1.3,\ 1.6,\ 3.5,\ 3.9\}~{\rm meV}.
\label{eq:mbb}
\end{equation}
All four lie below the reach of the ton-scale searches, so the
Majorana structure behind the masses is invisible to neutrinoless
double beta decay, and a signal at any level would falsify the
assignment as surely as a wrong ordering.
The two sectors then close one picture.  CP violation appears in the
quark triangle, where the tower carries its one odd exponent, and is
absent in the Dirac PMNS, where every form value is real.  Table~\ref{tab:leptons} collects the lepton
comparisons.

\begin{table}[t]
\caption{The lepton sector with no adjustable parameter.}
\label{tab:leptons}
\begin{ruledtabular}
\begin{tabular}{lll}
Quantity & Predicted & Measured\\
\hline
$\sin\theta_{13}/(m_2/m_3)$ & $\theta_2=0.86515$ & $0.8738\pm0.0128$\\
$\sin^2\theta_{12}$ & $0.3041$ [Eq.~(\ref{eq:solar})] & $0.3036\pm0.0064$\\
$\sin^2\theta_{23}$ & $0.4652$ & $0.470^{+0.017}_{-0.014}$\\
$\delta_{CP}$ & $\pi$ & $(212^{+26}_{-36})^\circ$\\
$\Delta m^2_{21}/\Delta m^2_{31}$ & $m_\mu/2m_\tau=0.02944$ & $0.02945\pm0.00044$\\
$m_1$ & $0.25$~meV & $\lesssim2$~meV (identity)\\
$\Sigma m_\nu$ & $0.0589\pm0.0003$~eV & $<0.064$~eV (DESI), $<0.055$~eV (with SPT-3G), both in $\Lambda$CDM\\
$m_{\beta\beta}$ & $\{1.3,1.6,3.5,3.9\}$~meV & below current reach\\
\end{tabular}
\end{ruledtabular}
\end{table}

\section{Why this is not a coincidence}
\label{sec:evidence}

Four relations hold at the per-mille level or better, and they
carry the argument.
\begin{itemize}
\item The lepton anchors of Eq.~(\ref{eq:anchor}) fix the base
through their ratio, Eq.~(\ref{eq:lepbase}), and then test the form
value $(2\theta_3)^4=20.95$ at the lattice-matched modulus to
$0.02\%$, at $-0.1\sigma$ and $+1.0\sigma$; the value coincides with
$\e^{-29/16}$ to $8\times10^{-5}$, so the test is of the number, and
the form-value reading rests on the recurrence of $\theta_3$ and
$\Imt$ across the postulates (Appendix~\ref{app:stats}).
\item The bottom-tau relation P1, $m_b/m_\tau=(2\e)^{-1/2}$, holds
at $0.55\%$ precision and $+0.4\sigma$.
\item The Cabibbo target P2, $|V_{us}|^2m_\mu/m_e=2/\e$, holds at
$0.7\%$ and $-0.5\sigma$.
\item The top anchor P4, $y_t=(\Imt)^{-1/2}$, holds at $0.4\%$ and
$-0.1\sigma$.
\end{itemize}
The four draw on disjoint measured inputs, the charged leptons, $m_b$,
$|V_{us}|$, and $m_t$, and compare each with a function of the one
base.  On the declared dictionary of Appendix~\ref{app:stats} their
one-standard-deviation budgets are $0.2\%$, $6\%$, $8\%$, and
$5\%$ (Appendix~\ref{app:stats}), whose product is of order
$10^{-6}$.  That product is the chance probability of the four
landings at the stated forms; the number of ratios examined before
these four were singled out is not assessed here, and the product
is offered as the scale of
the coincidence, not as a significance.
Everything at percent precision in the sections above is
subordinate to these four, and the tests of Sec.~\ref{sec:tests}
are what will decide whether the rest is structure.

\subsection{The symmetry is selected, not chosen}

Every modular flavor model chooses its finite group.  Here the
choice is motivated by a measured pattern.  The two to one pattern
with its mirrored ordering, read as charge counting with a uniform
shift between the axes, is realized at level four and at no level
that is not a multiple of four, Eq.~(\ref{eq:select}), and it lands on
level four with
the generations in the triplet and the mirror in the sign singlet.
No parameter enters the selection, and the group that emerges is the
one whose natural mixing textures, the trimaximal patterns, the
lepton data then require in Sec.~\ref{sec:leptons}.  The realized
assignment of Appendix~\ref{app:assignment} does not use the
wrap-around, as Sec.~\ref{sec:pattern} records, so the selection
motivates level four without deriving it, and the
evidence for the construction rests on what follows.

\subsection{The accounting discipline}

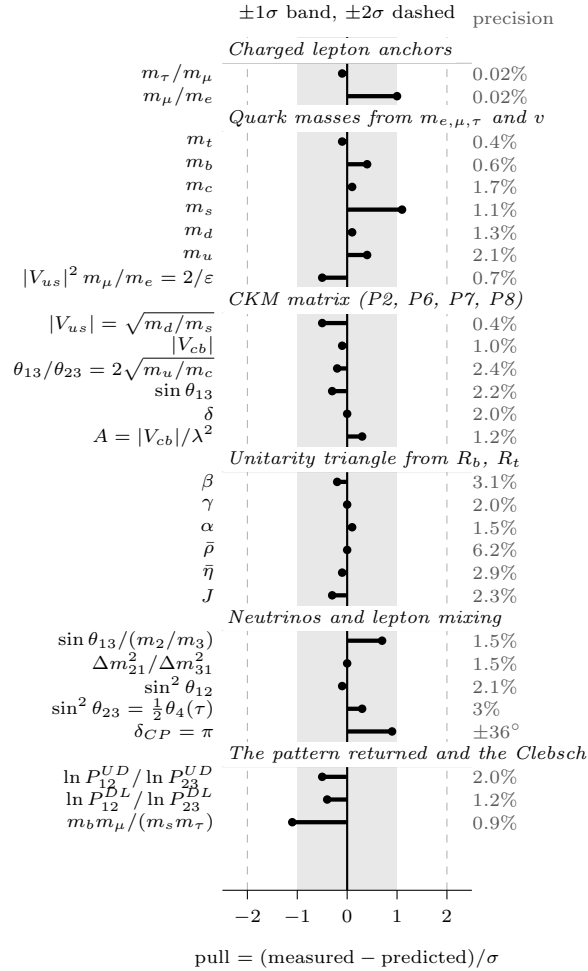
\begin{figure}[t]
\centering
\begin{tikzpicture}[x=0.66cm,y=-0.30cm,
  lab/.style={font=\scriptsize,anchor=east,inner sep=1pt},
  prec/.style={font=\scriptsize,anchor=west,inner sep=1pt,gray!70!black},
  grp/.style={font=\scriptsize\itshape,anchor=west,inner sep=1.5pt,fill=white},
  bar/.style={line width=1.6pt}]
\def\N{37}
\fill[gray!18] (-1,-0.6) rectangle (1,\N);
\draw[gray!60,dashed] (-2,-0.6) -- (-2,\N);
\draw[gray!60,dashed] ( 2,-0.6) -- ( 2,\N);
\draw[thick] (0,-0.6) -- (0,\N);
\draw (-2.5,\N) -- (2.5,\N);
\foreach \x in {-2,-1,0,1,2}{
  \draw (\x,\N) -- (\x,\N+0.4);
  \node[font=\scriptsize,anchor=north] at (\x,\N+0.5) {$\x$};}
\node[font=\scriptsize,anchor=north] at (0,\N+2.2)
  {pull $=(\text{measured}-\text{predicted})/\sigma$};
\node[font=\scriptsize,anchor=south] at (0,-0.9) {$\pm1\sigma$ band, $\pm2\sigma$ dashed};
\node[prec,anchor=south west] at (2.45,-0.9) {precision};
\newcommand{\row}[4]{%
  \node[lab] at (-2.6,#1) {#2};
  \draw[bar] (0,#1) -- (#3,#1);
  \fill (#3,#1) circle (1.6pt);
  \node[prec] at (2.45,#1) {#4};}
\newcommand{\grp}[2]{\draw[gray!50] (-2.5,#1+0.55) -- (2.5,#1+0.55);
  \node[grp] at (-2.5,#1) {#2};}
\grp{0}{Charged lepton anchors}
\row{1}{$m_\tau/m_\mu$}{-0.1}{$0.02\%$}
\row{2}{$m_\mu/m_e$}{1.0}{$0.02\%$}
\grp{3}{Quark masses from $m_{e,\mu,\tau}$ and $v$}
\row{4}{$m_t$}{-0.1}{$0.4\%$}
\row{5}{$m_b$}{0.4}{$0.6\%$}
\row{6}{$m_c$}{0.1}{$1.7\%$}
\row{7}{$m_s$}{1.1}{$1.1\%$}
\row{8}{$m_d$}{0.1}{$1.3\%$}
\row{9}{$m_u$}{0.4}{$2.1\%$}
\row{10}{$|V_{us}|^2\,m_\mu/m_e=2/\e$}{-0.5}{$0.7\%$}
\grp{11}{CKM matrix (P2, P6, P7, P8)}
\row{12}{$|V_{us}|=\sqrt{m_d/m_s}$}{-0.5}{$0.4\%$}
\row{13}{$|V_{cb}|$}{-0.1}{$1.0\%$}
\row{14}{$\theta_{13}/\theta_{23}=2\sqrt{m_u/m_c}$}{-0.2}{$2.4\%$}
\row{15}{$\sin\theta_{13}$}{-0.3}{$2.2\%$}
\row{16}{$\delta$}{0.0}{$2.0\%$}
\row{17}{$A=|V_{cb}|/\lambda^2$}{0.3}{$1.2\%$}
\grp{18}{Unitarity triangle from $R_b$, $R_t$}
\row{19}{$\beta$}{-0.2}{$3.1\%$}
\row{20}{$\gamma$}{0.0}{$2.0\%$}
\row{21}{$\alpha$}{0.1}{$1.5\%$}
\row{22}{$\bar\rho$}{0.0}{$6.2\%$}
\row{23}{$\bar\eta$}{-0.1}{$2.9\%$}
\row{24}{$J$}{-0.3}{$2.3\%$}
\grp{25}{Neutrinos and lepton mixing}
\row{26}{$\sin\theta_{13}/(m_2/m_3)$}{0.7}{$1.5\%$}
\row{27}{$\Delta m^2_{21}/\Delta m^2_{31}$}{0.0}{$1.5\%$}
\row{28}{$\sin^2\theta_{12}$}{-0.1}{$2.1\%$}
\row{29}{$\sin^2\theta_{23}=\tfrac12\theta_4(\tau)$}{0.3}{$3\%$}
\row{30}{$\delta_{CP}=\pi$}{0.9}{$\pm36^\circ$}
\grp{31}{The pattern returned and the Clebsch}
\row{32}{$\ln P^{UD}_{12}/\ln P^{UD}_{23}$}{-0.5}{$2.0\%$}
\row{33}{$\ln P^{DL}_{12}/\ln P^{DL}_{23}$}{-0.4}{$1.2\%$}
\row{34}{$m_bm_\mu/(m_sm_\tau)$}{-1.1}{$0.9\%$}
\end{tikzpicture}
\caption{Twenty-nine comparisons with measurement, with no
continuous parameter.  Each bar runs from zero to the pull,
$(\text{measured}-\text{predicted})/\sigma$, in the sense of the
deviation columns of Tables~\ref{tab:postulates}
and~\ref{tab:masses}.  The band is $\pm1\sigma$, the dashed lines
$\pm2\sigma$, and the column at right gives the relative precision
of each measurement.  The inputs are the three charged-lepton
masses and the electroweak scale.  Entries within a group share
inputs and are correlated, and the mass postulates were
identified against these data, so the figure shows consistency and
overdetermination; it is not a blind test, and the
per-mille entries at the top sit two orders of magnitude below
their accident budgets (Appendix~\ref{app:stats}).}
\label{fig:pulls}
\end{figure}

Every comparison in this paper carries an accident budget.  On the
bare lattice of half-integer powers of $\e$ and small powers of
$2$, a randomly placed value lands within half a standard deviation
of some point about one percent of the time at per-mille precision
and up to tens of percent at percent precision
(Appendix~\ref{app:stats}, which applies the budget postulate by
postulate).  Figure~\ref{fig:pulls} collects every
comparison of the paper as a pull against measurement.  The
per-mille anchors of Eqs.~(\ref{eq:anchor}), (\ref{eq:t1}),
and~(\ref{eq:t2}) sit far below their budgets.  The percent-level
identifications are carried by structure rather than by single
landings, through recurrence, through closure, and through the
control test below.

\subsection{One number, eight determinations}

\begin{figure}[t]
\centering
\begin{tikzpicture}
\begin{axis}[width=8.6cm,height=6.4cm,
  xmin=0.1810, xmax=0.1925, ymin=0.3, ymax=8.7,
  xlabel={$\e$},
  ytick={1,2,3,4,5,6,7,8},
  yticklabels={up hierarchy $m_c/m_u$ (P3),
    up hierarchy $m_t/m_c$ (P3),
    top Yukawa $\Imt=y_t^{-2}$ (P4),
    CKM ratio $|V_{cb}|/|V_{us}|$,
    light down $m_d/m_e$ (P5),
    Cabibbo target $|V_{us}|^2m_\mu/m_e$ (P2),
    bottom $m_b/m_\tau$ (P1),
    charged leptons $(m_em_\tau/m_\mu^2)^{2/3}$},
  ytick pos=left, yticklabel style={font=\scriptsize},
  tick label style={font=\small}, label style={font=\small},
  xticklabel style={/pgf/number format/fixed,
    /pgf/number format/precision=3},
  xtick={0.182,0.184,0.186,0.188,0.190,0.192},
  clip=false]
\addplot[thick] coordinates {(0.186667,0.3) (0.186667,8.7)};
\node[font=\scriptsize,anchor=south,inner sep=1pt]
  at (axis cs:0.186667,8.7) {$\e=14/75$};
\addplot[dashed,gray] coordinates {(0.1875,0.3) (0.1875,8.7)};
\node[font=\scriptsize,gray,anchor=south west,inner sep=1pt]
  at (axis cs:0.1875,0.35) {$3/16$};
\addplot[dashed,gray] coordinates {(0.185185,0.3) (0.185185,8.7)};
\node[font=\scriptsize,gray,anchor=south east,inner sep=1pt]
  at (axis cs:0.185185,0.35) {$5/27$};
\addplot[only marks,black,mark=*,mark size=1.7pt,
  error bars/.cd,x dir=both,x explicit,error bar style={black}]
  coordinates {
  (0.186643,8) +- (0.000040,0)
  (0.185860,7) +- (0.002078,0)
  (0.187365,6) +- (0.001332,0)
  (0.186466,5) +- (0.001854,0)
  (0.186270,4) +- (0.001863,0)
  (0.186405,3) +- (0.002591,0)
  (0.186769,2) +- (0.000846,0)
  (0.186988,1) +- (0.000918,0)};
\node[font=\scriptsize,anchor=west] at (axis cs:0.1872,8) {$\pm0.02\%$};
\node[font=\scriptsize,anchor=west] at (axis cs:0.1885,7) {$\pm1.1\%$};
\node[font=\scriptsize,anchor=west] at (axis cs:0.1892,6) {$\pm0.7\%$};
\node[font=\scriptsize,anchor=west] at (axis cs:0.1888,5) {$\pm1.0\%$};
\node[font=\scriptsize,anchor=west] at (axis cs:0.1886,4) {$\pm1.0\%$};
\node[font=\scriptsize,anchor=west] at (axis cs:0.1893,3) {$\pm1.4\%$};
\node[font=\scriptsize,anchor=west] at (axis cs:0.1880,2) {$\pm0.5\%$};
\node[font=\scriptsize,anchor=west] at (axis cs:0.1883,1) {$\pm0.5\%$};
\end{axis}
\end{tikzpicture}
\caption{The base recovered from eight independent relations of
the construction, against the rational value $14/75$ (solid) and
the nearest fractions of small denominator (dashed).  The routes
are named at left, the charged-lepton route of Eq.~(\ref{eq:base})
at the top.  The leading-order forms not shown there are
$\e=(m_\tau/m_b)^{2}/2$ for P1, $\e=2m_e/(|V_{us}|^{2}m_\mu)$ for P2,
and $|q^{1/4}|=\e$ for the top Yukawa coupling P4, with $\Imt$
eliminated by lattice matching in the light down route P5 and the
form value $\theta_3$ entering the two up-sector routes of P3.
Appendix~\ref{app:stats} states how the quark-mass ratio
uncertainties are propagated.  The strange quark supplies no rung
because its relation P5, $m_s=\tfrac12 m_\mu(\Imt)^{1/2}$, contains
no power of $\e$, and it tests the modulus instead.  The routes are
not all mutually independent, since $|V_{us}|$ enters two of them
and $m_c$ two others.}
\label{fig:baseladder}
\end{figure}
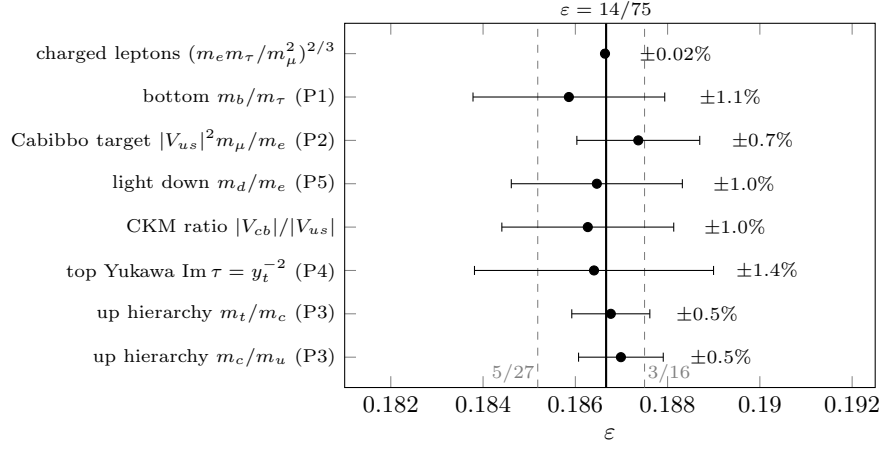

The base is overdetermined.  Table~\ref{tab:base} shows one
rational number recovered from the quark hierarchy, the top Yukawa
coupling, the CKM matrix, and the charged leptons, with the sharpest
route at $0.02\%$, and Fig.~\ref{fig:baseladder} widens the
recovery to eight relations spanning all three charged sectors and
the CKM matrix, every route within $0.6\sigma$ of $14/75$.  A
coincidence would have to repeat itself across these determinations
in every sector at once.

\subsection{The pattern returned}

The closed system also returns the pattern it was built on.  With
the five mass postulates and the lepton anchors, the four
independent double ratios are fixed numbers,
\begin{equation}
  P^{UD}_{12}=\frac{1}{16\sqrt2\,\theta_3^{2}} ,\quad
  P^{UD}_{23}=\frac{1}{4\theta_3^{3}(\Imt)^{1/2}} ,\quad
  P^{DL}_{12}=\frac{2}{\e} ,\quad
  P^{DL}_{23}=\left(\frac{\e\,\Imt}{2}\right)^{1/2} ,
\label{eq:closedP}
\end{equation}
so the two exponent ratios of Eq.~(\ref{eq:harmonic}) come out
$+2.007$ and $-2.057$, against the measured $+1.987\pm0.040$ and
$-2.066\pm0.024$, each within half a standard deviation.  The
lepton-side value is exactly $-2$ with the automorphy factor set
to one, and the whole measured departure from $-2$ is the single
factor $\Imt$ of P5.  Set to exactly $\mp2$, the relations become
polynomial in the masses.  The down-lepton one reads
$m_dm_s=m_em_\mu(m_b/m_\tau)^2$, which the construction dresses by
$\Imt$, and the up-down one reads $m_um_s^3m_t^2=m_c^3m_b^2m_d$,
which returns the top mass from the five lighter quarks,
\begin{equation}
  m_t=m_b\left(\frac{m_c}{m_s}\right)^{3/2}\left(\frac{m_d}{m_u}\right)^{1/2}
  =168.0\pm5.8~\text{GeV} ,
\label{eq:topback}
\end{equation}
against the measured $169.85\pm0.70$~GeV.

Both relations are equalities of geometric means.  The
down-lepton one says that the geometric mean of the two light
masses, in units of the third, is the same for down quarks and
charged leptons,
\begin{equation}
  \frac{\sqrt{m_dm_s}}{m_b}=\frac{\sqrt{m_em_\mu}}{m_\tau} ,
\label{eq:gm1}
\end{equation}
where the measured left side exceeds the right by the factor
$1.038$.  The construction does not predict equality.  P5 dresses
$m_d$ and $m_s$ each with $(\Imt)^{1/2}$, so their product carries
$\Imt$ and the left side of Eq.~(\ref{eq:gm1}) is predicted to
exceed the right by $(\Imt)^{1/2}=1.034$, within $0.4\%$ of the
measured factor.  Equation~(\ref{eq:gm1}) is the Golden Mass
Relation, proposed as a quark-lepton correlation without grand
unification in Refs.~\cite{Morisi:2011pt,King:2013hj} and derived
from $\Gamma_4\cong S_4$ modular symmetry, with calculable deviations
from the exact form, in Ref.~\cite{Chen:2023mwr}.  The present
construction fixes the deviation to the single automorphy factor
$(\Imt)^{1/2}$.  The up-down one says that the second
generation sits at the same displacement from the weighted
geometric mean of the first and third in both quark sectors,
\begin{equation}
  \frac{m_c}{(m_um_t^2)^{1/3}}=\frac{m_s}{(m_dm_b^2)^{1/3}} .
\label{eq:gm2}
\end{equation}

The same weighted mean returns in the neutrino sector, where the
harmonic condition of Eq.~(\ref{eq:m1}) places $m_2$ on
$(m_1m_3^2)^{1/3}$ with no displacement, and the base itself is a
displacement of the same kind, $m_\mu/\sqrt{m_em_\tau}=\e^{-3/4}$
being Eq.~(\ref{eq:base}) read as the position of the muon above the
geometric mean of its neighbors.  The construction does more than
fit the pattern that selected its symmetry.  It predicts the
pattern's departures from exactness.

\subsection{The exponents recur}

The same exponents appear where the rules link quantities.  The
Cabibbo exponent doubles into the down-quark mass ratio through the
Gatto--Sartori--Tonin relation.  The exponent $34$ appears in
$|V_{cb}|$ and, with $\sqrt2$, in $m_\mu/m_\tau$.  The exponent
$60$ is shared by $|V_{ub}|$ and $m_c/m_t$.  The closure exponent
$50$ appears three times with the binary prefactors
$2^{-1},2^{0},2^{+1}$, which carries, on the lattice, the base-free corollary
$m_s/m_b=4\,m_e/m_\mu$ at leading order.  The exact relations of the
construction give
\begin{equation}
  \frac{m_s}{m_b}=4\,\frac{m_e}{m_\mu}\times\frac{\sqrt{2\Imt}}{8\e}
  =0.98\times4\,\frac{m_e}{m_\mu}=0.01859 ,
\label{eq:msmb}
\end{equation}
and the measured ratio, $0.01877\pm0.00023$, lies between the
leading-order $0.01899$ and Eq.~(\ref{eq:msmb}), within one standard
deviation of each.  Recurrence of
this kind is what charge counting produces and noise does not, with
the caveat that the recurrences here are among relations identified
on the same data.

\subsection{The control test}

The framework also says where the lattice must be absent.  The
lepton mixings are set by ratios of theta constants, not by powers
of $\e$, so the nine PMNS magnitudes should not fit the quark
lattice $2^{k/2}\e^{n/18}$.  They do not.  The nearest lattice form
to each magnitude misses it by $0.49\%$ at the median, which is what
nine random numbers at this precision would give, $0.46\%$ on the
coverage of Appendix~\ref{app:stats}; the exponents the fit asks
for are not integers and none repeats; and the nine nearest forms,
assembled into a matrix, are not unitary, their squared norms
summing to $2.987$ where unitarity requires $3$.  The method finds
structure where the model
puts it, in the CKM matrix and the mass ladders, and finds pure
coverage where the model forbids it.  That is the behavior expected
of a charge-counting framework, not of an unconstrained fit.

\subsection{A selection-free closure}

The one relation adopted to close the CKM matrix was proposed by
Grossman and Ruderman in
2020~\cite{Grossman:2020qrp,Grossman:2022ehc}, before the data set
used here, and the
data have since moved from ten-percent to sub-percent agreement;
its companion is now returned by the construction rather
than adopted.  The phase prediction of Eq.~(\ref{eq:delta}) depends
on that provenance.  The mass postulates were identified against the
data and are labeled as inputs throughout, and their protection is
prospective, through the tests of the next section, on data the
identification never saw.

\section{Predictions and tests}
\label{sec:tests}

Every part of the model is testable, and most of the tests come
within the decade.
Table~\ref{tab:tests} lists the predictions and the measurements
that decide them.  Appendix~\ref{app:fits} restates the model as
a fixed point in the standard fit parameterizations for direct
use in global analyses.  Three are absolute.  The mass ordering is
normal.  The lepton Dirac phase is $\pi$, so any established
leptonic CP violation excludes the model.  Nothing appears in
neutrinoless double beta decay above $4$~meV.  One is the pattern
itself.  The two exponent ratios must stay at the closed-system
values $+2.007$ and $-2.057$ of Eq.~(\ref{eq:closedP}), where
improved light-quark masses test the departures from $\pm2$ at the
percent level.

\begin{table}[t]
\caption{Predictions and the measurements that decide them.}
\label{tab:tests}
\begin{ruledtabular}
\begin{tabular}{ll}
Prediction & Decided by\\
\hline
exponent ratios $+2.007$ and $-2.057$ & lattice $m_u$, $m_d$, $m_s$, $m_c$\\
$m_b/m_\tau=(2\e)^{-1/2}$ at per mille & improved $m_b$\\
$m_tm_s/(m_bm_c)=(\e\,\theta_3)^{-1}=5.008$ & lattice $m_c$, $\alpha_s$\\
$y_t^2\,\Imt=1$ at two per mille & top mass at $0.3$~GeV\\
$\gamma=65.2^\circ$, against $67.5^\circ$ for the quantized $\tau_2$ & sub-degree $\gamma$, LHCb Upgrade~II and Belle~II\\
tower vs.\ $\sqrt{m_d/m_s}$, split $0.26\%$ & $|V_{us}|$ at $0.1\%$\\
$|V_{ub}|^2|V_{us}|=|V_{cb}|^4$ returned at $0.6\%$ & $|V_{ub}|$ at $0.3\%$\\
$|V_{ub}|/|V_{td}|=\theta_2(\tau)/2=0.4326$, independent of $\lambda$ and $A$ & $|V_{ub}|$ and $|V_{td}|$ at $1\%$\\
reactor slope $\theta_2^2$ & JUNO splittings with $\theta_{13}$\\
$\sin^2\theta_{13}=\theta_2^2m_\mu/2m_\tau=0.02203$ & reactor $\theta_{13}$\\
$\Delta m^2_{21}/\Delta m^2_{31}=m_\mu/2m_\tau$ & JUNO $\Delta m^2_{21}$\\
normal ordering & JUNO standalone\\
$\delta_{CP}=\pi$, $J=0$ & DUNE and Hyper-Kamiokande\\
$\sin^2\theta_{23}=0.4652$, first octant & DUNE and Hyper-Kamiokande\\
$\sin^2\theta_{12}=0.3041$ & JUNO at $\pm0.002$\\
$\Sigma m_\nu=0.0589$~eV, $\Omega_\nu h^2=6.3\times10^{-4}$, two-sided & DESI, CMB-S4, within an assumed $\Lambda$CDM\\
$m_{\beta\beta}\le3.9$~meV & ton-scale $0\nu\beta\beta$\\
\end{tabular}
\end{ruledtabular}
\end{table}

\section{Ultraviolet completion}
\label{sec:uv}

\subsection{The class is known}

The insertion rules of this paper are the coupling dictionary of
magnetized toroidal compactifications.  In that setting Yukawa couplings are
Jacobi theta constants evaluated at the complex-structure modulus,
matter wavefunctions have half-integral modular weight, and
physical couplings are dressed by powers of
$(2\Imt)^{1/2}$~\cite{Cremades:2004wa}.  The metaplectic flavor
groups derive from the same magnetized
settings~\cite{Almumin:2021fbk}, and the level-four framework with
half-integral weights is developed in Ref.~\cite{Liu:2020msy}.
The completion problem is therefore the selection of a vacuum
within a known class.  What a completion must supply is the flux and
wrapping data that select the eight postulates within the charge
assignment of Appendix~\ref{app:assignment} and the multiplet
structure of Appendix~\ref{app:matrices}, the texture zeros behind the mixings, and
the second torus of Sec.~\ref{sec:cporigin} that carries the phase.
If the base is taken rational, as
the reference value $14/75$ does, the nome $q=(14/75)^4$ is rational
and the modulus $\Imt_1=(2/\pi)\ln(75/14)$ is transcendental,
whereas flux and nonperturbative stabilization select algebraic
values of $\tau$.  Nothing in the data requires rationality
(Sec.~\ref{sec:base}), so a completion is free to fix an algebraic
$\tau$ whose nome lies within $0.02\%$ of $(14/75)^4$; what it must
supply is that value, not its arithmetic character.

A different completion of the same
lattice, a hypercolor sector whose scalar subconstituents are
exchanged along a messenger chain at a confinement scale near
$10^{12}$~GeV, is developed in Ref.~\cite{Barger:2026sub}.  The two
readings share the lattice and the base and differ in what supplies
the coefficients, form values here and wavefunction overlaps
there.

\subsection{Where the phase can originate}
\label{sec:cporigin}

The construction places every form value on the imaginary axis, so
the couplings of Sec.~\ref{sec:rules} are real and the phase that
the closure of Sec.~\ref{sec:ckm} requires must come from the
completion.  The magnetized class says where, and the statement is
quantitative.  On a magnetized torus the Yukawa coupling is a theta
constant whose leading behavior is $\exp(i\pi a^2N\tau)$, with
$N=M_1M_2M_3$ the product of the three fluxes and
$a=(M_2I-M_1J+M_1M_2m)/N$ the theta characteristic, a rational
number fixed by the mode labels $I$ and $J$ and reduced to
$(-\tfrac12,\tfrac12]$~\cite{Cremades:2004wa}, so magnitude and
phase are carried by one exponent.  For any plaquette $P$ of four such
couplings, the double ratio that Sec.~\ref{sec:pattern} makes the
physical object, that exponent is $C=N\sum_\pm a^2$, real because the
$a$ are rational, so $P=e^{i\pi C\tau}$ and
\begin{equation}
  \arg P=-\frac{\Ret}{\Imt}\,\ln|P| \pmod{2\pi} .
\label{eq:phasemag}
\end{equation}

The phase of a rescaling-invariant combination is therefore its
suppression times the shape of the torus, and nothing else.  The
relation holds to leading order in the theta series, with corrections
of relative order $q^{N}$, so it is exact in the regime
$N\,\Imt\gg1$ where the fluxes generate hierarchy at all.
Scherk--Schwarz phases do not evade
this~\cite{Kobayashi:2016ahg}, since they enter each coupling
linearly in the mode indices and cancel from $P$, while
$\Ret$ multiplies $a^2$, whose cross term in $I$ and $J$ survives;
a non-vanishing $\Ret$ is
mandatory~\cite{Kobayashi:2016ahg}.\footnote{The same exponent
carries the charge bookkeeping.  With $q=e^{2\pi i\tau}$ the leading
term is $q^{a^2N/2}$, so the grading of
Appendix~\ref{app:assignment} is $n=4a^2N$, and the values $16$,
$12$, $8$, $6$, $2$, and $0$ of Eq.~(\ref{eq:gradings}) are
statements about $a^2N$ on the flavor torus.}

One torus cannot supply the phase.  The harmonic pattern fixes
$\ln|P|$ for the flavor torus, and the charged-lepton anchors of
Sec.~\ref{sec:base} hold only for $|\Ret|$ below about $0.02$, so
Eq.~(\ref{eq:phasemag}) caps the phase at $0.019\,|\ln|P||$, well
over an order of magnitude below $\delta$.  Equivalently, the phase
would require the plaquette to carry some thirty-six lattice units.

\begin{figure}[t]
\centering
\begin{tikzpicture}[font=\scriptsize]
\begin{scope}[shift={(0,0)},scale=0.95]
  \fill[gray!10] (0,0) -- (1.15,0) -- (1.15,1.229) -- (0,1.229) -- cycle;
  \draw[gray!45] (0,0) -- (1.15,0) -- (1.15,1.229) -- (0,1.229) -- cycle;
  \draw[->,thick,>=stealth] (0,0) -- (1.15,0);
  \draw[->,thick,>=stealth] (0,0) -- (0,1.229);
  \node[below] at (0.575,0) {$1$};
  \node[left] at (0,0.70) {$\tau_1$};
  \node[align=center] at (0.575,-0.90)
    {(a)\ \ $\tau_1=1.0685\,i$\\ rectangular: $z\to z^*$\\
     is a symmetry\\ hierarchy, base $\e$};
\end{scope}
\begin{scope}[shift={(3.05,0)},scale=0.95]
  \fill[gray!10] (0,0) -- (1.15,0) -- (1.55,1.187) -- (0.40,1.187) -- cycle;
  \draw[gray!45] (0,0) -- (1.15,0) -- (1.55,1.187) -- (0.40,1.187) -- cycle;
  \draw[dashed,gray!55] (0,0) -- (-0.40,1.187) -- (0.75,1.187) -- (1.15,0);
  \draw[->,thick,>=stealth] (0,0) -- (1.15,0);
  \draw[->,thick,>=stealth] (0,0) -- (0.40,1.187);
  \node[below] at (0.575,0) {$1$};
  \node[right] at (0.40,0.72) {$\tau_2$};
  \node[gray!70,above left=-1pt] at (-0.40,1.187) {$\tau_2^{*}$};
  \node[align=center] at (0.575,-0.90)
    {(b)\ \ $\tau_2=0.35+1.032\,i$\\ sheared: $z\to z^*$\\
     gives a different lattice\\ phase, $n=2$};
\end{scope}
\end{tikzpicture}

\vspace{1.6em}

\begin{tikzpicture}[scale=2.7,font=\scriptsize]
  \fill[gray!12] (-0.5,0.866) -- (-0.5,1.62) -- (0.5,1.62) -- (0.5,0.866)
    arc[start angle=60, end angle=120, radius=1] -- cycle;
  \draw[->,gray!70] (-0.78,0.62) -- (0.78,0.62) node[right,black] {$\Ret$};
  \draw[->,gray!70] (0,0.62) -- (0,1.80) node[above right=-2pt,black] {$\Imt$};
  \draw[very thick] (-0.5,0.866) -- (-0.5,1.70);
  \draw[very thick] (0.5,0.866) -- (0.5,1.70);
  \draw[very thick] (0.5,0.866) arc[start angle=60, end angle=120, radius=1];
  \draw[very thick] (0,1) -- (0,1.70);
  \node[below] at (-0.5,0.60) {$-\tfrac12$};
  \node[below] at (0.5,0.60) {$+\tfrac12$};
  \fill (0,1) circle (0.016);   \node[left=2pt] at (0,1) {$i$};
  \fill (0.5,0.866) circle (0.016); \node[below right=-1pt] at (0.5,0.866) {$\omega$};
  \fill (0,1.0685) circle (0.026);
  \node[left=4pt] at (0,1.0685) {$\tau_1$};
  \fill (0.35,1.032) circle (0.026);
  \node[above=2pt] at (0.35,1.032) {$\tau_2$};
  \node[align=center] at (0,0.36) {(c)\ \ the fundamental domain};
\end{tikzpicture}
\caption{Where the phase can originate.  (a) The flavor torus is
rectangular, so the reflection $z\to z^*$ that implements CP maps
its lattice to itself and every form value is real.
(b) The second torus is sheared, the
reflected lattice (dashed) is a different one, and the phase of
Eq.~(\ref{eq:phaselaw}) follows, with $n=2$ lattice units
contributed by its plaquette for a free shape ratio (the quantized
alternative of Eq.~(\ref{eq:quantized}) sits at $\tfrac14+1.0685\,i$
with $n=3$).  (c) The same two moduli in the fundamental
domain, whose heavy boundary, the imaginary axis, the lines
$\Ret=\pm\tfrac12$, and the arc $|\tau|=1$, is the CP-conserving
locus.  The anchors of Sec.~\ref{sec:base} hold $\tau_1$ on the axis;
$\tau_2$ must lie off the boundary, and the orbifold fixed points $i$
and $\omega$ are excluded.}
\label{fig:tori}
\end{figure}
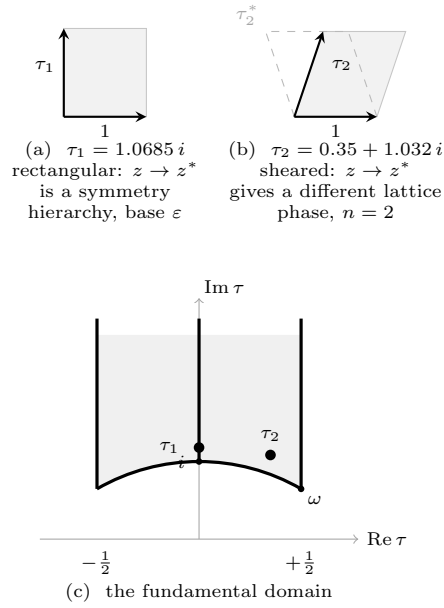

Two tori can.  Let the flavor structure sit on the first torus at
$\tau_1=1.0685\,i$, and let the second contribute a plaquette of $n$
lattice units, $|P_2|=\e^{-n}$, so that the observed double ratios
stay on the lattice with the charges of
Appendix~\ref{app:assignment} shifted by $n$ to compensate.  Writing
$|P_2|$ as a power of the flavor base is itself an assumption, that
the second torus suppresses in the same unit, and the bound $n\ge2$
below is conditional on it.  Then
Eq.~(\ref{eq:phasemag}) reads
\begin{equation}
  \delta=n\,\ln(1/\e)\,\frac{\Ret_2}{\Imt_2}\pmod{2\pi} ,
\label{eq:phaselaw}
\end{equation}
with $\ln(1/\e)=1.6784$, and the phase becomes a lattice statement.
Since $\Ret_2/\Imt_2<1/\sqrt3$ throughout the fundamental domain,
$n=1$ would require the ratio $0.678$ and is excluded, so the second
torus must shift the lattice by at least two units.  Figure~\ref{fig:tori} draws the pair.  For a free shape ratio the
minimal solution sits at $n=2$ with $\Ret_2/\Imt_2=0.339$, giving
$\delta=1.138$ at, for example, $\tau_2=0.35+1.032\,i$.

Three consequences follow, and one limitation.  A second torus that
leaves the magnitudes alone supplies no phase, since $\ln|P_2|=0$
forces $\arg P_2=0$ in Eq.~(\ref{eq:phasemag}), so the phase and the
mass lattice are not independent and the completion must absorb
exactly $n$ units in the charge assignment.  The second torus cannot
be a $\mathbb{Z}_3$, $\mathbb{Z}_4$, or $\mathbb{Z}_6$ orbifold, since
its fixed points $\tau_2=e^{i\pi/3}$ and $\tau_2=i$ lie on the
boundary of the fundamental domain, where the modular symmetry is
enhanced and generalized CP is exact~\cite{NovichkovCP}, so the
physical phase vanishes there by symmetry whatever the leading
plaquette phase of Eq.~(\ref{eq:phasemag}) reads; it is a $T^2$ or a
$T^2/\mathbb{Z}_2$, where $\tau_2$ is not stabilized by the geometry.
And $\tau_2$ must avoid the rest of the CP-conserving boundary,
$\Ret_2=0$, $\Ret_2=\pm\tfrac12$, and $|\tau_2|=1$, which the minimal
solution does at $|\tau_2|=1.090$, while the matter fields carry
zero-mode structure on more than one torus, the flavor-landscape
setting of Ref.~\cite{Abe:2014vfa}.  The limitation is that with $\tau_2$ unstabilized the pair
$(n,\Ret_2/\Imt_2)$ is fitted to $\delta$, not predicted.
What Eq.~(\ref{eq:phaselaw}) supplies is a mechanism, a quantization
law, and the bound $n\ge2$; $\delta$ itself is already a prediction of
Sec.~\ref{sec:ckm}, delivered by four magnitudes and unitarity with no
phase input.

The limitation can be removed by one further postulate, offered
here as a conjecture.  Let the two tori share the base,
$\Imt_2=\Imt_1$, which is what writing $|P_2|=\e^{-n}$ in the same
$\e$ already suggests, and let the second torus sit at the quarter
shift that level four singles out, $\Ret_2=\tfrac14$, under which
each unit of $T$-charge acquires the phase $\pi/2$.  With
$\ln(1/\e)=\tfrac{\pi}{2}\Imt_1$ from Eq.~(\ref{eq:tau}),
Eq.~(\ref{eq:phaselaw}) collapses to
\begin{equation}
  \delta=\frac{n\pi}{8} ,
\label{eq:quantized}
\end{equation}
the phase quantized in the same eighths that quantize the mixing
lattice.  Few values are allowed.  The values $n=1$ and $2$ give
$22.5^\circ$ and $45^\circ$, and $n=4$ gives $90^\circ$ with
$|V_{td}|=0.0101$, all excluded; $n=3$ gives
$\delta=3\pi/8=67.5^\circ$ at $\tau_2=\tfrac14+1.0685\,i$, with
$|\tau_2|=1.097$, interior to the domain and off every
CP-conserving locus.

With the three magnitudes of
Sec.~\ref{sec:ckm} it returns $|V_{td}|=0.00874$, so the
Grossman--Ruderman relation of P7 becomes an output, returned at
$3\%$, not an input, and the triangle comes out
right-angled, $\alpha=90.0^\circ$ and $\beta=22.5^\circ$ with
$\gamma=3\beta$ to $0.2^\circ$, against $\alpha=92.3^\circ$ and
$\gamma-3\beta=-2.4^\circ$ for the closure
[Eq.~(\ref{eq:anglevalues})] and
$\gamma-3\beta=-2.1\pm2.5^\circ$ from the UTfit angles.
Table~\ref{tab:quantized} sets the two readings side by side.  Current data prefer the closure, the
quantized phase sitting at $-1.7\sigma$ from the global-fit
$\delta$, $-1.8\sigma$ from the LHCb combination, $-1.7\sigma$ on
$\alpha$, and $+1.1\sigma$ on $J$, correlated pulls that disfavor it
at roughly the $1.7\sigma$ level without excluding it.  Neither
assumption is derived, the common base being a minimality choice
and the quarter shift being motivated by the level rather than by a
stabilization mechanism, and the two readings differ by $2.3^\circ$
in $\gamma$, which the sub-degree measurements of
Sec.~\ref{sec:ckm} will resolve.

\begin{table}[t]
\caption{The closure of Sec.~\ref{sec:ckm}, with P7 adopted, against
the quantized second torus of Eq.~(\ref{eq:quantized}) at $n=3$,
with P7 returned.  Both use the magnitudes $|V_{us}|$, $|V_{cb}|$,
and $|V_{ub}|$ of Table~\ref{tab:ckm}.  Measured angles and $J$ as in
Table~\ref{tab:triangle}.  The direct $\gamma$ is given as the LHCb
combination~\cite{LHCb:gamma2025} and as the average of direct
determinations in Ref.~\cite{PDG:2026}; $\beta$ is given as the
UTfit angle and as derived from the world average
$\sin2\beta=0.710\pm0.011$~\cite{PDG:2026}; $|V_{td}|$ is the
$\Delta m_d$ determination of Ref.~\cite{PDG:2026}, with its
ratio to $|V_{cb}|^3$ formed from the global-fit $|V_{cb}|$.}
\label{tab:quantized}
\begin{ruledtabular}
\begin{tabular}{llll}
Quantity & Closure & $\delta=3\pi/8$ & Measured\\
\hline
$\delta$ & $65.2^\circ$ & $67.5^\circ$ & $65.3\pm1.3^\circ$\\
$\gamma$ (LHCb) & $65.2^\circ$ & $67.5^\circ$ & $62.8\pm2.6^\circ$\\
$\gamma$ (PDG average) & $65.2^\circ$ & $67.5^\circ$ & $66.4^{+2.7}_{-2.8}{}^\circ$\\
$|V_{td}|$ & $0.00860$ (P7) & $0.00874$ & $0.0086\pm0.0002$~\cite{PDG:2026}\\
$|V_{td}|^2/|V_{cb}|^3$ & $1$ (input) & $1.033$ & $1.01\pm0.07$\\
$\beta$ (UTfit) & $22.5^\circ$ & $22.5^\circ$ & $22.4\pm0.7^\circ$\\
$\beta$ (from $\sin2\beta$) & $22.5^\circ$ & $22.5^\circ$ & $22.6\pm0.5^\circ$\\
$\alpha$ & $92.3^\circ$ & $90.0^\circ$ & $92.4\pm1.4^\circ$\\
$J$ & $3.11\times10^{-5}$ & $3.17\times10^{-5}$ & $(3.09\pm0.07)\times10^{-5}$\\
$\Ret_2/\Imt_2$ & $0.339$ ($n=2$, fitted) & $0.234$ ($n=3$) & \\
\end{tabular}
\end{ruledtabular}
\end{table}

\subsection{The phase of the second torus is the angle \texorpdfstring{$\gamma$}{gamma}}
\label{sec:torusUT}

The phase of Eq.~(\ref{eq:phaselaw}) is a definite angle of the
unitarity triangle.  In the realization of
Appendix~\ref{app:matrices} the one complex entry is $(M_u)_{13}$,
and with the left-handed doublet shared between the two sectors the
texture zeros leave exactly one rephasing invariant,
\begin{equation}
  \Phi=\arg\frac{(M_u)_{13}\,(M_d)_{22}}{(M_u)_{23}\,(M_d)_{12}} ,
\label{eq:invariant}
\end{equation}
a plaquette in the sense of Sec.~\ref{sec:pattern} that spans the
two sectors, of magnitude $0.364$ and phase $-\gamma$.  Its image in
the mixing matrix is the quartet
$-V_{ud}V_{ub}^{*}/(V_{cd}V_{cb}^{*})$, whose magnitude is the side
$R_b$ and whose argument is $\gamma$, so the phase the second torus
supplies is the angle $\gamma$ rather than a
parametrization-dependent $\delta$, and Eq.~(\ref{eq:phaselaw})
reads
\begin{equation}
  \gamma=n\,\ln(1/\e)\,\frac{\Ret_2}{\Imt_2} .
\label{eq:gammatorus}
\end{equation}
The tree-level determination of $\gamma$ in $B\to DK$ is therefore
a measurement of the shape of the second torus at fixed $n$, free of
the loop-level inputs that enter the global fit.  The current LHCb
combination, $\gamma=(62.8\pm2.6)^\circ$~\cite{LHCb:2021dcr,
LHCb:gamma2025}, gives $\Ret_2/\Imt_2=0.327\pm0.014$ at $n=2$, the
combined Belle and Belle~II value,
$\gamma=(75.2\pm7.6)^\circ$~\cite{Belle:2024gamma}, gives
$0.391\pm0.040$, against $0.339\pm0.007$ from the global-fit phase
of Eq.~(\ref{eq:delta}), and sub-degree precision at LHCb
Upgrade~II and Belle~II will fix the ratio to $\pm0.003$.

Two geometric statements follow.  The orientation of the triangle
is the sign of $\Ret_2$.  The CP image of the second torus,
$\tau_2\to-\tau_2^{*}$, is the dashed lattice of
Fig.~\ref{fig:tori}(b), and it reflects the apex through the real
axis, $\bar\eta\to-\bar\eta$, so the measured orientation
$\bar\eta>0$ selects the half of the fundamental domain in which
$\tau_2$ lies.  And the three angles are three readings of the one
shape ratio, since $\alpha$ and $\beta$ follow from $\gamma$ and the
sides.  The direct measurements of $\beta$ in $B\to J/\psi K_S$ and
of $\alpha$ in $B\to\pi\pi$ and $\rho\rho$ test whether the phase
has a single source, which is what one plaquette on one second
torus implies; an established failure of the directly measured
angles to close would require the completion to carry a second
phase-bearing plaquette.

The phase has a torus analogue of the Jarlskog conditions.  On the
invariant quartet of Eq.~(\ref{eq:invariant}) the squares of the
mode labels cancel around the plaquette and the surviving exponent
is a cross term proportional to $(I_1-I_2)$ times the difference of
the up and down right-handed structures on the second torus, so the
phase vanishes if $\Ret_2=0$, if the two left-handed generations
share a mode label there, or if the up and down sectors carry the
same flux, Higgs mode, and mode label on that torus, the
counterparts of $\delta=0$, of degenerate masses, and of $V=1$.
The bound $n\ge2$ is the quantitative form of the last two.

\subsection{Boundary data at any scale}
\label{sec:transport}

The identities hold at $\MZ$, and the construction transports to
any other scale by renormalization-group running, with the scale
dependence confined to a few factors.  The
scale-sensitive content reduces to three calculable numbers, one
universal QCD rescaling of the quark masses and the two top-Yukawa
distortions of $m_t$ and $m_b$, known to five loops; the reason is
that $d\ln P/d\ln\mu$ is a difference of anomalous dimensions in
which the universal piece of each sector cancels within that
sector's own pair of masses, and the only large Yukawa, the top's,
acts non-universally through $Y_uY_u^\dagger$ on the fields that
share the quark doublet with the top, so it distorts $m_t$ and $m_b$
and, at this order, nothing else.  The following relations are inert
under this transport, up to the per-mille electromagnetic running of
the lepton mass ratios: the lepton anchors and the base of
Eqs.~(\ref{eq:anchor}) and~(\ref{eq:lepbase}), the Cabibbo target P2,
the texture ratio P6, the Grossman--Ruderman relation of P7, the
form-value relation P8, the Gatto--Sartori--Tonin relation
Eq.~(\ref{eq:vus}), the light up step $m_c/m_u$ of P3, the eighteenths
tower of Eq.~(\ref{eq:tower}), and the $1$--$2$ double ratios
$P^{UD}_{12}$ and $P^{DL}_{12}$.  These survive at $10^{16}$~GeV
unchanged, and they are the boundary data any completion must
reproduce at its own scale.

Each remaining relation
carries a definite product of the three factors, with P5 the QCD
rescaling alone, the heavy up step $m_t/m_c$ of P3 the $m_t$
distortion alone, P1 and P4 one rescaling and one distortion each,
and the $2$--$3$ double ratios the distortions with the QCD factor
cancelled, the $1$--$3$ ratios drifting with them through
$P_{13}=P_{12}P_{23}$.

A completion may live high or low.  The data favor a low scale in
the specific sense that the transport factors equal one at the
electroweak scale, where every relation takes its exact form, and a
completion at its own scale must reproduce the same relations with
the three factors evaluated there.  The drifting Georgi--Jarlskog plaquette
makes the preference quantitative.  The form value $2/(\e\,\Imt)^{1/2}=3.17$
meets the data at $1.1\sigma$ at $\MZ$, while the same comparison at
$10^{9}$~GeV, where Standard Model transport carries the measured
plaquette to $2.88$ [Eq.~(\ref{eq:plaqrun})], would miss by roughly
ten standard deviations,
so the one scale-sensitive relation localizes the exact form at the
electroweak scale in a way the inert relations cannot.

\subsection{Pati--Salam as the natural gauge embedding}

\begin{figure*}[t]
\centering
\begin{tikzpicture}[
  box/.style={draw,rounded corners,align=center,font=\scriptsize,
    inner sep=4pt},
  arr/.style={->,thick},
  node distance=0.45cm and 0.6cm]
\node[box] (in)
  {\textbf{Inputs.}\quad $m_e,\ m_\mu,\ m_\tau$ at $\MZ$
   ($0.02\%$)\quad and\quad the electroweak scale $v$};
\node[box,below=of in]
  (bm) {base $\e=14/75$\qquad modulus $\tau=1.0685\,i$\qquad
   $\theta_2=0.86515$,\ \ $\theta_3=1.06969$,\ \ $\theta_4=0.93031$};
\node[box,below=of bm,xshift=-4.5cm,text width=7.4cm] (q)
  {\textbf{Quarks.}\ six masses from P1--P5
   ($\chi^2\simeq3$)\ $\to$\ CKM complete\ $\to$\ unitarity
   triangle, $\gamma=65.2^\circ$, $J=3.11\times10^{-5}$};
\node[box,below=of bm,xshift=+4.5cm,text width=7.4cm] (l)
  {\textbf{Leptons.}\ $\sin\theta_{13}=\theta_2(\tau)\,m_2/m_3$\
   $\to$\ harmonic spectrum, $\Sigma m_\nu=0.0589$~eV\ $\to$\
   $\delta_{CP}=\pi$, discrete $m_{\beta\beta}$};
\node[box,below=1.8cm of bm,text width=9.2cm] (ret)
  {the closed matrix returns the base,\quad
   $|V_{cb}||V_{ub}|/(|V_{us}||V_{td}|)\simeq\e^{3/2}=m_em_\tau/m_\mu^{2}$ \ at $0.13\%$};
\draw[arr] (in) -- (bm);
\draw[arr] (bm.south) -- (q.north);
\draw[arr] (bm.south) -- (l.north);
\draw[arr] (q.south) -- (ret.north);
\draw[arr] (l.south) -- (ret.north);
\end{tikzpicture}
\caption{The model end to end.  Three lepton masses fix the base
and the modulus.  The theta constants at that point deliver the
quark spectrum, the CKM matrix, and the neutrino sector, and the
closed matrix returns the base as a charged-lepton combination.
Fig.~\ref{fig:baseladder} shows the same base recovered by eight
independent routes across all three charged sectors and the CKM
matrix.}
\label{fig:flow}
\end{figure*}
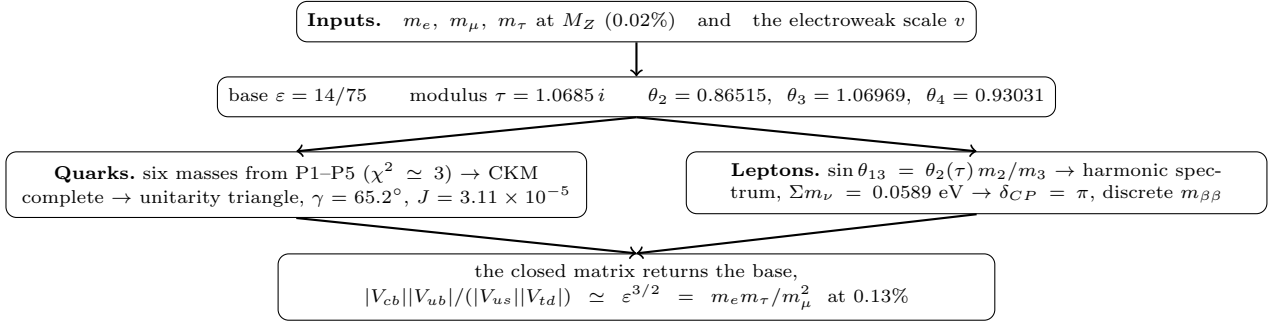

The model is a cross-sector statement that makes no reference to a
gauge group.
One lattice spans both sectors, and the lepton line differs from
the down line by one unit of $\mathbf{1}'$, the construction traced end
to end in Fig.~\ref{fig:flow}.  The equal participation of the
colorless leptons in the exponent lattice, with the same units and the
same precision as the quarks in Eq.~(\ref{eq:harmonic}), points any
dynamical completion at a flavor sector blind to color, and
Pati--Salam unification, with lepton number as the fourth
color~\cite{PatiSalam}, is the minimal gauge structure that makes
such statements natural; Pati--Salam groups are standard targets of
the magnetized settings above.  The extension supplies a multiplet
origin for the $\mathbf{1}'$ twist in the slot of the
Georgi--Jarlskog factors, right-handed neutrinos for the
seesaw~\cite{Minkowski:1977sc,Yanagida:1979as,GellMann:1979kx,
Mohapatra:1979ia}, and a left-right structure that motivates the
symmetric texture behind Eq.~(\ref{eq:vus}).

At the scales the
construction favors the natural mechanism is the inverse
seesaw~\cite{Mohapatra:1986aw,Mohapatra:1986bd}, as in the minimal
low-scale theory~\cite{FileviezPerez:2013zmv,Debnath:2026lfv}, with
the small lepton-number-violating mass carrying the quantized
phases behind Eq.~(\ref{eq:mbb}); the $SU(4)$ scale sits far above
the electroweak scale in any case, bounded by
$K_L\to\mu e$~\cite{FileviezPerez:2013zmv,Butterworth:2025qlu,Debnath:2026lfv},
and no light right-handed $W$ boson is implied.  Pati--Salam embeds
in $SO(10)$ and $E_6$, and nothing here excludes those embeddings;
the paper stays at the minimal group because a full unification
commits the matching to rational Clebsch factors where the
construction returns a form-value ratio, and because Pati--Salam
groups are standard targets of the magnetized class of
Sec.~\ref{sec:cporigin}, though a brane realization of the full
Higgs content the embedding needs is not established.

Two obligations come with the embedding, and both are quantitative.
The first is the Georgi--Jarlskog plaquette.  It measures
$m_bm_\mu/(m_sm_\tau)=3.136\pm0.027$ at $\MZ$, five standard
deviations above the rational Clebsch $3$, and the construction
returns it as the form-value ratio $\bigl(2/(\e\,\Imt)\bigr)^{1/2}$
of Eq.~(\ref{eq:closedP}).  The Clebsch acts at the matching scale,
and only the top-Yukawa distortion of $m_b$ moves the plaquette, the
QCD factor cancelling between $m_b$ and $m_s$ and the lepton ratio
inert (Sec.~\ref{sec:transport}),
\begin{equation}
  \frac{d\ln P}{d\ln\mu}=-\frac{3}{2}\,\frac{y_t^{2}}{16\pi^{2}}
  \ \text{(SM)},\qquad
  +\frac{y_t^{2}}{16\pi^{2}}\ \text{(MSSM)},
\label{eq:plaqrun}
\end{equation}
at one loop, with $P=m_bm_\mu/(m_sm_\tau)$.  Integrated with the
running top Yukawa coupling this carries the plaquette to about
$2.96$ at $10^{6}$~GeV and $2.88$ at $10^{9}$~GeV in the Standard
Model, while with superpartners near $1$~TeV the distortion
reverses sign above the threshold and the plaquette rises instead,
to about $3.17$, $3.25$, and $3.4$ at $10^{6}$, $10^{9}$, and
$2\times10^{16}$~GeV.  The obligation is therefore not that
Pati--Salam avoid the Clebsch but that its matching reproduce the
form-value ratio run to its own scale by Eq.~(\ref{eq:plaqrun}),
with the gauge group supplying representations and the modular
rules the coefficient.

The second is the charged-lepton rotation.  A left-handed
charged-lepton rotation $\theta$ in the $1$--$2$ block shifts the
solar angle of Eq.~(\ref{eq:solar}) by
$2\theta\,U_{12}U_{22}/\cos^2\theta_{13}=0.74\,\theta$, so the
measured $\sin^2\theta_{12}$ bounds it below $0.009$ against a
Cabibbo rotation of $0.2255$, a suppression by a factor of
twenty-five that the Pati--Salam Yukawa structure must deliver; a
minimal bidoublet with $Y_e=Y_d$ is excluded outright, and a lepton
texture of the down-sector form gives
$\theta^e_{12}=\sqrt{m_e/m_\mu}=0.069$, an order of magnitude over
the bound.  The rules supply a mechanism of the required size.
One insertion of $\mathbf{1}'$ shifts the $T$ charge by two, four
eighths on the fine lattice, so the same twist that separates the
lepton line from the down line, placed on the lepton $1$--$2$
transition, gives $\theta^e_{12}\approx\sqrt{m_e/m_\mu}\,\e^{2}=0.0024$
by charge counting, a residual shift of $0.74\times0.0024\approx0.002$
in $\sin^2\theta_{12}$, at the precision JUNO will reach.  Both
obligations are targets for the vacuum selection above.

\subsection{Strong CP}

If the phase the closure of Sec.~\ref{sec:ckm} requires enters
through entry phases proportional to the gradings, the same phases
feed the CKM phase and $\arg\det(Y_uY_d)$.  Every holomorphic factor
$q^{n/8}$ carries the phase $2\pi\rho\,n/8$ and the theta
coefficients are real, so $\bar\theta$ equals $(2\pi\rho/8)\,N_q$
plus a computable series of size $10^{-3}$ to $10^{-2}$, with $N_q$
the total $q$-charge of the two quark determinants, an integer the
infrared description leaves to the completion.  The assignment
$N_q=0$, the metaplectic level-four form of the non-anomalous weight
condition of
Refs.~\cite{Feruglio:2023uqy,Petcov:2024vph,Penedo:2024gtn}, makes
the determinant a real constant and sets $\bar\theta=0$ at tree
level for any displacement, so the quark CP phase and strong CP
conservation coexist without an axion; the realization of
Appendix~\ref{app:matrices} has real determinants by its texture and
is of this kind.  For generic charges a Peccei--Quinn
axion~\cite{Peccei:1977hh,Weinberg:1977ma,Wilczek:1977pj} relaxes
$\bar\theta$ without moving the CKM phase, its scale can ride the
$B{-}L$ breaking that serves the seesaw, and high-quality
realizations on Pati--Salam structure
exist~\cite{DiLuzio:2020ps,DiLuzio:2025vh,Gherghetta:2025ps}.  The
integer $N_q$ therefore decides at once whether the flavor sector
solves strong CP by itself and whether it comes with an axion
dark-matter candidate.

What the vacuum does not supply is a stabilizing symmetry for dark
matter.  Modular models protect a candidate by a $Z_2$ of
the modular weight~\cite{Nomura:2019jxj} or by the residual $Z_2$
of $S$ at $\tau=i$ and $Z_N$ of $T$ at the
cusp~\cite{Kobayashi:2021ajl}.  Here $\tau_1$ sits off the self-dual
point by $u=0.033$, so $Z_2^S$ protection fails at order
$u^2\sim10^{-3}$; the cusp $Z_4^T$ is broken at order $\e$ per unit
of charge, so charge counting would need thirty-five units against
the nine of the deepest operator of Eq.~(\ref{eq:gradings}); and the
weight route is closed because the dictionary of weight-one-half
theta constants fills every weight in half units, so no parity of
weights survives.  Any candidate the completion carries must be
stabilized by a parity the completion preserves.  Pati--Salam
supplies the standard one, $(-1)^{3(B-L)}$, when $B-L$ is broken by
a field of even charge, as a seesaw through a
$(\mathbf{10},\mathbf1,\mathbf3)$ does; an inverse seesaw through a
$(\mathbf4,\mathbf1,\mathbf2)$ breaks it, and what survives is the
matter parity under which the fermions and the singlets are odd and
every Higgs field even, a global symmetry accidental at the
renormalizable level, which a supersymmetric completion imposes as
R-parity with the singlets counted as matter.

\section{Summary}
\label{sec:summary}

The measured masses of the quarks and charged leptons follow a two
to one harmonic pattern with a mirrored ordering.  Read as charge
counting on a clock, the pattern is realized at level four and at no
level that is not a multiple of four, with its mirror the sign singlet of
$S_4$.  The charged-lepton masses fix the base
$\e=14/75$ and the modulus $\tau=1.0685\,i$ at $0.02\%$, and the
theta constants at that point determine the rest.  Four equations make the
argument, the measured pattern of Eq.~(\ref{eq:harmonic}), the
selection it forces in Eq.~(\ref{eq:select}), and the base and
modulus it fixes in Eqs.~(\ref{eq:base}) and~(\ref{eq:tau}).

Eight postulates, one of them empirical and the rest identified on
the data, then deliver the quark sector complete.  Six masses are
reproduced from three lepton masses and the electroweak scale.  The
CKM matrix follows, with every parameter fixed by mass ratios and
form values.  The unitarity triangle is built from the sides
$R_b=0.383$ and $R_t=0.908$ of the closed matrix, with the phase and
the Jarlskog invariant as closure defects, not inputs, the amounts
by which those sides fail to lay a triangle of unit base flat on the
real axis (Sec.~\ref{sec:ckm}).  The same rules deliver the
neutrinos, giving a reactor identity on a parameter-free slope, a
spectrum completed by the harmonic step, a CP-conserving phase, and
discrete Majorana observables.

The base is determined eight ways (Fig.~\ref{fig:baseladder}), the exponents recur across
sectors, the one adopted closing relation has independent
provenance and is equivalent at $0.04\%$ to a half-unit insertion of
the rules, and a
built-in control shows structure where the model puts it
and coverage where it forbids it (Fig.~\ref{fig:pulls}).  The completion belongs to a known
class, with Pati--Salam the natural gauge embedding
and the tree-level angle $\gamma$ a direct measurement of the shape
of the second torus that carries the phase,
and the test schedule runs through this decade.  JUNO, DUNE,
Hyper-Kamiokande, the flavor factories, and cosmology each test part
of the model.

\begin{acknowledgments}
V.B. gratefully acknowledges support from the U.S. Department of
Energy, Office of Science, Office of High Energy Physics, under
Award Number DE-SC0017647 and from the William F. Vilas Estate.
\end{acknowledgments}

\appendix

\section{Operator charges and one leading-order assignment}
\label{app:assignment}

This appendix takes the disclosure of Sec.~\ref{sec:rules} as
far as the postulates permit.  It derives the operator charges that
any modular-invariant realization must reproduce, exhibits the
field assignment those charges force at leading order, and states
what the assignment does not fix.

\subsection{The operator gradings}

Write each mass as a physical coupling and expand it at the
harmonic modulus.  The leading power of the nome, in eighths of the
metaplectic unit $q^{1/8}=\sqrt\e$, defines the grading of the
operator.  Normalizing each column to the third-generation quark
operator of its sector, the nine relations of
Table~\ref{tab:postulates} and Eq.~(\ref{eq:anchor}) give
\begin{equation}
\begin{array}{lccc}
 & \text{gen 1} & \text{gen 2} & \text{gen 3}\\
\hline
\text{up} & 16 & 8 & 0\\
\text{down} & 16 & 8 & 0\\
\text{lepton} & 18 & 8 & 1\\
\end{array}
\label{eq:gradings}
\end{equation}
with the coefficient content of each entry listed in
Fig.~\ref{fig:insertions}.  For example the down-quark entries are
$m_d/m_b=\sqrt2\,(2\theta_3)^8(\Imt)^{1/2}\e^{8}$ and
$m_s/m_b=(\sqrt2/2)(2\theta_3)^4(\Imt)^{1/2}\e^{4}$, verified to
the deviations of Table~\ref{tab:masses}.  Two features of
Eq.~(\ref{eq:gradings}) organize everything below.  The up and
down columns are identical and even, which is why the up-down
double ratios of Eq.~(\ref{eq:closedP}) contain no power of $q$.
The lepton column differs from the down column by $(2,0,1)$,
whose single odd entry is the half unit of P1, and the difference
carries over unchanged to the right-handed charges of
Eq.~(\ref{eq:forced}), since the common left-handed subtraction
leaves it untouched.

\subsection{The forced charges}

The $T$-charge of an operator is its grading modulo eight, since
$T$ acts on $q^{n/8}$ with the phase $e^{2\pi in/8}$.  Charges add
across the fields in an operator, so with the left-handed doublets
$Q$ and $L$ in the triplet $\mathbf{3}$ at the quarter-charges
$(3,1,0)$ of Sec.~\ref{sec:pattern}, that is, at $(6,2,0)$ in
eighths, the right-handed charges are forced by
Eq.~(\ref{eq:gradings}),
\begin{equation}
\begin{aligned}
 u^c,\,d^c&:\ (16,8,0)-(6,2,0)\equiv(2,6,0)\bmod8 ,\\
 e^c&:\ (18,8,1)-(6,2,0)\equiv(4,6,1)\bmod8 .
\end{aligned}
\label{eq:forced}
\end{equation}

In quarter-units the right-handed quarks sit at $(1,3,0)$, the
charge multiset $\{0,1,3\}$ of the triplet $\mathbf{3}$ once more,
with the first two generations interchanged on the clock relative
to the left-handed map.  The right-handed charged leptons sit at
$(2,3,\tfrac12)$.  The first two components are integral and
embeddable in $\mathbf{3}'$, whose multiset is $\{1,2,3\}$, and the
third carries the half-integral charge $\tfrac12$, a $T$-phase of
order eight, which no representation of $\widetilde S_4$ contains,
since $T^4=1$ there~\cite{Liu:2020msy}.  Its home is the
metaplectic cover at level eight, of order
$768$~\cite{Liu:2020msy}, equivalently one component of the
level-four weight-one-half doublet at the halved modulus, with its
partner left to the completion to project or pair.  Every component of an
irreducible representation has charges of one parity, so no
irreducible $e^c$ can serve all three generations.  The reducible
split of $e^c$, with the third generation alone on the metaplectic
cover, is the representation-theoretic content of P1, one
$\theta_2(\tau)$ between the $b$ and $\tau$ operators, and it is
forced by the data, not chosen.

\subsection{What charges do not fix}

The charges determine the gradings only modulo eight, while
Eq.~(\ref{eq:gradings}) uses the unreduced integers.  A charge-zero
operator generically opens at $q^{0}$, and opening at $q$ or $q^{2}$
requires the lower coefficients of its form to vanish, which is a
statement about modular weight, since higher-weight spaces contain
forms of higher vanishing order.  Representative forms with the
correct charge and opening exist at every entry, for example
$\theta_2(\tau)^{16}$ of weight $8$ for the up and down $(1,1)$
operators, $\theta_2(\tau)^{8}$ of weight $4$ for the $(2,2)$
operators, $\theta_2(\tau)^{12}$ and $\theta_2(\tau)^{6}$ for the
first two lepton operators, and $\theta_2(\tau)$ itself, of weight
one half, for the $\tau$ operator.  Their leading coefficients,
powers of $2$, are not the postulated ones, so the coefficients of
Fig.~\ref{fig:insertions}, the $(2\theta_3)^4$ per lepton step and
the $2^{3/2}\theta_3$ and $2^{1/2}\theta_3^{2}$ on the up steps,
are selections within the weight-$k$ spaces, whose dimension
$2k+1$ leaves room for them, and not consequences of the charges.

The automorphy factors of P4 and P5 fix total operator weights of
$-1$ and $+1$ through the dressing of Eq.~(\ref{eq:automorphy}).  The
split of each total weight between the holomorphic form and
the K\"ahler normalization is completion data; the magnetized torus
can supply such splits~\cite{Cremades:2004wa}, and the same weights
may instead be the transport factors that P4 and P5 carry in
Sec.~\ref{sec:transport}, which a completion at its own scale would
reproduce as scale effects rather than as normalizations.
Finally, the CKM tower of Eq.~(\ref{eq:tower}) is quantized in eighteenths
of $\e$, which are not multiples of the charge unit
$\e^{1/2}$, so the mixings are texture conditions, the
symmetric zero behind Eq.~(\ref{eq:vus}) and the closure of
Eq.~(\ref{eq:GR}), that the left-handed assignment must produce as
vanishing entries rather than as charges.

The assignment of this appendix is the unique one at leading order
given $Q$ and $L$ in the triplet.  What remains open is the form
selection at each weight, the K\"ahler split of the automorphy
weights, the partner of the metaplectic $e^c_3$, and the texture
zeros, which together constitute the vacuum-selection problem of
Sec.~\ref{sec:uv}.

\section{Statistical procedures}
\label{app:stats}

Three quantitative claims in the text rest on three Monte Carlo
procedures, the accident budget of Sec.~\ref{sec:evidence}, the
null test of Sec.~\ref{sec:pattern}, and the control test on the
PMNS magnitudes.  This appendix specifies each so that the
look-elsewhere exposure can be assessed.  Every number quoted here
was recomputed for this paper's inputs.

Quark-mass ratio uncertainties, which enter the deviations quoted
throughout and the base determinations of
Fig.~\ref{fig:baseladder}, are propagated with the common
$\alpha_s(\MZ)=0.1180\pm0.0009$ dependence of the running masses
treated as fully correlated and the residual single-mass
uncertainties independent, which is the structure of the sampling
in Ref.~\cite{Antusch:2025rqp}, and each $\alpha_s$ sensitivity is
anchored to that reference as the quadrature difference between its
input and output uncertainties, consistent within the rounding of
the published errors with direct five-loop running.  The
correlation tightens $m_c/m_u$ from $2.7\%$ to $2.0\%$ and leaves
$m_t/m_c$ at $1.8\%$.

\subsection{The form family and the accident budget}

The candidate forms against which a measured value is compared
constitute a fixed dictionary declared in advance.  For mass
ratios it is the lattice $2^{k/2}\,\e^{n/2}$ with integer $n$ and $|k|\le2$.  For mixing magnitudes it is the finer
lattice $2^{k/2}\,\e^{n/18}$ with the same prefactors.  Theta-function
values enter only through the stated insertions and are not
scanned.  The accident budget assesses a match on this dictionary.
Drawing values log-uniformly over the working range and recording
the fraction that lands within half a standard deviation of some
point of the mass family gives $0.6\%$ at per-mille precision,
$1.2\%$ at two per mille, $6\%$ at one percent, and $18\%$ at
three percent.  The per-mille anchors of Eqs.~(\ref{eq:anchor}),
(\ref{eq:t1}), and~(\ref{eq:t2}) therefore sit one to two orders
of magnitude below their budgets, while any single percent-level
match is worth little on its own, which is why the percent-level
identifications in the text are carried by recurrence and closure
rather than by single landings.

The dictionary produces such matches on demand, and two examples
make the budget concrete.  Numerically $|V_{ub}|\approx m_c/m_t$ holds
at $0.9\%$ and $|V_{td}|\approx2m_d/m_c$ at $0.7\%$ with the masses of
Table~\ref{tab:data}, and each reduces to a coefficient sitting near
a power of the base, and the first is the statement
$2^{3/2}\theta_3\approx\e^{-2/3}$, true to $1.2\%$.  Neither is a
relation of the construction, which fixes
$|V_{ub}|/(m_c/m_t)=\e^{-2/3}/(2^{3/2}\theta_3)=1.011$, not
one, so the framework rejects these near-identities at a computable
level that sub-percent $|V_{ub}|$ data can test.  The exact relations
of the text are distinguished from such accidents by recurrence,
closure, and the absence of any residual coefficient.

A second kind of accident is the trade between a power of the base
and a coefficient.  The same measured number admits readings with
different exponents and compensating dictionary factors, and the
data do not choose among them.  The Cabibbo angle is the clearest
case.  It is $\e^{8/9}$ on the tower, $4\sqrt2\,\theta_3^{2}\e^{2}$
from P2 with the anchors, $\e^{-1/9}\,\e$ with the ninth read as a
coefficient, $(4\theta_3^{3})^{1/8}\e=(m_u/m_t)^{1/8}$ to one
percent, and $(m_\mu/m_e)^{1/4}(m_\mu/m_\tau)$ to two tenths of a
percent, and the readings agree because of identities of the base
such as $2^{18}\theta_3^{13}\e^{8}=0.93$ and
$m_\mu^{26}=1.001\,m_e^{9}m_\tau^{17}$, which are coincidences of
$\e=14/75$ of the same kind as $2^{3/2}\theta_3\approx\e^{-2/3}$
above, together with $\theta_3(\tau)=1.001\,\Imt$.  The sharpest of
them concerns the anchors themselves.  At the harmonic modulus
$2\theta_3(\tau)=\e^{-29/64}$ to $2\times10^{-5}$, so the dressing
$(2\theta_3)^{4}=20.949$ of Eq.~(\ref{eq:anchor}) is
$\e^{-29/16}=20.950$ to $8\times10^{-5}$, well inside the $0.02\%$
of the lepton data, and the two anchors can be read as the
sixteenths $m_\mu/m_e=\e^{-51/16}$ and $m_\tau/m_\mu=\e^{-27/16}$,
with no theta constant, at $+1.3\sigma$ and $+0.4\sigma$ against the
$+1.0\sigma$ and $-0.1\sigma$ of the form-value reading.  The anchors
therefore test a number at $0.02\%$ rather than its origin as a form
value.  What selects the form-value reading over a fractional
lattice is recurrence: the same $\theta_3$ enters P3 and P8, and the
modulus the form value fixes enters P4 and P5 as $\Imt$, so the
reading is carried by the closures of Sec.~\ref{sec:evidence}
rather than by the anchor precision alone, and a fractional lattice
that reproduced the anchors would still have to supply those
recurrences.

\subsection{The budget applied postulate by postulate}

Table~\ref{tab:budget} applies the budget to each identification of
the paper at its own precision.  Two dictionaries are used.  The
declared mass family is $2^{k/2}\e^{n/2}$ with $|k|\le2$.  The
form-value family adds the factors the postulates actually use,
$2^{k/2}\theta_3^{b}\theta_4^{c}(\Imt)^{w/2}\e^{n/2}$ with $|k|\le3$,
$0\le b\le3$, $c\in\{0,1\}$, and $w\in\{-1,0,1\}$, one hundred
sixty-eight prefactors on each rung.  For each target the table gives
the number of dictionary elements within one standard deviation of
the measured value and the coverage, the fraction of numbers drawn
log-uniformly over six $e$-folds around the target that land within
one standard deviation of some element.  Two conclusions follow.  On
the declared family the per-mille relations P1, P2, and P4 have
budgets of five to eight percent each, while every percent-level
identification has a budget of order twenty percent.  On the
form-value family, the one from which the coefficients of P1, P3, P5,
and P8 were drawn, the coverage is complete at percent precision, so
no single one of those identifications is significant on its own,
and the solar rotation $\theta_{12}^\nu$, fixed by the data only to
$\pm45\%$, is matched by many elements of either family.  The
evidence for the construction therefore rests on the per-mille
anchors, on the closures that return relations not adopted, and on
the prospective tests of Table~\ref{tab:tests}, not on the
percent-level landings individually.

\begin{table}[t]
\caption{The accident budget applied to each identification.  For
each target, the measured value with its relative uncertainty, and
for each dictionary the number of elements within one standard
deviation of it and the coverage defined in the text.  The anchor
row is the form value $(2\theta_3)^4$
tested by Eq.~(\ref{eq:anchor}); it lies on neither dictionary, and
its coverage entries give the chance that a random value at its
precision would have landed on one.}
\label{tab:budget}
\begin{ruledtabular}
\begin{tabular}{lcccc}
Identification & Measured & \multicolumn{2}{c}{Declared family} & \multicolumn{1}{c}{Form-value family}\\
 & (rel.\ unc.) & hits & coverage & hits / coverage\\
\hline
anchor $(2\theta_3)^4$ & $20.95$ ($0.02\%$) & 0 & $0.2\%$ & 0 / $8\%$\\
P1 $m_b/m_\tau$ & $1.640$ ($0.55\%$) & 1 & $6\%$ & 1 / $87\%$\\
P2 $|V_{us}|^2m_\mu/m_e$ & $10.67$ ($0.7\%$) & 1 & $8\%$ & 3 / $96\%$\\
P3 $m_t/m_c$ & $271.6$ ($1.8\%$) & 0 & $21\%$ & 4 / $100\%$\\
P3 $m_c/m_u$ & $505.5$ ($2.0\%$) & 1 & $24\%$ & 7 / $100\%$\\
P4 $y_t$ & $0.967$ ($0.4\%$) & 0 & $5\%$ & 4 / $79\%$\\
P5 $m_d/m_e$ & $5.545$ ($1.3\%$) & 0 & $15\%$ & 8 / $100\%$\\
P5 $m_s/m_\mu$ & $0.5231$ ($1.1\%$) & 0 & $13\%$ & 4 / $100\%$\\
P8 $|V_{ub}|$ & $0.00370$ ($2.2\%$) & 0 & $26\%$ & 6 / $100\%$\\
$(\Delta m^2_{21}/\Delta m^2_{31})/(m_\mu/m_\tau)$ & $0.500$ ($1.5\%$) & 1 & $18\%$ & 4 / $100\%$\\
$\theta_{12}^\nu$ & $0.015$ ($45\%$) & 5 & $100\%$ & 176 / $100\%$\\
\end{tabular}
\end{ruledtabular}
\end{table}

\subsection{The null test of the harmonic pattern}

Hierarchy-matched unstructured spectra are generated by
multiplying each of the nine measured masses of
Table~\ref{tab:data} by an independent factor drawn log-uniformly
from $[e^{-0.35},e^{0.35}]$, which preserves the hierarchy while
erasing any lattice structure.  For each draw the two exponent
ratios of Eq.~(\ref{eq:harmonic}) are formed, and a draw counts as
a success when both fall at least as close to $+2$ and $-2$ as the
measured values, $|r_{UD}-2|\le0.013$ and $|r_{DL}+2|\le0.066$.
In $4\times10^{5}$ draws the success rate is $0.14\%$, while the
sign reversal alone occurs in $99.9\%$ of draws, so the mirrored
ordering is generic to the hierarchy and the two to one precision
is not.  The rate is sub-percent for any reasonable window,
$0.28\%$ at $\pm0.25$ and $0.06\%$ at $\pm0.50$.  The test assesses
the observed proximity to $\pm2$; it does not assess the choice of
$2$ among the small rational ratios that would have been read as a
pattern had the data landed elsewhere, $1$, $\tfrac32$, or $3$, and
that factor of a few should be borne in mind.

\subsection{The control test on the PMNS magnitudes}

Each of the nine PMNS magnitudes, built from the oscillation
inputs of Sec.~\ref{sec:leptons} with the NuFIT~6.1
normal-ordering angles, the 207-day solar angle, and
$\delta_{CP}=212^\circ$, is fitted with the best form of the
mixing family above.  The coverage baseline quoted in
Sec.~\ref{sec:evidence} is the median best-form residual for
numbers drawn log-uniformly over the span of the magnitudes, and
the unitarity audit sums the squared norms of the nine best forms
against the exact $3$ of any unitary matrix.  The one lepton
quantity the dictionary does match is the reactor identity of
Eq.~(\ref{eq:identity}), where the exponent is one and the residual
coefficient is the form value $\theta_2(\tau)$ that the construction
supplies rather than a number chosen to fit, which is the
distinction the budget is meant to enforce.

\section{A parameterization card for global fits}
\label{app:fits}

The predictions of the construction can be tested inside existing
global analyses without new code, and this appendix restates
them in fit conventions for that purpose.  The model has no
continuous parameter.  Once the three charged-lepton masses at
$\MZ$ are supplied, every entry below is a number, so the
appropriate exercise is a hypothesis test of a fixed point, not
a parameter estimation.  The constants are
\begin{equation}
  \e=\frac{14}{75},\qquad \Imt=1.0685\ \ [\text{Eq.~(\ref{eq:tau})}] ,
\label{eq:fitconst}
\end{equation}
with the theta values $\theta_2=0.86515$, $\theta_3=1.06969$, and
$\theta_4=0.93031$ of Sec.~\ref{sec:base}, and the quark relations
hold for $\MS$ running quantities at $\MZ$.

In the quark sector the card is stated in the standard (PDG)
parameterization.  Writing the two renormalization-stable mass
ratios as $x=\sqrt{m_d/m_s}$ and $y=2\sqrt{m_u/m_c}$, the four
parameters are
\begin{equation}
\begin{aligned}
  \sin\theta_{12}&=x , &
  \sin\theta_{13}&=\tfrac12\theta_3^{2}\,\e^{3} ,\\
  \sin\theta_{23}&=\sin\theta_{13}/y , &
  \cos\delta&=\frac{(s_{12}s_{23})^{2}+(c_{12}c_{23}s_{13})^{2}-s_{23}^{3}}{2\,s_{12}s_{23}c_{12}c_{23}s_{13}} ,
\end{aligned}
\label{eq:quarkcard}
\end{equation}
the closed forms of Eqs.~(\ref{eq:vus}), (\ref{eq:P8main}),
(\ref{eq:vcb}), and~(\ref{eq:GR}).  With the mass relations of
Table~\ref{tab:masses}, $x$ and $y$ are themselves functions of
$\e$, and the equivalent lattice form is the pure-power card of
Eqs.~(\ref{eq:tower}) and~(\ref{eq:wolfenstein}),
$(\sin\theta_{12},\sin\theta_{23},\sin\theta_{13})
=(\e^{8/9},\e^{17/9},\e^{10/3})$ with
$R_b=\e^{5/9}$ and $R_t=\e^{1/18}$.  The two forms agree at the
few-per-mille level, and their split is itself a scheduled test,
the tower against $\sqrt{m_d/m_s}$ at $0.26\%$ in
Table~\ref{tab:tests}.  In the lepton sector, in the PDG
convention with normal ordering,
\begin{equation}
\begin{aligned}
  \sin^{2}\theta_{13}&=\theta_2^{2}\,
    \frac{\Delta m^{2}_{21}}{\Delta m^{2}_{31}} ,\qquad
  \sin^{2}\theta_{23}=\tfrac12\,\theta_4=\tfrac12-\e^{2} ,\\
  \sin^{2}\theta_{12}&=\frac{1-3\sin^{2}\theta_{13}}
    {3\cos^{2}\theta_{13}}
    -2\sin\theta_{12}\cos\theta_{12}\,\e^{5/2} ,
\end{aligned}
\label{eq:leptoncard}
\end{equation}
with $\delta_{CP}=\pi$, Majorana phases quantized at $0$ or
$\pi$, and the spectrum closed by
Eqs.~(\ref{eq:splitratio}) and~(\ref{eq:m1}).  The slope form of the
reactor identity is
exact up to an $m_1^{2}$ correction below the per-mille level.
Table~\ref{tab:fitcard} collects the card with its central
values.

\begin{table}[t]
\caption{The model as a fixed point in fit conventions.  Quark
entries are $\MS$ statements at $\MZ$, and lepton entries assume
normal ordering in the PDG convention.  Central values as in
Tables~\ref{tab:ckm}, \ref{tab:triangle}, and~\ref{tab:leptons}.}
\label{tab:fitcard}
\begin{ruledtabular}
\begin{tabular}{lll}
Fit parameter & Model expression & Value\\
\hline
$\sin\theta_{12}$ & $\sqrt{m_d/m_s}$, lattice $\e^{8/9}$ & $0.2255$\\
$\sin\theta_{23}$ & $\sin\theta_{13}/(2\sqrt{m_u/m_c})$, lattice $\e^{17/9}$ & $0.04197$\\
$\sin\theta_{13}$ & $\tfrac12\theta_3^{2}\,\e^{3}$, lattice $\e^{10/3}$ & $0.003721$\\
$\delta$ & Eq.~(\ref{eq:quarkcard}) & $1.138$\\
$(\lambda,A)$ & $(\e^{8/9},\,\e^{1/9})$ & $(0.2255,\,0.825)$\\
$(\bar\rho,\bar\eta)$ & $R_b(\cos\gamma,\sin\gamma)$, Eq.~(\ref{eq:angles}) & $(0.161,\,0.348)$\\
$J$ & closed matrix & $3.11\times10^{-5}$\\
\hline
$\sin^2\theta_{13}$ & $\theta_2^{2}\,\Delta m^2_{21}/\Delta m^2_{31}$ & slope $0.7485$\\
$\sin^2\theta_{23}$ & $\tfrac12-\e^{2}$ & $0.4652$\\
$\sin^2\theta_{12}$ & Eq.~(\ref{eq:leptoncard}) & $0.3041$\\
$\delta_{CP}$ & CP conserving & $\pi$\\
$\Delta m^2_{21}/\Delta m^2_{31}$ & Eq.~(\ref{eq:splitratio}) & $0.02944$\\
$m_1$ and $\Sigma m_\nu$ & $m_2(m_2/m_3)^2$ & $0.25$~meV, $0.0589$~eV\\
$m_{\beta\beta}$ & phases in $\{0,\pi\}$ & $\{1.3,1.6,3.5,3.9\}$~meV\\
\end{tabular}
\end{ruledtabular}
\end{table}

Three remarks for the fitting groups.  First, the recommended
statistic is the $\Delta\chi^2$ of the fixed point against the
unconstrained minimum in the group's own likelihood, fixing
$(\lambda,A,\bar\rho,\bar\eta)$ or the four standard parameters in
a CKM fit and the four oscillation entries in a neutrino fit, and the
ordering and the octant are absolute discriminants.  Second, the
theory uncertainty of the card is negligible at current
precision.  The lepton masses enter at $0.02\%$, the difference
between $\e=14/75$ and its measured value of Eq.~(\ref{eq:base})
moves entries at the $10^{-4}$ level, and the only spread of
consequence is the few-per-mille split between the mass-ratio and
lattice forms noted above, which sub-percent data will resolve,
not obscure.  Third, the four CKM parameters and the sum
rules P2, P6, P7, and P8, together with Eq.~(\ref{eq:vus}), are
renormalization-inert within the Standard Model
(Sec.~\ref{sec:uv}), so the quark card may be imposed at any
scale, while the mass formulas of Table~\ref{tab:masses} hold at
$\MZ$ with the running of Sec.~\ref{sec:uv}.

\section{Mass matrices that realize the insertions}
\label{app:matrices}

Appendix~\ref{app:assignment} derives the operator gradings and the charges they
force, and stops there. This appendix supplies one pair of quark mass matrices
that carries those gradings, reproduces the predicted spectrum of
Table~\ref{tab:masses} and every entry of Table~\ref{tab:ckm}, and then
separates what the charge assignment already delivers from what the vacuum
selection must supply. The pair is a realization and not a derivation.

\subsection{Setting}
\label{app:setting}

With a single Higgs doublet at $\langle H^0\rangle=v/\sqrt2$, $v=248.40$~GeV, the
$\MS$ value at $\MZ$ of Ref.~\cite{Antusch:2025rqp} used throughout,
the mass matrices are $M_{u,d}=Y_{u,d}(\tau)\,v/\sqrt2$, with each $Y_{ij}$ the
canonically normalized effective coupling of Sec.~\ref{sec:rules}, a product
of theta constants times the dressing $(2\,\Imt)^{w/2}$ of
Eq.~(\ref{eq:automorphy}). In a supersymmetric realization with two doublets,
$v_u=v\sin\beta$ and $v_d=v\cos\beta$ with $\tan\beta=v_u/v_d$ and
$v^2=v_u^2+v_d^2$, the down-sector entries below carry the factor
$v/v_d=1/\cos\beta$ and the up-sector entries the factor
$v/v_u=1/\sin\beta$; the top anchor P4 uses the full $v$, so the numbers quoted
correspond to $v_u\simeq v$, i.e., $\sin\beta\simeq1$, so that the
down-sector Yukawas are enhanced by $1/\cos\beta\simeq\tan\beta$.

\subsection{The matrices}

In GeV,
\begin{widetext}
\begin{equation}
M_d=\begin{pmatrix}
0 & 0.012262 & 0\\
0.011699 & 0.050406 & 0\\
0 & 0 & 2.857
\end{pmatrix},
\qquad
M_u=\begin{pmatrix}
0.0012260 & 0 & 0.6317\,e^{-i\,65.2^{\circ}}\\
0 & 0.62445 & 7.1326\\ 0 & 0 & 169.77
\end{pmatrix},
\label{eq:matrices}
\end{equation}
equivalently, in Yukawa couplings $Y=\sqrt2\,M/v$,
\begin{equation}
Y_d=\begin{pmatrix}
0 & 6.98 & 0\\ 6.66 & 28.70 & 0\\ 0&0&1627
\end{pmatrix}\times10^{-6},
\qquad
Y_u=\begin{pmatrix}
0.00698 & 0 & 3.596\\ 0 & 3.555 & 40.61\\ 0&0&966.5
\end{pmatrix}\times10^{-3}.
\label{eq:yukawas}
\end{equation}
\end{widetext}
Each mixing element has a single source. The Cabibbo angle comes from the down-sector
$1$--$2$ block, $|V_{cb}|$ from $(M_u)_{23}/(M_u)_{33}$, $|V_{ub}|$ from
$(M_u)_{13}/(M_u)_{33}$, and the phase from the argument of that one entry.
Since $M_u$ is triangular and $M_d$ is triangular apart from its
$1$--$2$ block, $\det M_u$ and $\det M_d$ are real, so the texture
sets $\bar\theta=0$ at tree level by itself, the case $N_q=0$ of
Sec.~\ref{sec:uv}.
Table~\ref{tab:matrixoutputs} lists the outputs.

\begin{table*}
\caption{Singular values and mixing parameters of Eq.~(\ref{eq:matrices})
against the predictions of Tables~\ref{tab:masses} and \ref{tab:ckm}.
The last two rows are consequences of the first six, not inputs. Model
values are those of the closure of Sec.~\ref{sec:ckm}, with P8 in place.}
\label{tab:matrixoutputs}
\begin{ruledtabular}
\begin{tabular}{lcc}
Quantity & Eq.~(\ref{eq:matrices}) & Model \\
\hline
$(m_u,m_c,m_t)$ & $(1.226~\mathrm{MeV},\,0.6239,\,169.92~\mathrm{GeV})$ & Table~\ref{tab:masses}\\
$(m_d,m_s,m_b)$ & $(2.701,\,53.11~\mathrm{MeV},\,2.857~\mathrm{GeV})$ & Table~\ref{tab:masses}\\
$\sin\theta_{12}$ & $0.22550$ & $0.2255$\\
$\sin\theta_{23}$ & $0.04198$ & $0.04197$\\
$\sin\theta_{13}$ & $0.003721$ & $0.003721$\\
$\delta$ & $1.1380$ & $1.138$\\
$|V_{td}|$ & $0.008600$ & $0.008599$\\
$J$ & $3.112\times10^{-5}$ & $3.11\times10^{-5}$\\
\end{tabular}
\end{ruledtabular}
\end{table*}

\subsection{What the charges already deliver}

With the left-handed doublets at $(6,2,0)$ eighths and the right-handed quarks
at the minimal lift $(10,6,0)$ of Eq.~(\ref{eq:forced}), the grading
$n_{ij}=l_i+r_j$ predicts the magnitude $\e^{\,n_{ij}/2}$ of each
entry relative to the third-generation operator. Five of the six nonzero magnitudes of Eq.~(\ref{eq:matrices}) land on that
lattice, four with the coefficients of Fig.~\ref{fig:insertions} and one with the
dressing of P8; Table~\ref{tab:latticeaudit} records the audit.

\begin{table*}
\caption{Each entry of Eq.~(\ref{eq:matrices}), normalized to the
third-generation entry of its sector, against the grading
$n_{ij}=l_i+r_j$ in eighths that the charges of Eq.~(\ref{eq:forced})
permit. Only the last row lies off the lattice.}
\label{tab:latticeaudit}
\begin{ruledtabular}
\begin{tabular}{lccl}
Entry & Value$/m_3$ & $n_{ij}$ & Reading \\
\hline
$(M_u)_{11}$ & $4.90\,\e^{8}$ & 16 & $4\theta_3^3\,\e^{8}$, P3\\
$(M_u)_{22}$ & $3.03\,\e^{4}$ & 8 & $2^{3/2}\theta_3\,\e^{4}$, P3\\
$(M_d)_{22}$ & $14.53\,\e^{4}$ & 8 & $\tfrac{1}{\sqrt2}(2\theta_3)^4(\Imt)^{1/2}\e^4$, P5\\
$(M_d)_{12},(M_d)_{21}$ & $101.4\,\e^{6},\;96.8\,\e^{6}$ & 12 & $(2\theta_3)^6(\Imt)^{1/2}\e^{6}$, split by $\pm m_d/2m_s$\\
$(M_u)_{13}$ & $\tfrac12\theta_3^2\,\e^{3}=0.5721\,\e^{3}$ & 6 & neutral dressing, P8\\
$(M_u)_{23}$ & $\lambda\,\e=\e^{17/9}$ & 2 or 10 & off the lattice, see text\\
\end{tabular}
\end{ruledtabular}
\end{table*}

The $1$--$2$ row gives a result. The grading the charges assign to the
$(1,2)$ entry, twelve eighths, is the mean of the two diagonal gradings sixteen
and eight, so the entry sits at the geometric mean of the two diagonal masses
with no further input, and a texture zero at $(M_d)_{11}$ then returns
$m_d=(M_d)_{12}(M_d)_{21}/(M_d)_{22}$ at the value P5 assigns it. The
Gatto--Sartori--Tonin relation~(\ref{eq:vus})
\cite{Gatto:1968ss} is therefore compatible with the charge assignment of
Appendix~\ref{app:assignment} rather than in competition with it, and the single piece of information it adds is the vanishing of the $(1,1)$
entry.

The third-generation entries are of a different kind.  With the lower-left
entries zero, the left rotations of $M_u$ are its upper entries, so
$(M_u)_{23}/m_t=|V_{cb}|$ and $(M_u)_{13}/m_t=|V_{ub}|$ to the precision of
Table~\ref{tab:matrixoutputs}, and the entries sit at
$0.69\sqrt{m_cm_t}$ and $1.38\sqrt{m_um_t}$, not at the geometric
means of the masses they join; $\sqrt{m_c/m_t}=0.061$ is the textbook value
of $|V_{cb}|$ that the data reject by a factor of $1.4$.  The charges say
the same.  A geometric-mean entry has grading $(n_{aa}+n_{bb})/2$, while
charge counting gives $n_{ab}=l_a+r_b$, and the two agree only where
$l-r$ is the same for both generations; with $l-r=(-4,-4,0)$ that holds
for the $1$--$2$ entries, which carry the Gatto--Sartori--Tonin block, and
fails for every entry that involves the third generation, which carry the
mixings $|V_{cb}|$ and $|V_{ub}|$.

The $1$--$3$ row gives a second result, and the grading there is not free.
Since $n_{13}-n_{33}=l_1-l_3$, which the triplet assignment $(3,1,0)$ of
Sec.~\ref{sec:pattern} fixes at six eighths, and $n_{33}=0$ by the third-generation
normalization, the $(1,3)$ grading is six in either quark sector and for any
lift of the right-handed charges. Vanishing orders step by eight, so the
available values are six and fourteen and never seven. The coefficient on $\e^3$ is the neutral dressing
$\tfrac12\theta_3^2=0.5721$ of P8, one power of the charge-zero form per unit of
the step and the inverse of the leading coefficient of $\theta_2$, with the
caveat on dictionary density that Sec.~\ref{sec:ckm} records. The same entry carries
the phase, and an equivalent texture places it in $M_d$ with
the phase shifted by $\pi$; the two give identical CKM matrices and differ only
in the right-handed rotations, which the Pati--Salam completion of Sec.~\ref{sec:uv}
would have to decide.

\subsection{What the vacuum selection must supply}

Three conditions in Eq.~(\ref{eq:matrices}) lie outside the charges. The
coefficients are completion data throughout, as Appendix~\ref{app:assignment} notes, but these three are conditions on magnitudes that no charge assignment supplies.
Two of the zeros follow from the representation content of the
weight spaces and two do not, as the enumeration below shows.

\emph{Texture zeros.} The entries $(M_d)_{11}$ and $(M_u)_{12}$ vanish, as do
the $1$--$3$ and $2$--$3$ entries of the sector that does not carry them. The
first is the Gatto--Sartori--Tonin zero. The second is the sharper condition,
since a $1$--$2$ rotation of the charge-allowed size $\sqrt{m_u/m_c}=0.044$
would move $|V_{us}|$ by twenty percent and its residual must lie below
$10^{-3}$, a suppression of at least a factor of forty below the charge value.

\emph{The $2$--$3$ coefficient.} The ratio $(M_u)_{23}/(M_u)_{33}$ equals
$\lambda\e=\e^{17/9}$, which is the stable Wolfenstein ratio
$A\lambda=\e$ of Eq.~(\ref{eq:wolfenstein}) and is therefore consistent with
the eighteenths tower of Eq.~(\ref{eq:tower}). It is not a multiple of the
charge unit $\e^{1/2}$. Charge counting offers $\e=0.187$ or
$\e^{5}=2\times10^{-4}$, and the measurement falls between them, so the
required coefficient is $\lambda=0.225$, not unity. A scan of the dictionary $2^{k/2}\theta_3^{\,b}\e^{n/2}$ of Appendix~\ref{app:stats}
returns nothing closer than $1.4\%$, a shift of $1.4\sigma$ in $|V_{cb}|$. This is
the one matrix element at which the rules of Sec.~\ref{sec:rules} supply no
coefficient; the closure of Sec.~\ref{sec:ckm} fixes its value through P6 and P8
without explaining it.

\emph{The $1$--$2$ asymmetry.} A symmetric block $[[0,x],[x,y]]$ with singular
values $m_d$ and $m_s$ gives $\tan\theta_{12}=\sqrt{m_d/m_s}$ and hence
$\sin\theta_{12}=0.2200$, while Eq.~(\ref{eq:vus}) is a statement about the
sine and predicts $0.2255$. The difference is $2.4\%$, and $0.2200$ sits $6.4\sigma$ below the
measured $0.2251\pm0.0008$, so a realization that keeps the $(1,1)$ zero
must break the symmetry of the block at the level
$(M_d)_{12}/(M_d)_{21}=1.048\simeq1+m_d/m_s$, which is the asymmetry carried by
Eq.~(\ref{eq:matrices}). The charges permit it, since $n_{12}=n_{21}=12$ fixes
the two gradings but not the two coefficients. An up-sector rotation of
$-0.0055$ would serve equally well.

\subsection{The weight spaces enumerated}
\label{app:enumeration}

The zeros of Eq.~(\ref{eq:matrices}) cannot be obtained from modular
weight, and the enumeration that shows this also fixes what the
finite group can and cannot supply.  Every level-four form is a
polynomial in the weight-one-half doublet $A=\theta_3(2\tau)$,
$B=\theta_2(2\tau)$, on which $T$ acts as $\mathrm{diag}(1,i)$ and
$S$ as $\tfrac{1}{\sqrt2}\bigl(\begin{smallmatrix}1&1\\1&-1
\end{smallmatrix}\bigr)$ up to phase, so the weight-$k$ space is
$\mathrm{Sym}^{2k}$ of the doublet, of dimension $2k+1$, and its
decomposition into irreducible multiplets of the finite group these
two matrices generate is a finite computation.
Table~\ref{tab:weightspaces} lists the content through weight ten
in a unitary basis, each multiplet labeled by its $T$-charges in
units of $q^{1/4}$.  The even weights contain the $S_4$ multiplets of
Table~\ref{tab:charges}, the odd weights the double-cover triplets
$\hat{\mathbf 3}$ with shifted charge sets, and the half-integral
weights the metaplectic doublets and quartets.  The decomposition
reproduces the known cases, the weight-four singlet being
$A^8+14A^4B^4+B^8\propto E_4$ and the weight-two content being
$\mathbf 2\oplus\mathbf 3$.

\begin{table}[t]
\caption{Multiplet content of the level-four forms of weight $k$,
in unitary bases, from the decomposition of $\mathrm{Sym}^{2k}$ of
the weight-one-half doublet.  Unhatted multiplets are those of
$S_4$ with the charges of Table~\ref{tab:charges}; hatted ones
have the charge set shown, in units of $q^{1/4}$.  A prefix $m$
denotes multiplicity.}
\label{tab:weightspaces}
\begin{ruledtabular}
\begin{tabular}{ll|ll}
$k$ & content & $k$ & content\\
\hline
$\tfrac12$ & $\hat{\mathbf 2}_{01}$ &
$5$ & $2\,\hat{\mathbf 3}_{012}\oplus\hat{\mathbf 3}_{023}\oplus\hat{\mathbf 2}_{13}$\\
$1$ & $\hat{\mathbf 3}_{012}$ &
$6$ & $2\,\mathbf 3\oplus\mathbf 3'\oplus\mathbf 2\oplus\mathbf 1\oplus\mathbf 1'$\\
$\tfrac32$ & $\hat{\mathbf 4}_{0123}$ &
$7$ & $2\,\hat{\mathbf 3}_{012}\oplus2\,\hat{\mathbf 3}_{023}\oplus\hat{\mathbf 2}_{13}\oplus\hat{\mathbf 1}_{1}$\\
$2$ & $\mathbf 3\oplus\mathbf 2$ &
$8$ & $2\,\mathbf 3\oplus2\,\mathbf 3'\oplus2\,\mathbf 2\oplus\mathbf 1$\\
$3$ & $\hat{\mathbf 3}_{012}\oplus\hat{\mathbf 3}_{023}\oplus\hat{\mathbf 1}_{1}$ &
$9$ & $3\,\hat{\mathbf 3}_{012}\oplus2\,\hat{\mathbf 3}_{023}\oplus\hat{\mathbf 2}_{13}\oplus\hat{\mathbf 1}_{1}\oplus\hat{\mathbf 1}_{3}$\\
$4$ & $\mathbf 3\oplus\mathbf 3'\oplus\mathbf 2\oplus\mathbf 1$ &
$10$ & $3\,\mathbf 3\oplus2\,\mathbf 3'\oplus2\,\mathbf 2\oplus\mathbf 1\oplus\mathbf 1'$\\
\end{tabular}
\end{ruledtabular}
\end{table}

Two facts about the components follow from the enumeration.  A
component of charge $c$ in a multiplet that occurs once at its
weight opens at $q^{c/4}$ and never higher, since its coefficient
on $A^{2k-c}B^{c}$ is fixed by $S$-covariance and is nonzero; a
charge-zero component always contains $A^{2k}$.  When a multiplet
occurs $m$ times, one direction in the $m$-dimensional space of
couplings reaches $q^{(c+4(m-1))/4}$, at the cost of $m-1$ fixed
ratios among independent invariants.  The triplets $\mathbf 3$ and
$\hat{\mathbf 3}_{012}$ occur with multiplicities $\lfloor k/4\rfloor+1$
and $\lfloor(k+3)/4\rfloor$, so a charge-zero component first
reaches $q^{1}$ at weight five or six and $q^{2}$ at weight nine or
ten.

Against Eq.~(\ref{eq:matrices}), with $Q$ in the triplet of charges
$\{0,1,3\}$ and the right-handed charges of Eq.~(\ref{eq:forced}),
column $j$ of either sector couples through a multiplet with
charges $\{0,2,1\}$, $\{2,0,3\}$, and $\{3,1,0\}$ for $j=1,2,3$, and
the entries have the gradings
\begin{equation}
n=\begin{pmatrix}16&12&6\\ 12&8&2\\ 10&6&0\end{pmatrix}.
\label{eq:gradingmatrix}
\end{equation}

Two zeros are automatic: $(M_u)_{12}=(M_u)_{21}=0$, since charge two
is absent from $\mathbf 3$, and the down $(1,3)$ and $(2,3)$ from a
weight-zero $b^c$.  Three conclusions follow for the rest.
(i) Zeros from weight do not exist.  The
valence bound, $n_{ij}\le4k_{ij}$ for a form of weight $k_{ij}$ to
open at $q^{n_{ij}/8}$, is correct, but an entry that violates it
does not vanish; it opens at the lowest order its charge allows,
since the symmetry fixes gradings only modulo eight.  The three
diagonal quark entries carry charge zero and open at $q^{0}$ in
every multiplet through weight four, and the down $(1,1)$ entry is
the charge-zero component of $\hat{\mathbf 3}_{012}$, present at
every odd weight, so the Gatto--Sartori--Tonin zero is not exact in
this framework and can be pushed only to $q^{m-1}$ by $m-1$ tuned
ratios.  (ii) The first two columns require odd weight, since their
charge sets occur only there, and the third requires even or
half-integral weight; the right-handed quarks are therefore
reducible, as the texture requires, while the two sectors may share
weights as well as charges; what separates them is the Higgs
coupling, $H_u$ against $H_d$, and nothing in the modular
assignment.  (iii) At weight $\tfrac32$ the third up
column is a single quartet with no freedom, and its components
stand in the ratio
$(M_u)_{13}:(M_u)_{23}:(M_u)_{33}=(B/A)^{3}:\sqrt3\,B/A:1=0.052:0.65:1$
at the harmonic modulus, against the $0.0037:0.042:1$ of
Eq.~(\ref{eq:matrices}), up to Clebsch--Gordan factors of order
one; the $2$--$3$ coefficient is not $\lambda$ against unity but
$\lambda$ against $2\sqrt3$.  The matrices of
Eq.~(\ref{eq:matrices}) are therefore a realization the charges
permit, not one they force, as Sec.~\ref{app:setting} states.

\end{document}